\documentclass[ reprint,
showpacs,  
aps,
pra,
floatfix,
]{revtex4-1}
\usepackage{amsmath,amsfonts,amsthm}
\usepackage{dsfont}
\usepackage{braket}
\usepackage{graphicx}
\usepackage{epstopdf}
\usepackage{dcolumn}
\usepackage{bm}
\usepackage{float}
\usepackage[caption=false]{subfig}
\usepackage{color}
\usepackage{float}
\usepackage{hyperref}

\begin{document}

\preprint{APS/123-QED}

\title{Dynamical crossover in the transient quench dynamics of short-range transverse field Ising models}
\author{Ceren B. Da\u{g}}
\email{cbdag@umich.edu}
\author{Kai Sun}
\affiliation{Department of Physics, University of Michigan, Ann Arbor, Michigan 48109, USA}
\date{\today}

\begin{abstract}
Dynamical detection of quantum phases and phase transitions (QPT) in quenched systems with experimentally convenient initial states is a topic of interest from both theoretical and experimental perspectives. Quenched from polarized states, longitudinal magnetization decays exponentially to zero in time for the short-range transverse-field Ising model (TFIM) and hence, has a featureless steady state regime, which prevents it from exhibiting dynamical phase transitions of type-I. In this paper, we ask whether the transient regimes of such non-equilibrium processes probed by single-site observables, that is magnetization per site, could encode information about the underlying QPT. The decay rates of time-dependent and single-site observables exhibit a dynamical crossover that separates two dynamical regions, ordered and disordered, both of which have distinct nonequilibrium responses. We construct a dynamical order parameterlike quantity that exhibits a scaling law in the vicinity of the crossover. Our results reveal that scaling law exponent in short times in the close vicinity of the dynamical crossover is significantly different than the one predicted by analytical theory for long times. When integrability is strongly broken, the crossover boundary turns into a region that separates two other dynamical regions that act like dynamically-ordered and -disordered regimes.
\end{abstract}

\maketitle


\section{Introduction}

Criticality, defined under Landau paradigm \cite{landau1937theory}, is one of the milestones in our understanding of matter, providing us a framework to classify microscopically diverse phenomena in a handful of universality classes with their associated critical exponents \cite{Zinn-Justin:572813,sachdev2001quantum}. Building on this physical principle, the studies of dynamical criticality, phase transitions and crossovers could range from dynamical detection of equilibrium criticality  \cite{PhysRevLett.95.105701,PhysRevLett.99.130402,doi:10.1080/00018732.2010.514702,RevModPhys.83.863,PhysRevLett.115.245301,PhysRevLett.110.135704,Heyl_2018,Mori_2018,PhysRevLett.120.130601,2017Natur.551..601Z,PhysRevLett.121.016801,PhysRevLett.123.140602,PhysRevLett.123.115701,PhysRevB.100.195107,PhysRevB.101.245148}, to non-equilibrium phase transitions that might not necessarily originate from an equilibrium transition \cite{PhysRevLett.103.056403,PhysRevB.88.201110,PhysRevLett.110.136404,PhysRevB.95.024302,Mori_2018}. A commonly applied protocol in some of these studies is a sudden quench, which results in a nontrivial time evolution of either an observable, e.g. an (equilibrium) order parameter (OP) \cite{Mori_2018,PhysRevLett.120.130601,PhysRevLett.121.016801,PhysRevLett.123.115701}, or Loschmidt echo \cite{Mori_2018,PhysRevLett.110.135704,Heyl_2018,PhysRevLett.120.130601,PhysRevB.96.134427} when the system is quenched from an initial state that is not an eigenstate of the evolution Hamiltonian. A popular choice of initial state in the current works on quench dynamics is a polarized state, due to its relevant convenience to prepare in quantum simulators \cite{Mori_2018,2017Natur.551..601Z,PhysRevLett.120.130601,PhysRevB.96.134427,PhysRevLett.121.016801,PhysRevLett.123.115701}. Dynamical phase transitions of type-I (DPT-I) is defined when the quench dynamics equilibrate either to a thermal or a prethermal value in long times, and hence long-time average of the time-dependent observable could act like a dynamical OP, demonstrating a phase boundary. Although DPT-I is well-defined for magnetization of the long-range transverse field Ising model (TFIM) \cite{PhysRevB.95.024302,PhysRevLett.120.130601}, there is no persistent dynamic order for short-range TFIM, simply because the steady state regime of one-point observables, and likewise two-time correlators, is featureless \cite{PhysRevLett.109.247206,Calabrese_2012,PhysRevLett.106.227203}. The featureless steady-state for magnetization originates from the fact that this observable decays exponentially in time as both analytically and numerically studied in the integrable TFIM \cite{PhysRevLett.78.2220,PhysRevLett.96.136801,PhysRevLett.102.127204,PhysRevLett.106.227203,PhysRevLett.109.247206,Calabrese_2012}. In fact exponential decay is also shown to exist in the XXZ model for magnetization \cite{PhysRevLett.102.130603}. Therefore, one cannot dynamically detect equilibrium quantum phase transitions (QPT) of short-range TFIM quenched from polarized states by focusing on the steady-state regime of the magnetization dynamics.

Recently higher order observables are shown to exhibit steady-state regimes with a persistent dynamic order in the quench dynamics of short-range TFIM \cite{PhysRevLett.121.016801,PhysRevLett.123.115701}. Ref.~\cite{PhysRevLett.121.016801} proposed measuring out-of-time-order correlators (OTOC) of an arbitrary single-site observable (longitudinal magnetization per site) both for integrable and nonintegrable short-range TFIM to access such steady-state regimes. Later Ref.~\cite{PhysRevLett.123.115701} showed that two-point nearest-neighbor correlators (averaged over space) could signal a dynamical phase transition in short-range TFIM, albeit the dynamical critical point shifts from the equilibrium QPT to favor disorder when integrability is broken.

Motivated by the recent research interests in finding dynamical probes of equilibrium QPT in short-range Hamiltonians \cite{PhysRevE.96.022110,PhysRevLett.121.016801,PhysRevLett.123.115701}, in this paper we ask whether the transient regimes of short-range TFIM quenched from polarized states and probed by single-site local observables, that is magnetization per site, could encode information about the underlying equilibrium QPT. We stress that we focus on transient regimes of dynamics and single-site observables, instead of steady-state regimes and global observables. Let us first note that transient probes of QPT could prove useful in laboratory implementations of quantum many-body systems, given that it might be challenging to reach steady-state regimes in experimental setups that are naturally coupled to an environment and experiences decoherence \cite{PhysRevA.100.013622}. Both Refs.~\cite{PhysRevA.100.013622} and \cite{PhysRevLett.124.043001} utilized transient signatures of the underlying QPT in the experiments on spinor condensates, e.g. the amplitude and time of the first dip of an oscillatory nonequilibrium response. Furthermore, quantum simulators are ideal testbeds to study the properties and potential of single-site observables, which require only minimal resources for measurement with technologies like quantum gas microscope \cite{2009Natur.462...74B}. Ref.~\cite{PhysRevLett.121.016801} demonstrated that OTOC of single-site observables could be useful to probe the equilibrium QPT, however probing OTOCs in laboratory requires sophisticated protocols such as reversing the overall sign of the Hamiltonian to realize backward time evolution \cite{PhysRevA.94.040302,PhysRevA.99.052322}, or equally sophisticated alternative methods \cite{PhysRevA.94.062329,2016arXiv160701801Y,PhysRevX.9.021061}.
 
In our work, we focus on the single-site observables in both open-boundary and closed chains of TFIM. An open-boundary chain is experimentally more relevant, whereas the results for a periodic TFIM could be obtained by utilizing a mapping to noninteracting fermions. Open-boundary chain simulations are performed via time-dependent density-matrix renormalization group (t-DMRG). Via utilizing the representation of noninteracting fermions, we could easily reach hundreds of sites in the integrable TFIM, and compare the crossover dynamics of small and large system sizes. A crossover in integrable TFIM probed by single-site observables was analytically predicted in Ref.~\cite{Calabrese_2012} for large times in the space-time limit. This crossover separates two distinct nonequilibrium responses where the observable decays exponentially without and with oscillations in the dynamically-ordered and -disordered regimes, respectively. Here we reveal that the scaling predicted by the analytic theory for long times ($\beta=1/2$) in the dynamically-ordered regime, significantly changes for short times ($\beta=1$) in the close vicinity of the crossover which coincides with the equilibrium QPT in the integrable TFIM, $h_c=1$. As one moves away from the vicinity of the crossover, the analytically predicted exponent is recovered, which suggests a smooth crossover between short and long time dynamics. In the dynamically-disordered regime, we find that the analytical prediction conjectured in Ref.~\cite{Calabrese_2012} is not the only possible description of the dynamics for short times and small system sizes, e.g. $N=48$ spins. Additionally, the angular frequency has a correction for short times and small system sizes, while we recover the analytically predicted exponent $\delta=1/2$ for long times when we increase the system size to $N=192$ spins.

We also find that the scaling in the vicinity of the crossover in the dynamically-ordered regime can be described by a logarithmic function regardless of the system size and the temporal regime, e.g. short or long time dynamics. We show that this is consistent with the analytical predictions. Logarithmic form eventually becomes useful in proposing a dynamical OP-like quantity in the vicinity of the crossover. This proposal eases the experimentation of crossover physics discussed in our paper.

We note that the location of the crossover corresponds to the TFIM Hamiltonian that exhibits the fastest decay in the set of all Hamiltonians $H(h)$ across both sides of the equilibrium QPT, in particular for short times. Given that observables cannot show divergent decay in short-range interacting systems due to lightcone bounds, it is reasonable that all decay rates are finite. Hence our data suggests a link between the fastest decay and the equilibrium QPT, confirming Ref.~\cite{PhysRevLett.102.130603}. We use this observation to mark the boundary between dynamically-ordered and crossover regions in the nonintegrable TFIM. We break the integrability by introducing next-nearest neighbor coupling to TFIM and study how the quench dynamics for single-site magnetization behave. After modeling the quench dynamics, we notice that three quantitatively distinct dynamical regimes emerge for the nonintegrable TFIM. The crossover of the integrable TFIM enlarges into a region around the equilibrium QPT and separates two other dynamical regimes which act as -ordered and -disordered regimes of the integrable TFIM. This means that the nonintegrable TFIM exhibits a dominant trend of exponential decay in its dynamically-ordered regime; and a dominant trend of oscillatory exponential decay in its dynamically-disordered regime. We study the relevant decay rate and find that breaking integrability results in a smooth crossover, a minimum, at $h_c=2.278\pm0.001$. The associated scaling exponent of the dynamical order parameterlike quantity reads $\beta\sim 2$, consistent with the smooth crossover of the decay rates.

In Sec.~II, we introduce the models and our methods. Then in Sec.~III and~IV we focus on the dynamical crossover of the integrable and nonintegrable TFIMs, respectively. We conclude in Section V.

\section{Methods}

In this paper, we work with TFIM with both nearest-neighbor (NN) and next-nearest-neighbor (NNN) couplings,
\begin{eqnarray}
H = -J \sum_r \sigma_r^z \sigma_{r+1}^z - \Delta  \sum_r \sigma_r^z \sigma_{r+2}^z + h \sum_r \sigma_r^x, \label{Hamiltonian}
\end{eqnarray}
where $\sigma_r^{\alpha}$ are spin$-\frac{1}{2}$ Pauli spin matrices. TFIM preserves its gapped long range Ising ground state even when the interactions (or nonintegrability) $\Delta$ are introduced, although the transition boundary shifts to favor order as $\Delta$ increases. For all data in the paper, we fix $J=1$ as the energy scale. Specifically we focus on the integrable model $\Delta/J=0$ and nonintegrable model with $\Delta/J=-1$. 

Since an open-boundary chain is more experimentally relevant, we study the open-boundary TFIM with matrix product states (MPS \cite{ITensor}). To reproduce the decay dynamics of an arbitrary site in a periodic chain we focus on the longitudinal magnetization in the middle of the chain $\sigma_{N/2}^z$ (Appendix A). Hence the observable's decay is similar to the decay of total magnetization given that total magnetization is $M=1/N\sum_r\sigma_r^z$. We also study an arbitrary site on a periodic TFIM to utilize the mapping to noninteracting fermions and increase the system size for the integrable TFIM. To calculate single-site dynamics in noninteracting fermions, we make use of the cluster theorem similar to Ref.~\cite{Calabrese_2012}. See Appendix B for the details of the mapping in quench dynamics and the limitations due to cluster theorem. In both open-boundary and closed chains, we focus on the single-site dynamics of Eq.~\eqref{Hamiltonian} quenched from a polarized state $\Ket{\psi_0}=\Ket{\uparrow \uparrow ... \uparrow}$: $C(t)=\Bra{\psi_0}\sigma^z_{i}(t)\Ket{\psi_0}$ where $i=N/2$ for open-boundary chains.

In DPT-I, one studies steady-state regime where the dynamics is expected to become independent of the time. Since such steady-state regimes might exhibit oscillatory behavior, typically due to finite-size effects in small systems, often times averaging over a an interval of time is employed \cite{PhysRevLett.110.135704,PhysRevLett.121.016801,PhysRevLett.123.115701}. Averaging over a long interval of time also makes the dynamic OP to be less sensitive to where a temporal cutoff is applied in the steady-state regime. This is because oscillations could alter the dynamic OP if one only uses the value at the temporal cutoff. As a result, exact location of the temporal cutoff is not significant in the construction of the dynamical OP based on DPT-I as long as the temporal cutoff is in steady-state regime. A valid temporal cutoff that can be utilized in studying DPT-I is a system-size dependent cutoff, $t\sim \alpha N$ where the interval of time-averaging is proportional to the system size \cite{PhysRevLett.120.130601} up to a coefficient $\alpha$.

\begin{figure}
\centering{\includegraphics[width=0.45\textwidth]{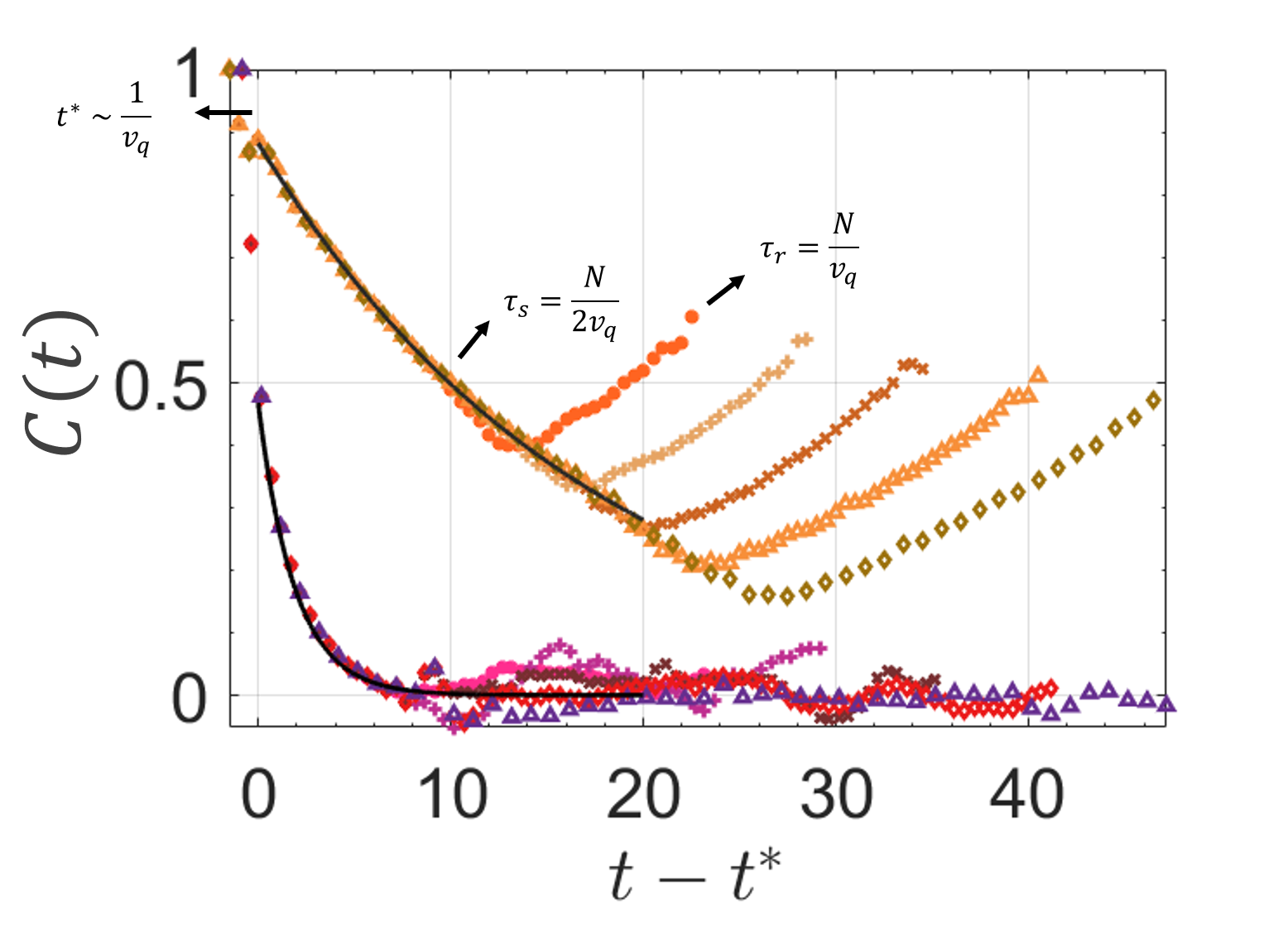}}\hfill 
\caption{$C(t)$ for $h/J=0.5$ upper curves with orange tones and $h/J=0.9$ lower curves with red tones. Each set of curves have system sizes between $N=24$ (dots) and $N=48$ (diamonds) denoted by different markers. $\tau_s$ and $\tau_r$ are separation and revival timescales (see text). x-axis is shifted by $t^*$, the reference time where the exponential decay starts.}
\label{Fig1}
\end{figure}

This temporal cutoff does not work for one-point observables in short-range models, because as already mentioned before, these observables are featureless in their steady state regime, meaning that they decay exponentially to zero. If one were to use a cutoff $t\sim \alpha N$, we would simply observe a vanishing dynamic OP for one-point observables \cite{PhysRevLett.120.130601,PhysRevLett.121.016801} (see Appendix C). This observation aligns with the fact that one cannot construct DPT-I for magnetization in short-range TFIM. Hence, motivated on working in the transient regime, we turn our attention to the decay rates of the initial magnetization, which is known to exhibit a cusplike feature at the QPT for the XXZ model \cite{PhysRevLett.102.130603}. In order to extract the exponential decay in the thermodynamic limit with finite-size systems, which are the only experimentally relevant systems, we find the lightcone bounds \cite{lieb1972,Calabrese_2005,Calabrese_2012} on the magnetization per site for the finite sizes under study. The dynamics that remain in the lightcone exhibit exponential decay and show finite-size effects exponentially suppressed in the system size \cite{2020arXiv200912032W}. Fig.~\ref{Fig1} shows the open-boundary integrable TFIM dynamics for $h/J=0.5$ (orange tones) and $h/J=0.9$ (red tones) for system sizes $N=24:6:48$. In the lightcone, data for different system sizes collapse on each other while each separation point is roughly marked by $\tau_s=N/2v_q$ where $v_q$ is the maximum quasi-particle velocity $v_q=\text{max}|d\epsilon(h,k)/dk|=2J \text{min}(h,1)$ \cite{sachdev2001quantum,Calabrese_2005,Calabrese_2012}. $\tau_s$ is the time for the excitations caused by the quench to reach the end of the chain, and hence $\tau_s$ probes the size of the chain. When the chosen bulk spin is not in the middle of the chain its coefficient changes $\tau_s=a/v_q$ where $N/2 \leq a < N$. Revival timescale is marked by $\tau_r=N/v_q$, which is the time for the excitations to reflect back from the boundary to the middle of the chain. The timescale $t^*$ is the short-distance cutoff of the temporal axis defined by the lattice constant divided by velocity $t^*\sim v_q^{-1}$. Here, $t^*$ ($\tau_s$) serves as the ultraviolet (infrared) cutoff, below (above) which the physics is dominated by non-universal microscopic details (finite-size effects). Thus, we focus on the (intermediate) time range $t^*<t<\tau_s$, where data of different system sizes collapse on each other and universal behavior arises as shown in Fig.~\ref{Fig1} with an exponential decay \cite{PhysRevLett.78.2220,PhysRevLett.96.136801,Calabrese_2012,PhysRevLett.102.127204}. The time interval that remains in the lightcone effectively simulates the decay in the thermodynamic limit.

For our periodic chain results, we are always confined to the intermediate time range due to the application of cluster theorem (Appendix B).

\section{Dynamical Crossover in the integrable TFIM}

Integrable TFIM hosts a crossover at $h_c=1$ that separates two dynamical regimes. In the dynamically-ordered regime, single-site observables exhibit an exponential decay in time, whereas in the dynamically-disordered regime we observe an oscillatory exponential decay. We will systematically study the short-time nonequilibrium response of single-site observables in the integrable TFIM in this section. In the following, we focus on the dynamically-ordered regime.

\subsection{Decay rates}

\begin{figure}
\centering{\includegraphics[width=0.5\textwidth]{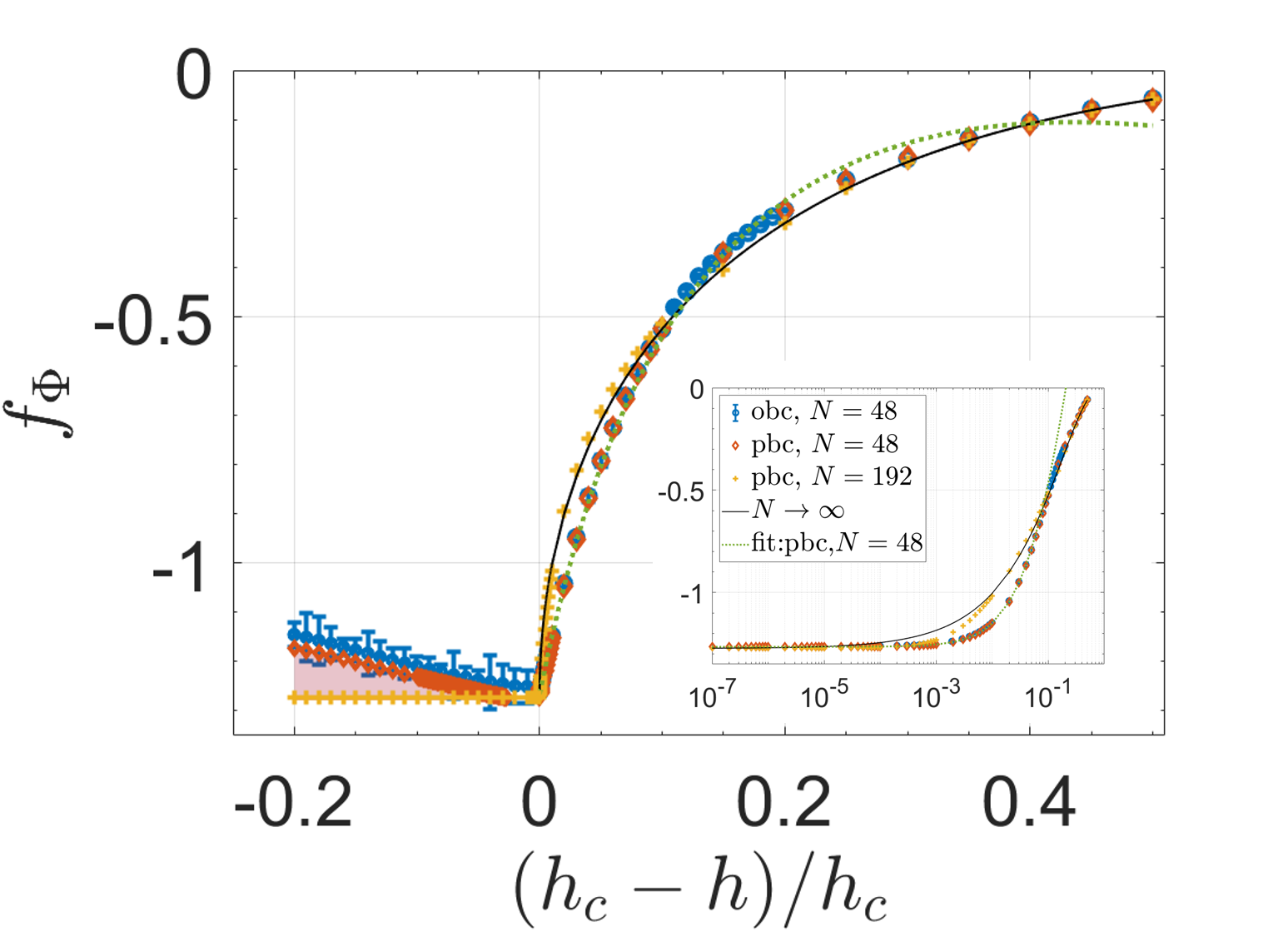}}\hfill 
\caption{Decay rates $f_{\Phi}$ for integrable TFIM in dynamically-ordered $(h_c-h)/h_c > 0$ and -disordered $(h_c-h)/h_c < 0$ regimes. The inset focuses on the dynamically-ordered regime in a semilog plot. Blue-circles, red-diamonds and yellow pluses are data for system sizes $N=48$ with open boundaries (obc), periodic boundaries (pbc) and $N=192$ with periodic boundaries, respectively. Green-dotted line is the logarithmic fit function for the data $N=48$ (pbc) (see text), whereas the black-solid line is the analytic result for the thermodynamic limit. In the disordered regime, the shaded region is the uncertainty for system size $N=48$ due to short time evolution.}
\label{fig3a}
\end{figure}

Bounded by the lightcone, we find the decay rates of magnetization per site around the crossover at $h_c=1$. Fig.~\ref{fig3a} shows how these decay rates $f_{\Phi}$ change with the reduced control parameter $h_n=(h_c-h)/h_c$ for system sizes $N=48$ with both open (blue-circles) and periodic (red-diamonds) boundary conditions and $N=192$ with periodic boundary condition (yellow-pluses). A cusplike feature is observed in Fig.~\ref{fig3a}, similar to the XXZ model in Ref.~\cite{PhysRevLett.102.130603}.
The dynamically-disordered regime will be explained in a following subsection.

We first note that our two methods explained in previous section match perfectly for $N=48$. Thus, one could measure the middle spin in an open-boundary chain and reproduce the results for an arbitrary site in a periodic chain. By increasing the system size to $N=192$, we observe a convergence to the prediction by the analytic theory for the thermodynamic limit (black-solid) for $h_n > 10^{-2}$. Hence, two remarks follow: (i) Although we work with data bounded by the lightcone, the data for small system sizes, e.g. $N=48$ still experiences finite-size effects \cite{2020arXiv200912032W}, because the simulation time is restricted by the system size in the lightcone. (ii) In the close vicinity of the crossover, $h_n < 10^{-2}$, even the large systems, e.g. $N=192$ diverge from the analytic prediction (see the inset in Fig.~\ref{fig3a}).

The analytic prediction is calculated based on the space-time limit derivation given in Ref.~\cite{Calabrese_2012}. For a quench from a polarized state, the asymptotic late-time scaling reads
\begin{eqnarray}
C(t) &\sim & \left(1+\sqrt{1-h^2}\right)^{1/2} \exp \left( t f_{\Phi}^{\infty}(h) \right), \\
f_{\Phi}^{\infty}(h) &=& -\frac{4}{\pi}\bigg(h+\sqrt{1-h^2}\bigg [ \arcsin\left(\sqrt{\frac{1-h}{2}}\right) \notag \\
&-& \arcsin\left(\sqrt{\frac{1+h}{2}}\right) \bigg] \bigg). \label{analytic}
\end{eqnarray}

For the numerics close to crossover, we propose a logarithmic fit function $f_{\Phi}=\log(\gamma h_n^{\beta} \exp(-h_n/\Lambda)+C_0)$ where $\gamma$ and $\beta$ are free parameters to be found and $\Lambda$ is the exponential cutoff coefficient which is explained below. We note that such a model for the decay rate is intuitive and describes the data in a large interval $0 < h_n \lesssim 0.4$, not only in the close vicinity of the crossover. The constant $C_0$ points to the observation that the decay rate is never infinite, however the largest at the crossover. Hence the system thermalizes the quickest at the crossover. Further $C_0$ is not a free parameter, but fixed by the data itself at the crossover. Furthermore, analytical prediction for thermodynamic limit at late times gives $C_0=\exp(-4/\pi)$.
Data follows $\log(\gamma h_n^{\beta}+C_0)$ in the vicinity of the crossover, while introducing an exponential cutoff \cite{doi:10.1137/070710111} to the model lets us describe a bigger region of $h_n$ as well as providing a definition for `vicinity of the crossover', $h_n\ll  \Lambda$. For example, the fit parameters for the decay rates of system size $N=48$ depicted in the main panel of Fig.~\ref{fig3a} (green-dotted line) are $\beta=1.05$ and $\Lambda=0.37$, meaning that the vicinity of the crossover could be defined as $h_n \lesssim 0.03$. Indeed by using the interval of $h_n\lesssim 0.03$, one can precisely determine the scaling exponent as $\beta=1$ in the fit function $\log(\gamma h_n^{\beta}+C_0)$ (green-dotted line in the inset of Fig.~\ref{fig3a}).

We note that the logarithmic function is consistent with the analytical expression Eq.~\eqref{analytic} in the vicinity of the crossover. This can be seen from the series expansions of Eq.~\eqref{analytic} and the logarithmic fit function. The series expansion of Eq.~\eqref{analytic} in the vicinity of the crossover is,
\begin{eqnarray}
f_{\Phi}^{\infty}(h_n \rightarrow 0) \sim -\frac{4}{\pi} +2 \sqrt{2 h_n} - \frac{4 h_n}{\pi} + \cdots, 
\end{eqnarray}
while the series expansion for the logarithmic fit function follows
\begin{eqnarray}
f_{\Phi}(h_n \rightarrow 0) \sim \log(C_0) +\frac{\gamma}{C_0} h_n^{\beta} + \mathcal{O}(h_n^{2\beta}).
\end{eqnarray}
Therefore, in the close vicinity of the crossover the analytic prediction could be written as the logarithmic fit function with the parameters of $C_0=\exp(-4/\pi)$, $\beta=1/2$ and $\gamma/C_0=2\sqrt{2}$, resulting in $\gamma=2\sqrt{2} \exp(-4/\pi)$. In the next subsection, we will see the use of logarithmic fit function in experimentation. However now let us show how it could be helpful in extracting the scaling exponent in the close vicinity of the crossover in the numerical data.

\begin{figure}
\centering{\includegraphics[width=0.5\textwidth]{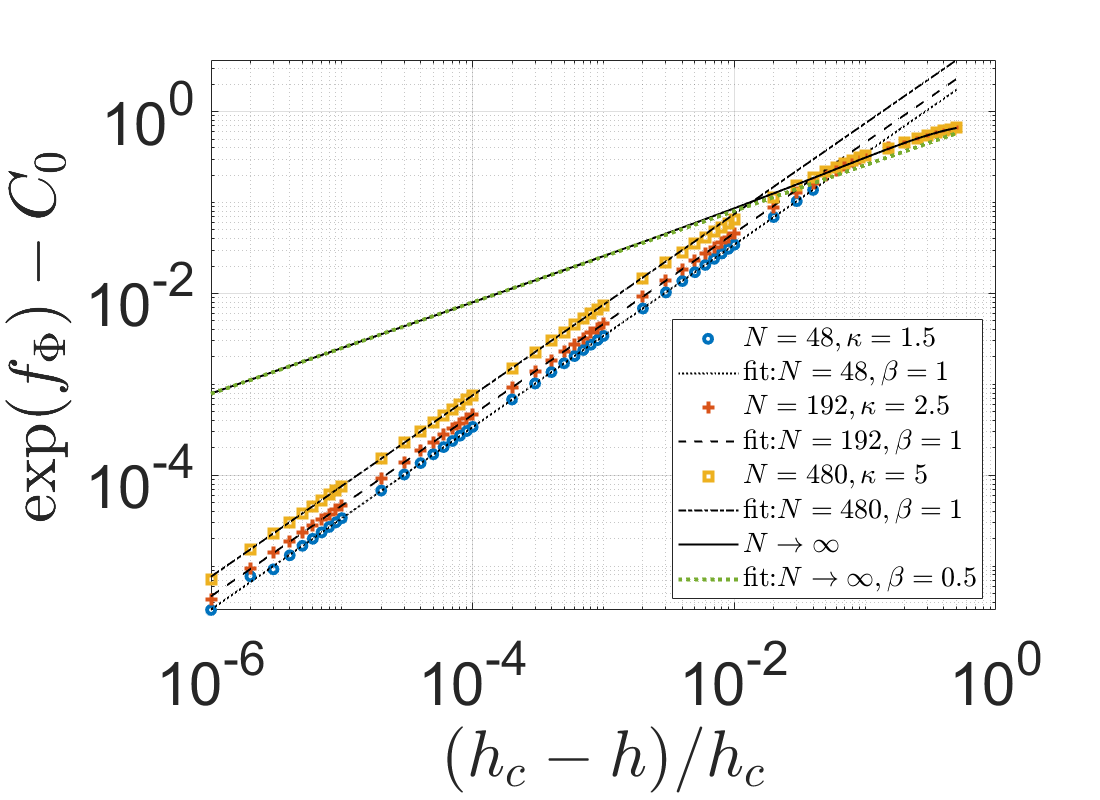}}\hfill 
\caption{Scaling of the decay rate function $\exp(f_{\Phi})-C_0$ for integrable TFIM in dynamically-ordered $(h_c-h)/h_c > 0$ regime. Blue-circles, red-pluses and yellow squares are data for system sizes $N=48$, $N=192$ and $N=480$ all with periodic boundaries, respectively. Black-solid line is the analytic prediction for the thermodynamic limit and late times, whereas the green-dotted line is the fit to the analytic expression in the vicinity of the crossover with $\beta=1/2$. The dotted, dashed and dotted-dashed lines are the fits to the numerical data in the close vicinity of the crossover with $\beta=1$.}
\label{fig3A}
\end{figure}

To extract the scaling exponent in the close vicinity of the crossover, we define a decay rate function $\exp(f_{\Phi}(h_n))-C_0 = \gamma h_n^{\beta}$. Fig.~\ref{fig3A} shows the decay rate function of both the numerical data and the analytical expression (black-solid). The green-dotted line is the fit to the analytical expression in the vicinity of the crossover with the expected scaling exponent of $\beta=1/2$ and coefficient $\gamma=2\sqrt{2} \exp(-4/\pi)$. The blue-circles, red-pluses and yellow-squares depict the data for system sizes $N=48,192,480$ all of which exhibit a scaling exponent of $\beta=1$. Note that we choose the ultraviolet cutoff $t^* = \kappa v_q^{-1}$ where $\kappa$ for each data set is given in the legend. Small coefficient $\kappa$ implies that we focus on early-time behaviour. 

\begin{figure}
\centering{\includegraphics[width=0.5\textwidth]{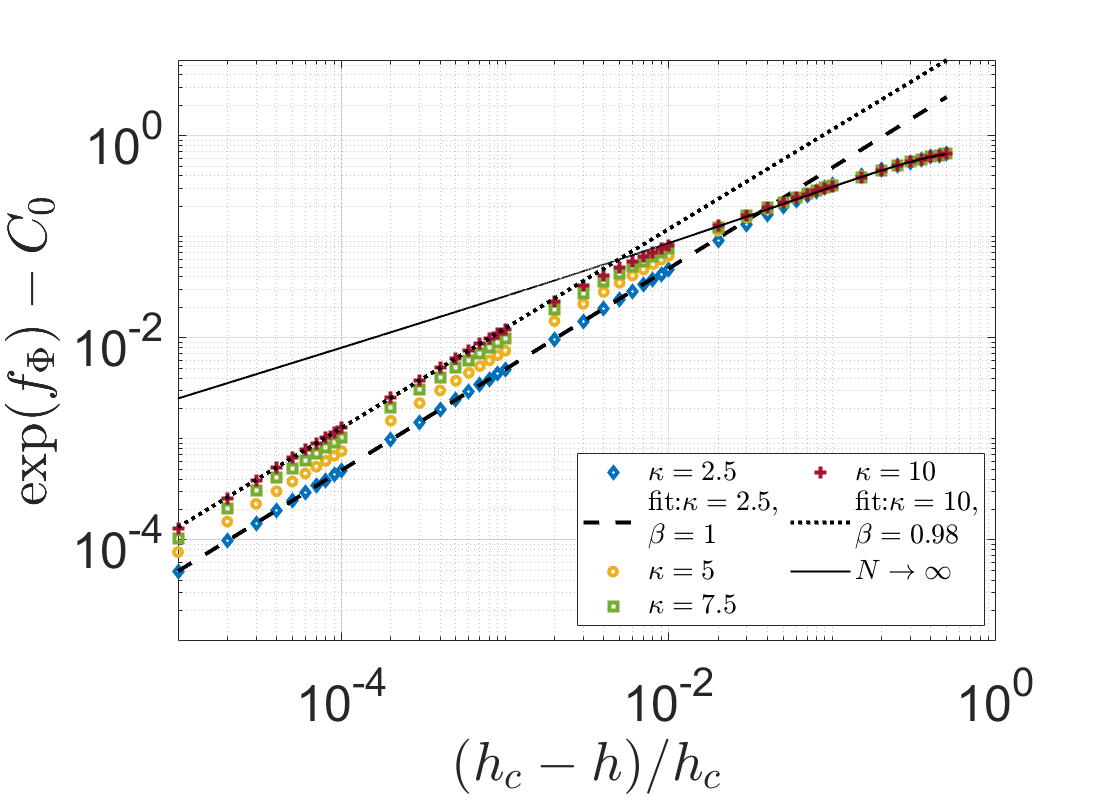}}\hfill 
\caption{Scaling of the decay rate function $\exp(f_{\Phi})-C_0$ for integrable TFIM in dynamically-ordered $(h_c-h)/h_c > 0$ regime for system size $N=480$ for different $\kappa$. Blue-diamonds, yellow-circles, green squares and red-pluses stand for $\kappa=2.5,5,7.5,10$, respectively. Black-solid line is the analytic prediction for the thermodynamic limit for comparison with the fits (dotted and dashed lines).}
\label{fig4A}
\end{figure}

As a result, regardless of system size we observe that early-time scaling exponent $\beta=1$ is significantly different than the late-time scaling exponent of $\beta=1/2$ in the close vicinity of the crossover. As we move further away from the vicinity of the crossover, the decay rate function at any system size converges to the prediction by analytical expression. Hence we observe a smooth crossover between different scaling exponents in Fig.~\ref{fig3A}, whose exact location depends on $\kappa$. To visualize the dependence on $\kappa$, we plot Fig.~\ref{fig4A} where the system size is fixed to $N=480$ for different $\kappa$. As we increase $\kappa$, we move the location of the crossover between analytical late-time and numerical early-time behaviors, to smaller $h_n$.

\begin{figure}
\centering
\subfloat[]{\label{fig5A}\includegraphics[width=0.24\textwidth]{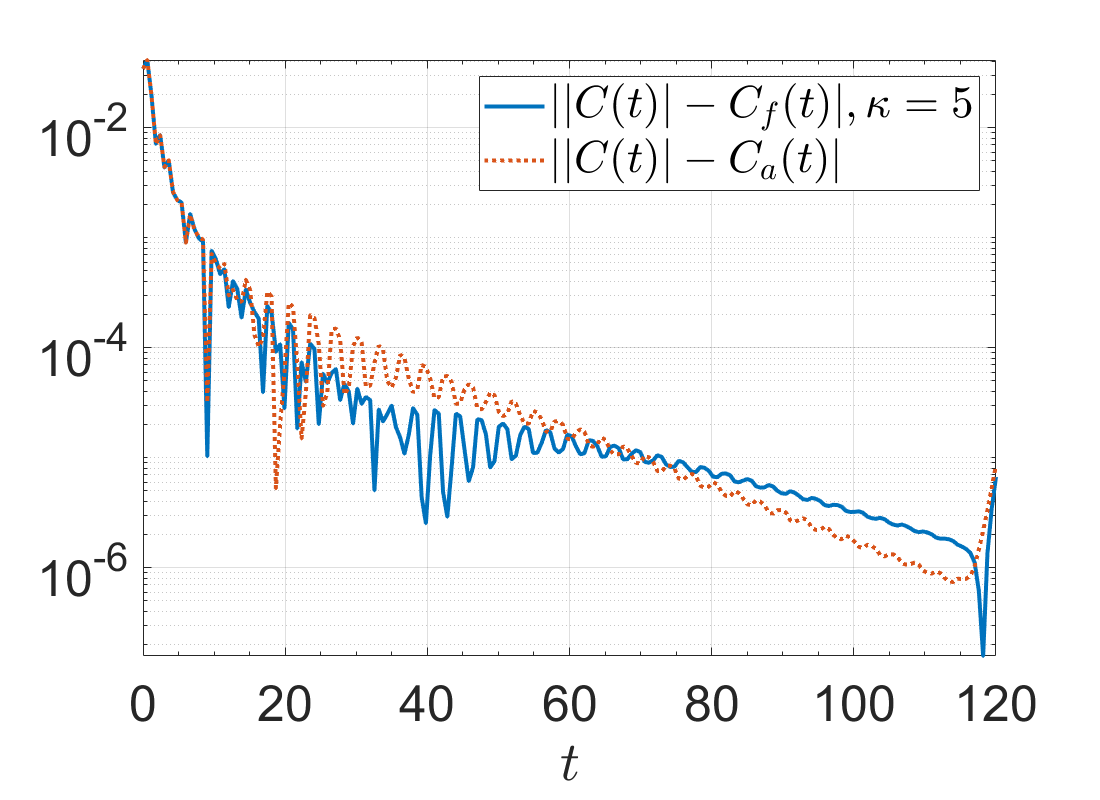}}
\hfill 
\subfloat[]{\label{fig5B}\includegraphics[width=0.24\textwidth]{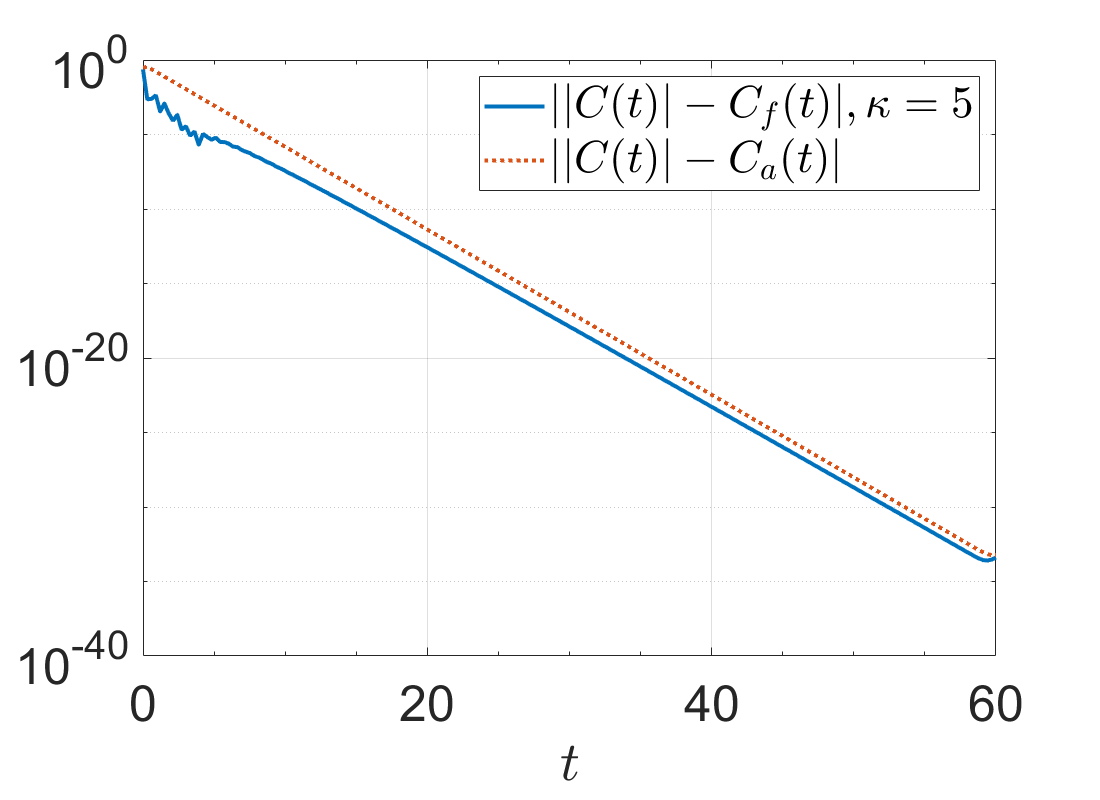}}\hfill 
\caption{The error functions, $|C(t)-C_f(t)|$ (solid) and $|C(t)-C_a(t)|$ for the fit function and the analytical expression (dotted), respectively, for a system size of $N=480$ at (a) $h=0.5$ and (b) $ h=9.9\times 10^{-6}$. The fit function is calculated with $\kappa=5$.}
\end{figure}

We observe that the numerical data mostly follows analytical prediction when $h_n$ is sufficiently away from the crossover, resulting in a nonequilibrium response where early-time behaviour does not really differ from the analytical prediction. Hence, one can probe analytical prediction by observing early-time behaviour. Fig.~\ref{fig5A} shows the difference between data and its fit function $C_f(t)$ which is named as an error function $|C(t)-C_f(t)|$ at $\kappa=5$ for a system size $N=480$, and similarly the difference between the data and its analytical prediction $|C(t)-C_a(t)|$ at $h=0.5$. At early times $t < 20/J$, fit function and the analytical expression are equally successful in predicting the data. In time interval $20/J < t < 60/J$, fit function is slightly better than the analytical expression while for later times $t>60/J$ the opposite is true, as expected. 

\begin{figure*}
\centering 
\subfloat[]{\label{fig6A}\includegraphics[width=0.33\textwidth]{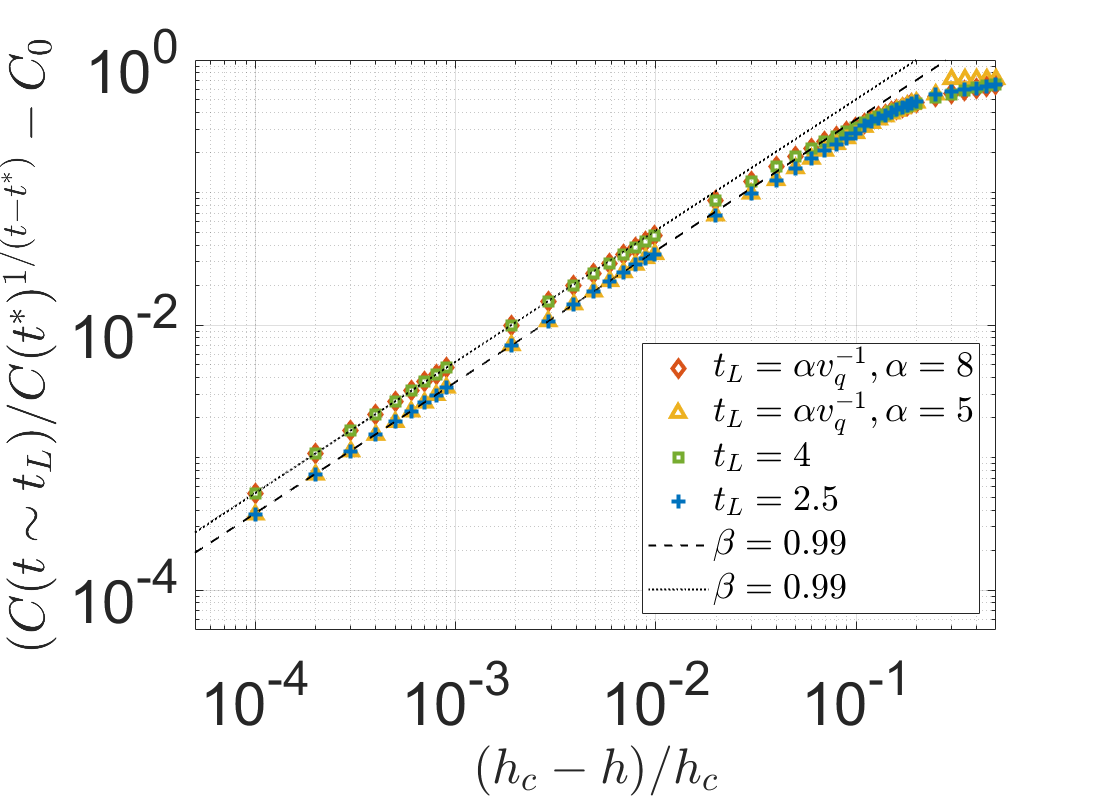}}\hfill
\subfloat[]{\label{fig6B}\includegraphics[width=0.33\textwidth]{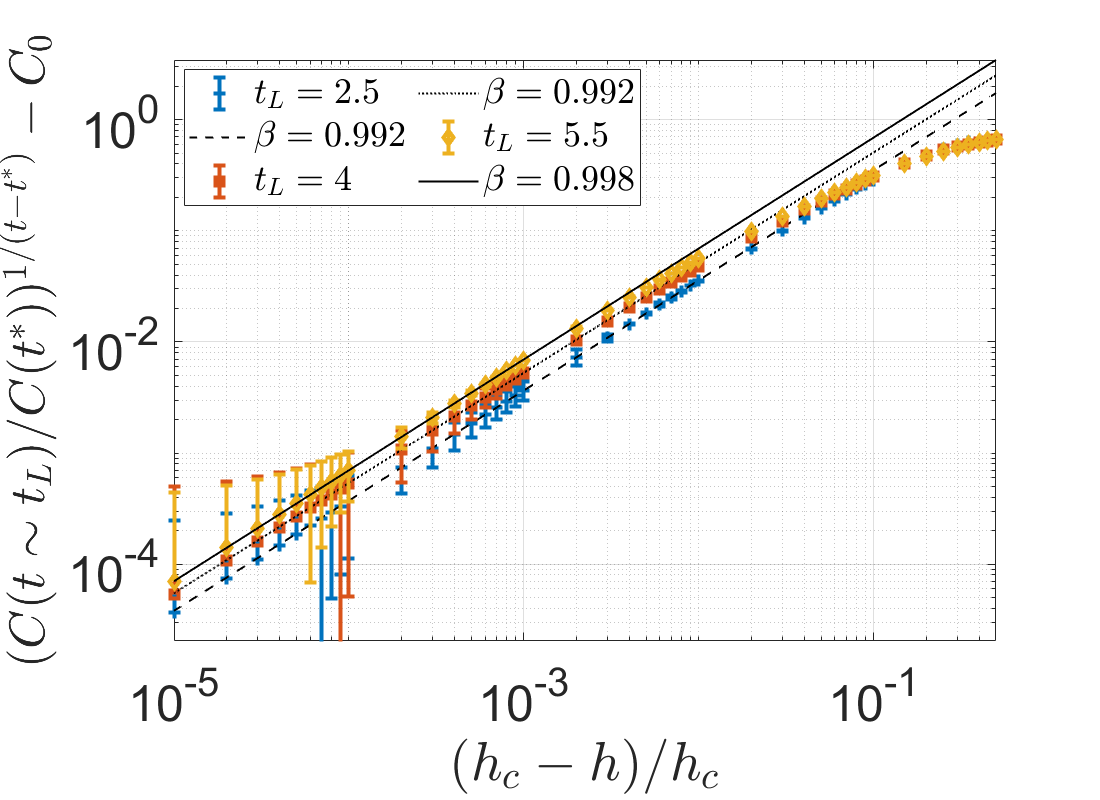}}\hfill
\subfloat[]{\label{fig6C}\includegraphics[width=0.33\textwidth]{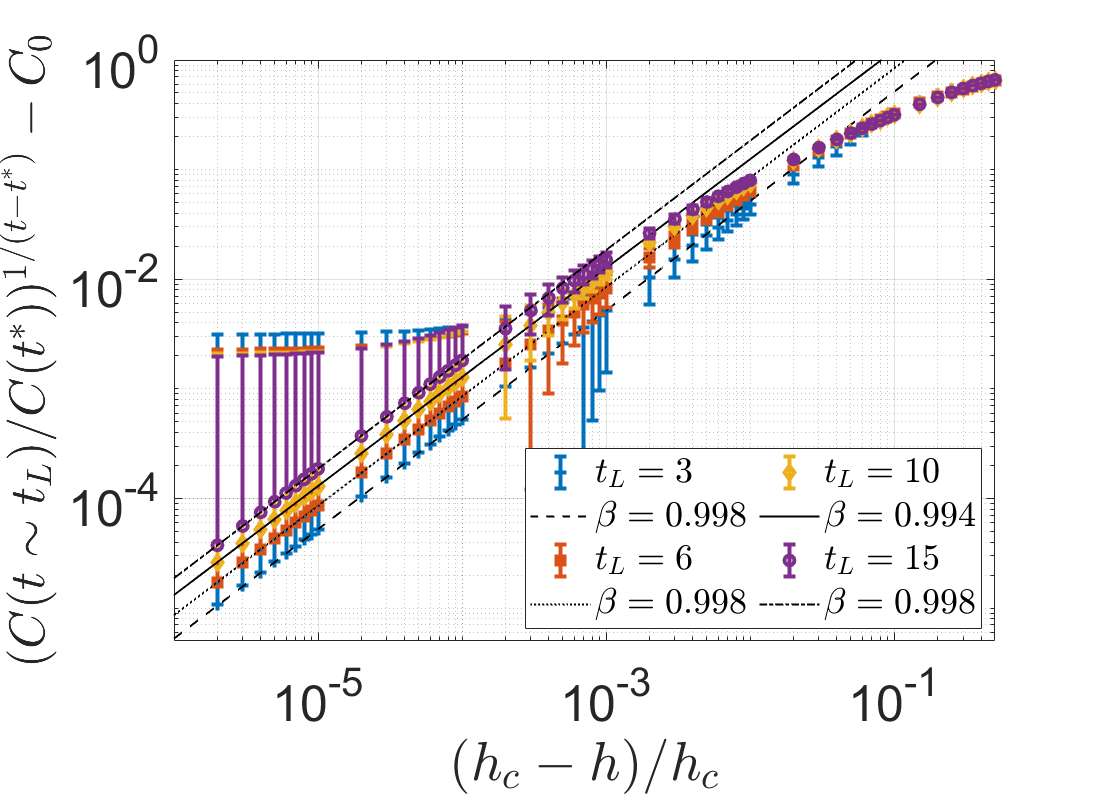}}\hfill 
\caption{The scaling in the vicinity of the crossover for integrable TFIM with respect to reduced control parameter $h_n$ for (a) $N=48$ with open boundary condition (b) $N=48$ and (c) $N=480$ with periodic boundary condition. y-axis is rescaled correctly to obtain the scaling (see text). The temporal cutoffs are (a) either fixed at $t_L=2.5,4$ or parametric with $\alpha=5,8$; fixed at (b) $t_L=2.5,4,5.5$ and (c) $t_L=3,6,10,15$. The solid, dotted, dashed and dotted-dashed lines are the fits in the vicinity of the crossover, all giving $\beta \sim 1$ for all subfigures. Error bars are explained in Appendix F.}
\label{Fig2}
\end{figure*}

At the other end where numerical data exhibits a distinct scaling exponent of $\beta=1$, crossover physics at $h_c=1$ take over with diverging relaxation time \cite{doi:10.1080/00018732.2010.514702,PhysRevLett.95.105701} and one cannot reach late-time behavior in accessible times for any system size that we studied. Fig.~\ref{fig5B} shows the error functions at $h=9.9\times 10^{-6}$ where the fit function is always better to predict the data than the analytical expression in a time interval of $t<60/J$. This suggests that in the close vicinity of the crossover, the analytical expression fails for accessible times and likewise we observe a scaling exponent of $\beta=1$ instead of $\beta=1/2$. 

\subsection{Constructing a dynamical order parameterlike quantity in the dynamically-ordered regime}

One can measure the decay rates of magnetization at each transverse field and probe the crossover between dynamically-ordered and -disordered regimes in TFIM. Alternatively, we aim to find a rescaling of the decaying observable $C(t)$ that can render the rescaled observable a quantity that acts like a dynamical OP in the ordered regime right in the vicinity of the crossover. One can see this procedure as a way to construct a dynamical order parameterlike quantity with the correct rescaling that is originated from the scaling behaviour of the decay rates in the vicinity of the crossover. In DPT-I, the observable naturally acts as a dynamical OP in a nonzero valued steady-state. We find that for magnetization per site in short-range TFIM one needs to correctly rescale the observable to construct a quantity alike. This, in the end, presents an alternative way of extracting the scaling exponent in an experiment, which is less laborious than measuring the decay rates directly.

Similar to how a dynamical OP in DPT-I is constructed by first choosing a temporal cutoff, we consider two different temporal cutoffs applied at a time either (i) fixed $t_L \sim$ constant or (ii) parametric $t_L = \alpha v_q^{-1}$ where $\alpha$ is chosen so that the dynamical response remains in the lightcone, e.g. $t_L \leq \tau_s$. Note that for all temporal cutoffs, $t_L \geq t^*$ holds. Eventually the rescaled dynamical OP-like quantity should not depend on how we choose our temporal cutoff. Furthermore, while one can average the observable for a time between the ultraviolet cutoff $t^*$ and the temporal cutoff $t_L$, this would complicate the functional form of the rescaling needed and it would require more data to compute/measure. Hence, we simply measure the observable $C(t)$ at time $t_L$ dictated by the fixed or parametric temporal cutoff.

Let us rewrite the observable in the vicinity of the crossover by substituting the logarithmic fit function for the decay rates in,
\begin{eqnarray}
C(t)&=&C(t^*)\exp(f_{\Phi}(t-t^*)) \notag \\
&=&C(t^*)(\gamma h_n^{\beta} +C_0)^{t-t^*}.\label{scaling}
\end{eqnarray}
The scaling of the decay rate as a function of $h_n$ reveals a scaling for the observable in the vicinity of the crossover. This expression points out to the correct form of rescaling for the observable to make the procedure independent of the temporal cutoff. Hence the correct rescaling for the observable reads,
\begin{eqnarray}
\left(\frac{C(t)}{C(t^*)}\right)^{1/(t-t^*)}-C_0 = \gamma h_n^{\beta}, \label{rescaling}
\end{eqnarray}
leading us to define a dynamical OP-like quantity,
\begin{eqnarray}
C'(h_n)=\left(\frac{C(t)}{C(t^*)}\right)^{1/(t-t^*)}-C_0,
\end{eqnarray}
which is strictly valid in the vicinity of the crossover. Hence, one can probe the exponent by simply computing $(C(t_L)/C(t^*))^{1/(t_L-t^*)}-C_0$ which requires data points at cutoffs $t^*$ and $t_L$ only, assuming $C_0$ is fixed by numerical prediction.

Figs.~\ref{Fig2} show how the dynamical OP-like quantity $C'(h_n)$, that is constructed based on different cutoffs, scales with $h_n$ in the vicinity of the crossover for $N=48$ with open boundaries in (a), periodic boundaries in (b) and for $N=480$ with periodic boundaries in (c). The colors yellow, red and blue correspond to cutoffs chosen at fixed $t_L=2.5,4,5.5$ and at $\alpha=5,8$ for parametric $t_L=\alpha/v_q$. All data at different temporal cutoffs exhibit the same exponent $\beta\sim 1$. The differences between different temporal cutoffs are detailed in Appendix C. The error bars originate from the uncertainty in time (Appendix F). Since Fig.~\ref{fig6A} is measured at temporal cutoffs, while the Figs.~\ref{fig6B} and \ref{fig6C} are not, there is no error bars for Fig.~\ref{fig6A}. In our data, the temporal uncertainty increases as we increase the system size, which explains the biggest error bars in Fig.~\ref{fig6C}. Therefore, by measuring exactly at the temporal cutoffs these error bars tend to vanish away.

\subsection{Dynamically-disordered regime}

The analytical prediction for the dynamically-disordered regime reads $f_{\Phi}^{\infty}=-4/\pi$ in the nonequilibrium response
\begin{eqnarray}
C(t) &\sim & (1+\cos(2\omega t+\alpha)+\cdots)^{1/2} \exp(tf_{\Phi}^{\infty}), \\
\omega(h) &=& 2\sqrt{1+h^2-2}, \label{disorderedOmega}
\end{eqnarray}
where $\cdots$ means that there are subleading terms and $\alpha$ is an unknown constant. We work with a slightly simplified version of this analytical conjecture: $C(t)= \gamma \cos(\omega t) \exp(tf_{\Phi})$, which also appears in Ref.~\cite{PhysRevLett.102.130603}.

\begin{figure}
\centering
\subfloat[]{\label{fig7a}\includegraphics[width=0.24\textwidth]{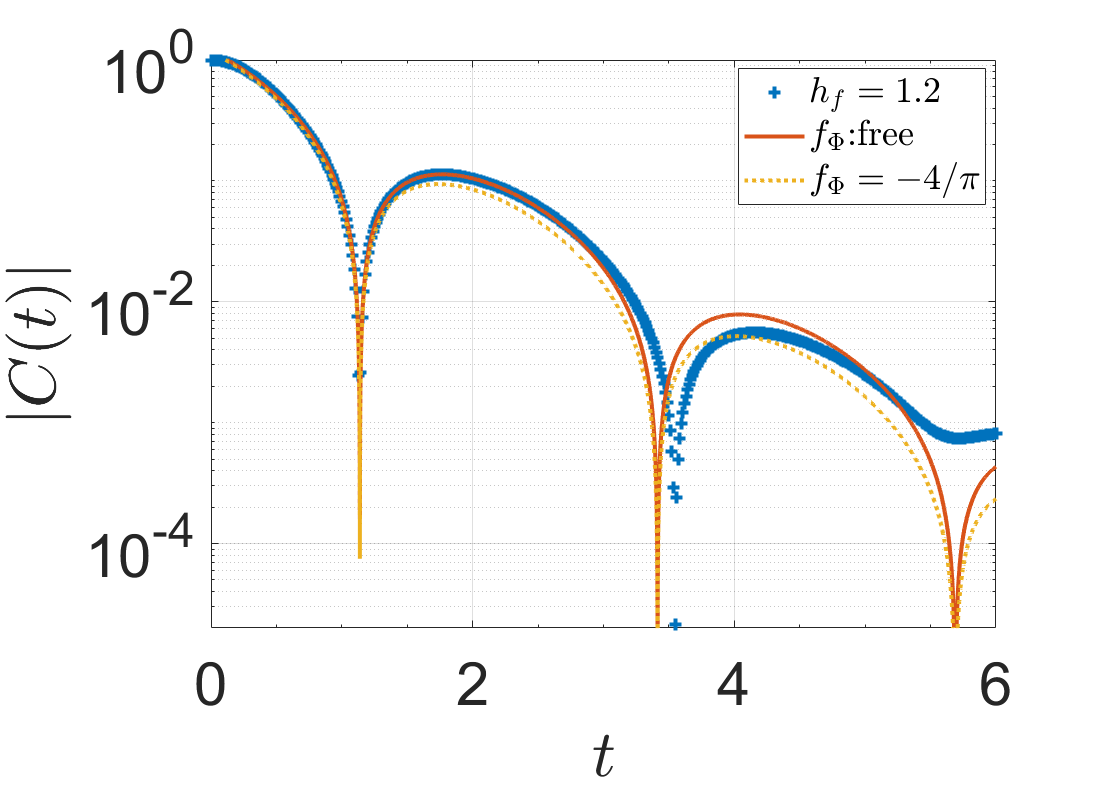}} \hfill
\subfloat[]{\label{fig7b}\includegraphics[width=0.24\textwidth]{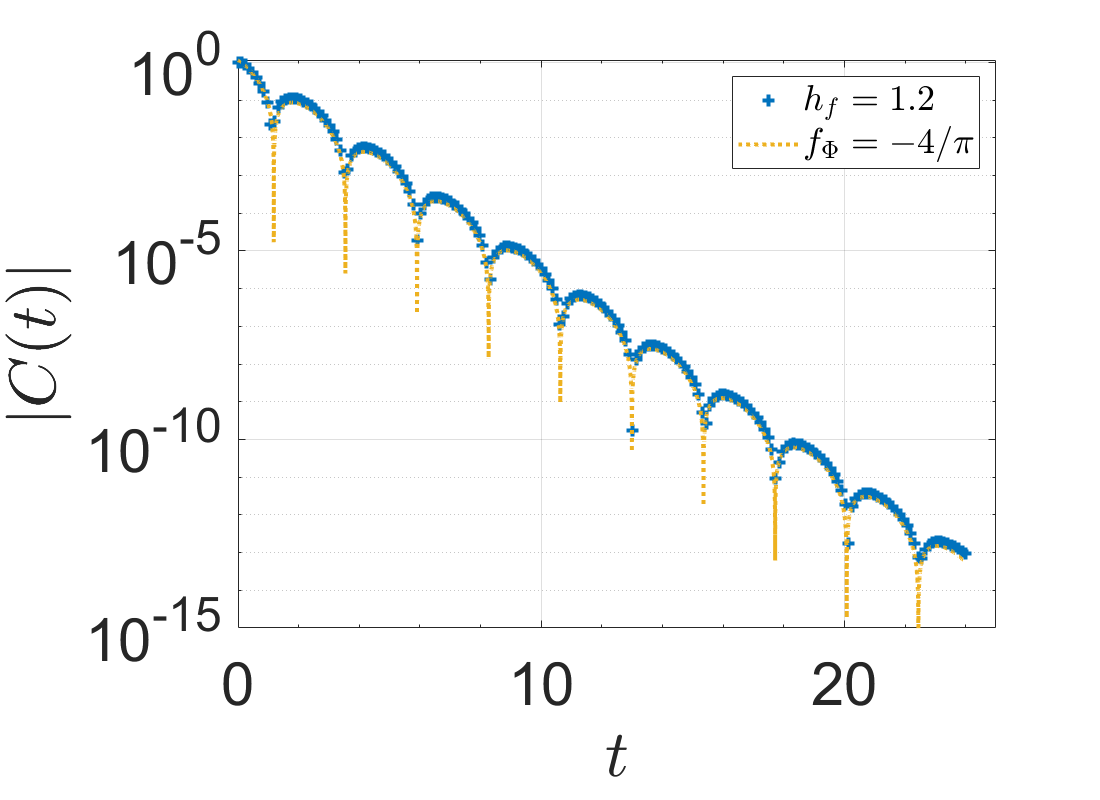}} \hfill \caption{The real time dynamics for the dynamically-disordered regime at $h_f=1.2$ in (a) short-times for $N=48$ and (b) long-times for $N=192$. (a) The short-time dynamics is fitted with analytically predicted $f_{\Phi}=-4/\pi$ (yellow-dotted) and numerically the best match (red-solid) which keeps $f_{\Phi}$ as a free parameter. (b) For large systems, the analytical prediction matches the data excellently.}
\label{Fig7}
\end{figure}

\begin{figure}
\centering
\subfloat[]{\label{fig8a}\includegraphics[width=0.24\textwidth]{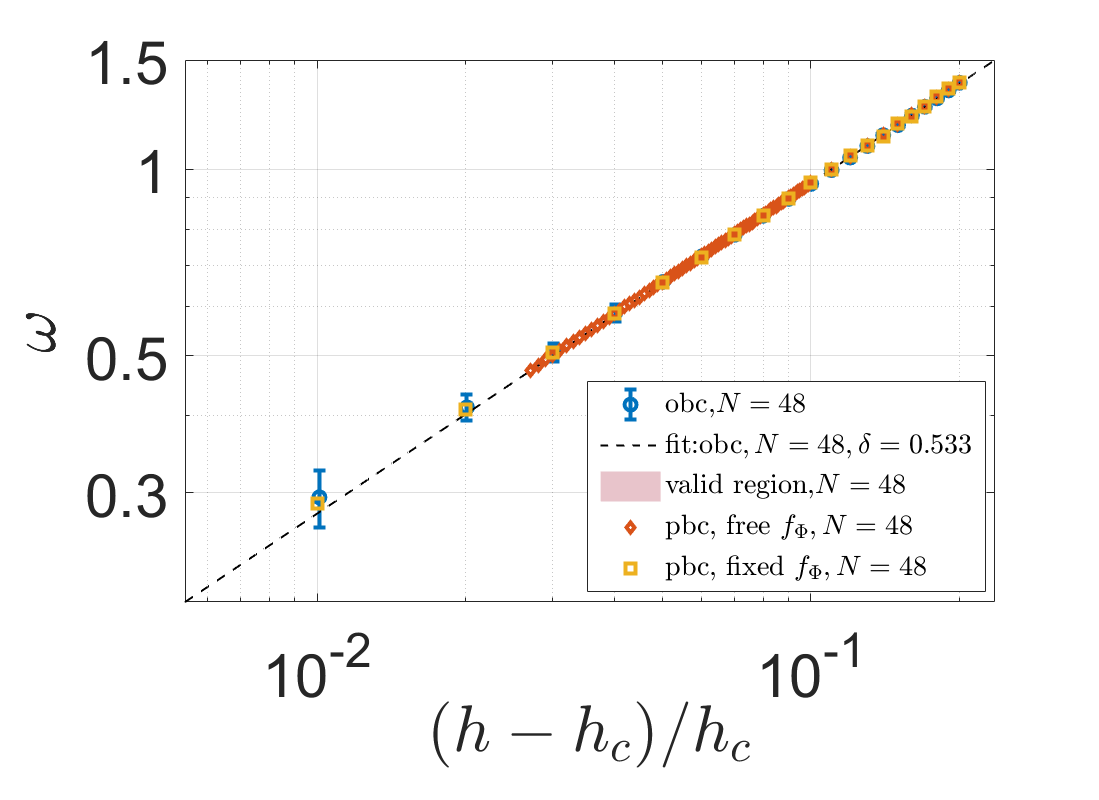}} \hfill
\subfloat[]{\label{fig8b}\includegraphics[width=0.24\textwidth]{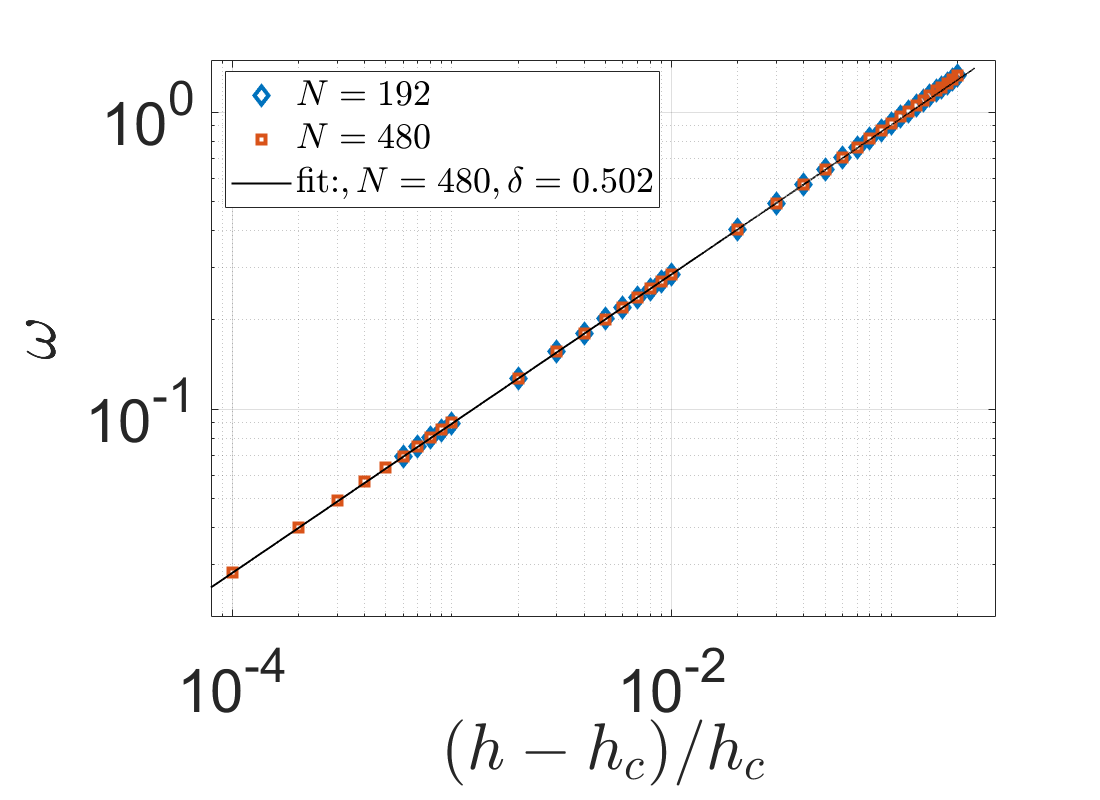}} \hfill \caption{Angular frequency scales with $-h_n$ with a power-law exponent of (a) $\delta\sim 0.53$ for small system sizes and (b) $\delta \sim 0.5$ for larger system sizes in the disordered regime in the vicinity of the crossover. In (a) we plot the scaling of $\omega$ for both cases where we either fix $f_{\Phi}=-4/\pi$ (orange-squares) or let it be a free parameter (red-diamonds). The area in between is shaded which is very negligible.}
\label{Fig8}
\end{figure}

We first focus on short times and small system sizes, e.g. $N=48$ and observe that in this limit, the dynamics could be equally well described by alternative expression to the analytical prediction. Fig.~\ref{fig7a} shows the dynamical response for $h_f=1.2$ where we fit two different curves to the data. The yellow-dotted line is the fit originated from the analytical expression where we fix the decay rate to $f_{\Phi}=-4/\pi$. We let $f_{\Phi}$ be a free parameter in the red-solid line. Therefore the latter performs slightly better than the former, especially for $t<4/J$. We plot the scaling of the decay rates in the dynamically-disordered regime for the latter in Fig.~\ref{fig3a} which turns out to be linear both for periodic and open boundary calculations. We also shade the area between the linear scaling and the constant line at $f_{\Phi}=-4/\pi$ to emphasize the uncertainty in the decay rates for the dynamically-disordered regime for short times and small system sizes. The corresponding scaling for the angular frequency $\omega$ is plotted in Fig.~\ref{fig8a} which is $\delta \sim 0.533$ for both cases where we either fix the decay rate $f_{\Phi}=-4/\pi$ or let it be a free parameter. The shaded area in between is negligible. 

When we increase the system size to $N=192$, we reach longer times and the analytical expression becomes the best fit for the general trend of the data, Fig.~\ref{fig7b}, as expected. In this case, the decay rate is constant at $f_{\Phi}=-4/\pi$ as can be seen in Fig.~\ref{fig3a}. The corresponding scaling for the angular frequency $\omega$ approaches to $\delta \sim 0.5$ as can be calculated from the series expansion of the analytical expression Eq.~\eqref{disorderedOmega} in the close vicinity of the crossover, $\omega(h_n\rightarrow 0) \sim 2 \sqrt{2} (-h_n)^{1/2} + \mathcal{O}((-h_n)^{3/2})$. The numerical demonstration of $\delta=0.5$ is shown in Fig.~\ref{fig8b} with system sizes $N=192,480$.

In conclusion, one observes corrections to the exponents $\delta_{\infty}=1/2$ and $\beta_{\infty}=0$ in the dynamically-disordered regime for short times, resulting in $\delta \sim 0.533$ and $\beta\sim 1$.

\section{Dynamical crossover in the nonintegrable TFIM}

\begin{figure*}
\centering
\subfloat[]{\label{fig4a}\includegraphics[width=0.33\textwidth]{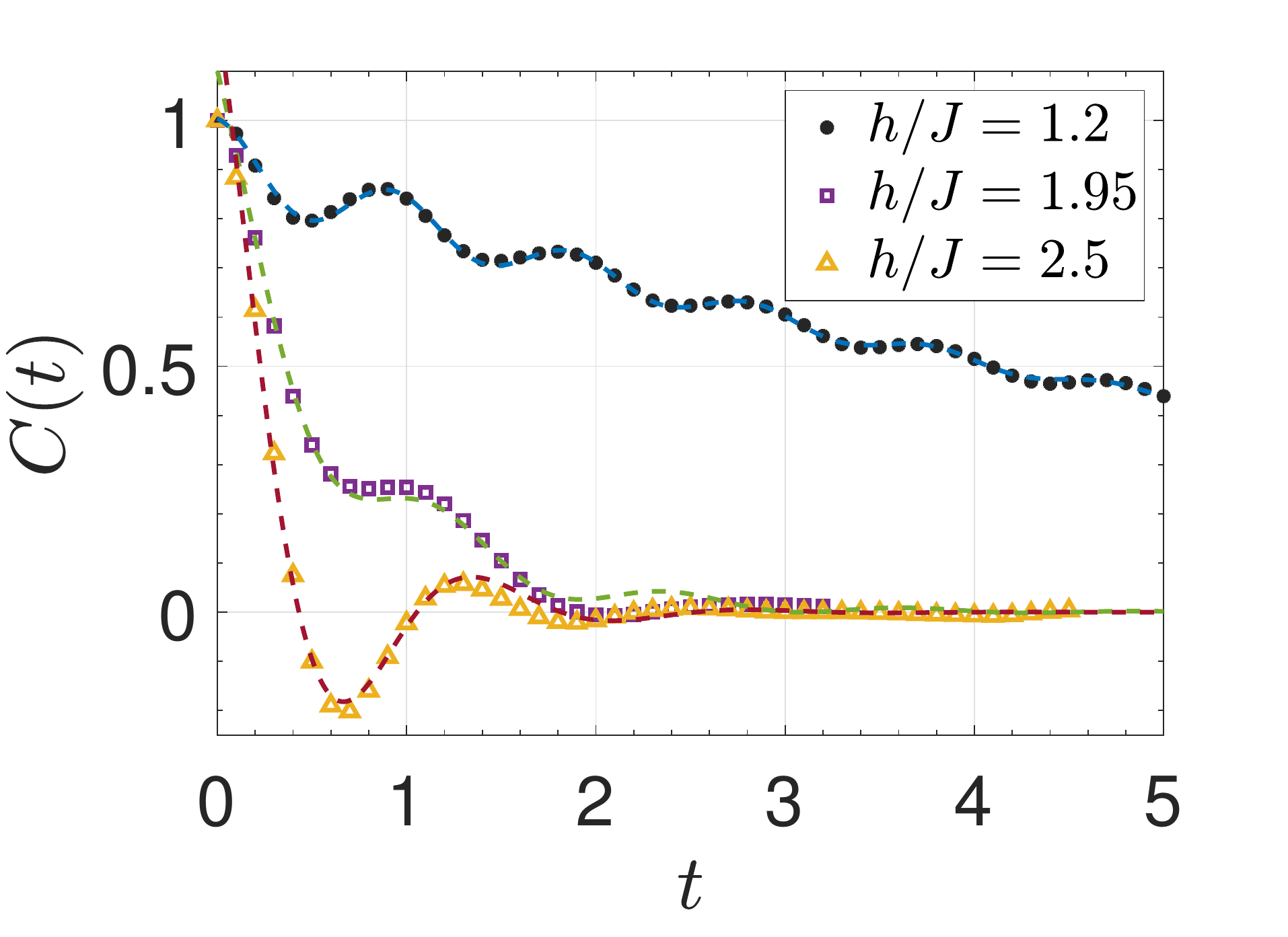}} \hfill
\subfloat[]{\label{fig4b}\includegraphics[width=0.33\textwidth]{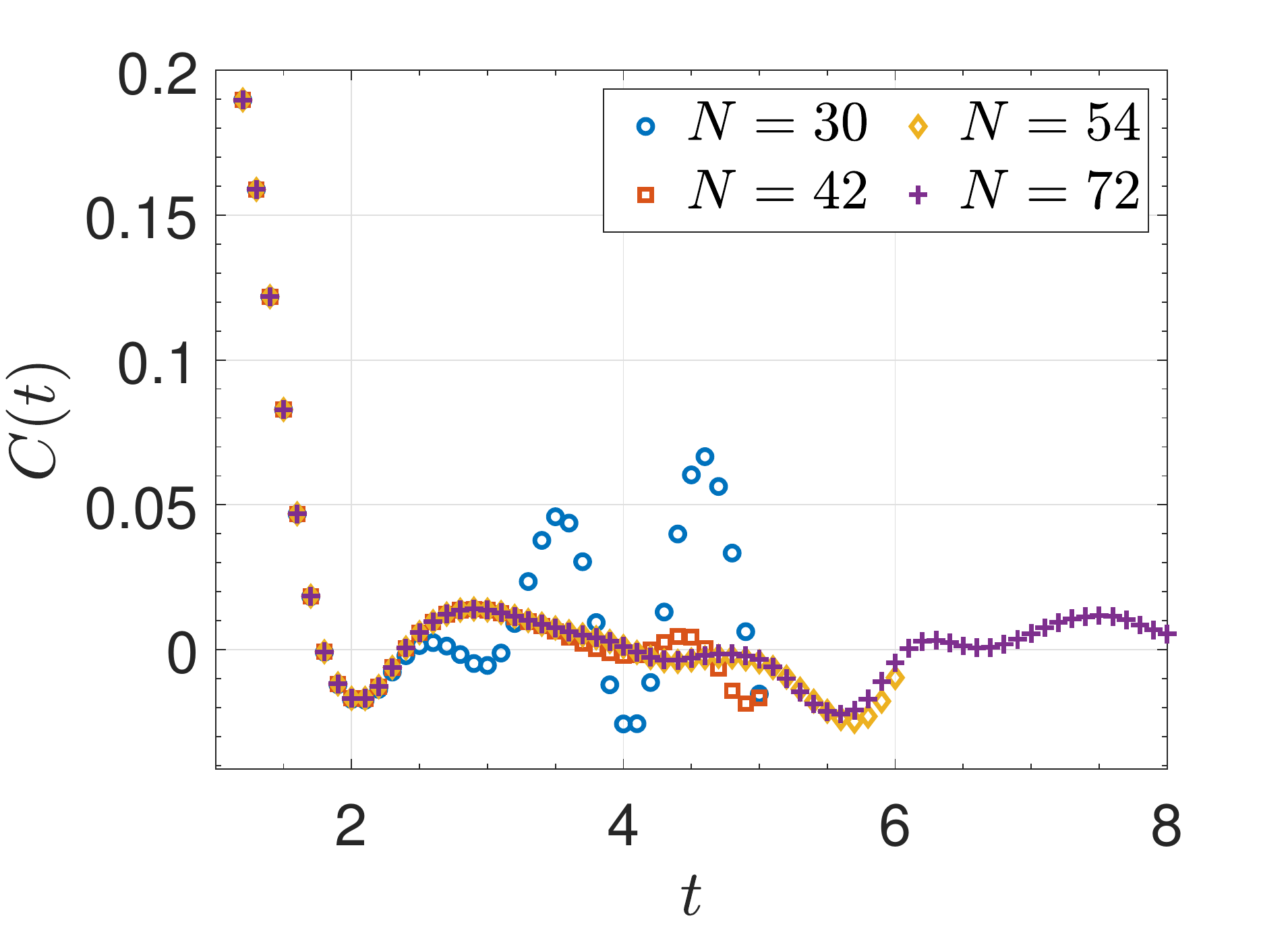}} \hfill
\subfloat[]{\label{figS10a}\includegraphics[width=0.33\textwidth]{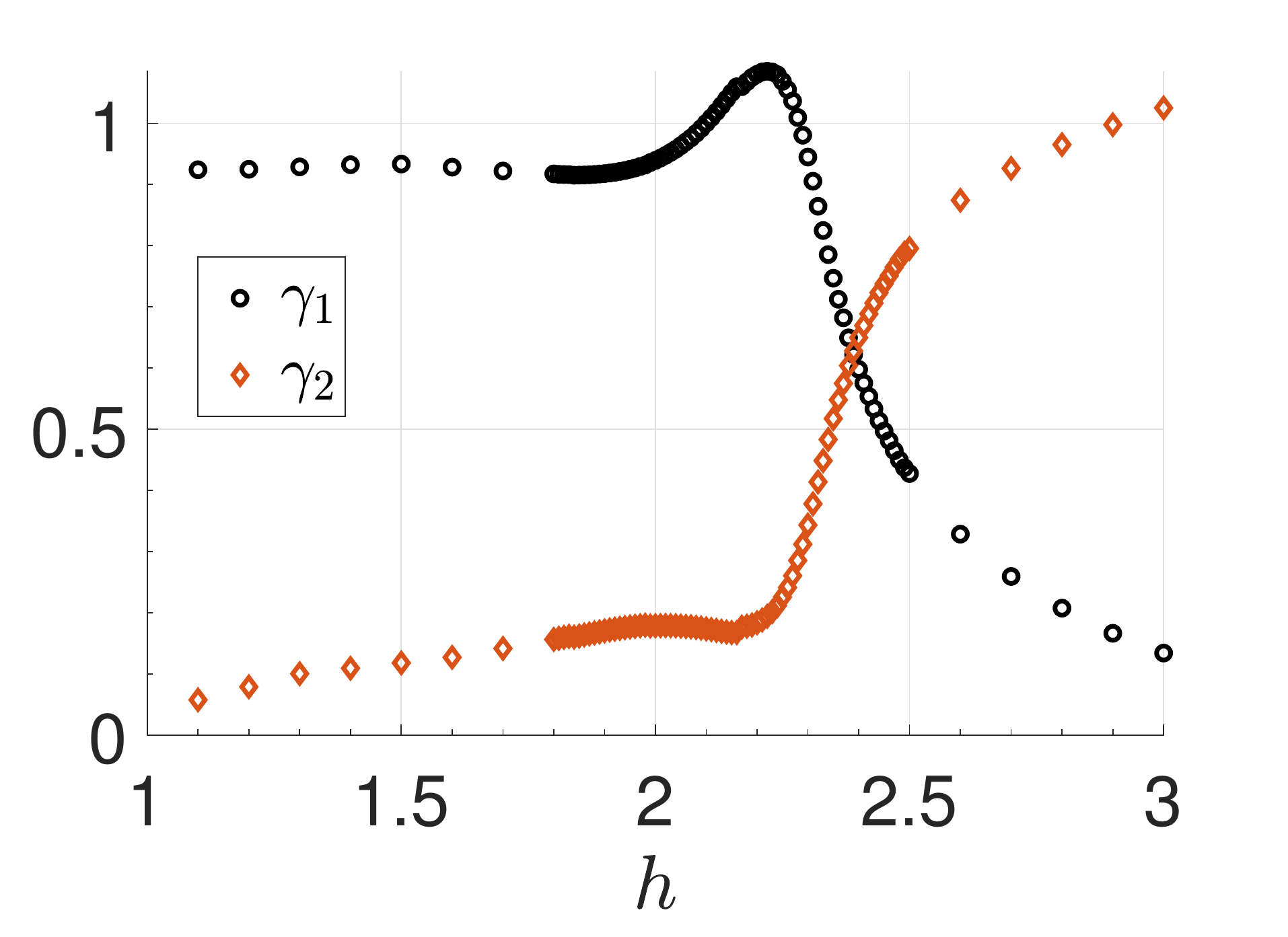}}
\caption{(a) Nonintegrable TFIM with $\Delta/J=-1$ for different $h/J=1.2,1.95,2.5$ and dashed lines are the fit function predictions for dynamical responses. (b) One-point observable for nonintegrable TFIM $\Delta/J=-1$ at $h/J\sim 2$ with respect to time for different system sizes between $N=30-72$. (c) The coefficients $\gamma_1$ (black-circles) and $\gamma_2$ (orange-diamonds) of the fit function for the dynamics of nonintegrable TFIM at $\Delta/J=-1$.}
\label{Fig4}
\end{figure*}

Having studied the dynamical crossover observed in the transient regime for a noninteracting model, we now turn our attention nonintegrable TFIM.

We break the integrability of the model by taking $\Delta/J=-1$ in Eq.~\ref{Hamiltonian}, which hosts an equilibrium QPT at $h_c\sim 2.46$ (Appendix E). Fig.~\ref{fig4a} shows the sophisticated dynamical response of this model calculated with MPS for different $h$ values in the lighcone determined by data ranging from $N=24$ to $N=42$. Lightcones are determined similarly by studying the separation timescales $\tau_s$ of different system sizes. Fig.~\ref{fig4b} shows the presence of well-defined $\tau_s$ timescales for a range of different system sizes at $h/J=2$.

\subsection{Fit function for the nonintegrable TFIM}

An important difference from the noninteracting model is the oscillations existing in both dynamically-ordered and -disordered regimes. Hence, we first aim to approximately model the dynamical response. Since oscillations are present at every $h/J$, a fit function that can reproduce the important features of the dynamics is,
\begin{eqnarray}
C(t)=\gamma_1 \exp(f_{\Phi,1}t)+\gamma_2 \exp(f_{\Phi,2}t) \cos \omega t. \label{minimal}
\end{eqnarray}
The dashed lines in Fig.~\ref{fig4a} show how well the fit function can describe the dynamics. The first and the second terms are analogous terms for the dynamically-ordered and -disordered regimes of the integrable TFIM, respectively. Thus, an immediate observation is that there seems no sharply distinct dynamical regimes as in integrable TFIM. We study the parameters $\gamma_1$, $\gamma_2$ (depicted in Fig.~\ref{figS10a}) $f_{\Phi,1}$ (depicted in Fig.~\ref{fig5}), $f_{\Phi,2}$ and $\omega$ (in Appendix G) as a function of transverse field.

By studying $\gamma_{1,2}$, coefficients of the terms, we first notice that the non-oscillatory term is dominant to the oscillatory term in the region $h \lesssim 2.3$. The opposite is true for $h \gtrsim 2.6$. Hence, even though there are not two distinct fit functions that describe two distinct regimes like in integrable TFIM, there are two limits of one fit function that exhibit distinct enough features. This behaviour seems to stem from the sharp crossover in the integrable model. This is because, the fit function reduces to one term only where $\gamma_2=0$ in dynamically-ordered regime and $\gamma_1=0$ in the -disordered regime. In this sense, with the fit function integrable and nonintegrable models are quantitatively connected to each other. Note that $\gamma_{1,2}$ intersects at a location very close to the equilibrium QPT and this is where both terms are equally significant in the nonequilibrium response. Therefore, one can separate the entire region roughly into three: (1) $h\lesssim 2.3$ where the dynamics can be approximated by only the non-oscillatory term, and hence acts like the dynamically-ordered regime in the integrable TFIM. (3) $h \gtrsim 2.6$ where the dynamics can be approximated by only the oscillatory term, and hence acts like the dynamically-disordered regime in the integrable TFIM. (2) The intermediate crossover region where both terms are important. 

\begin{figure*}
\centering
\subfloat[]{\label{fig5}\includegraphics[width=0.33\textwidth]{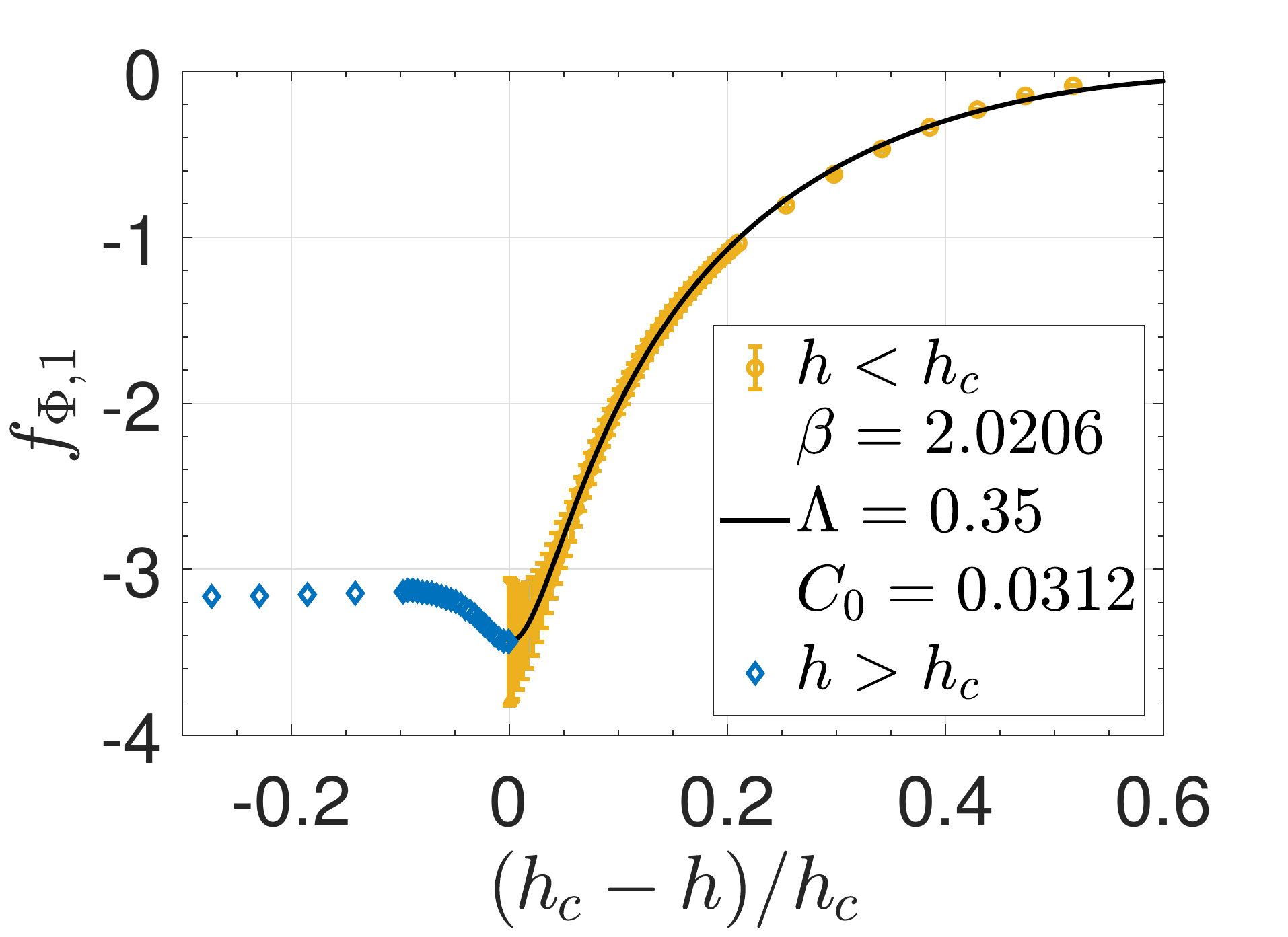}} \hfill
\subfloat[]{\label{fig6}\includegraphics[width=0.33\textwidth]{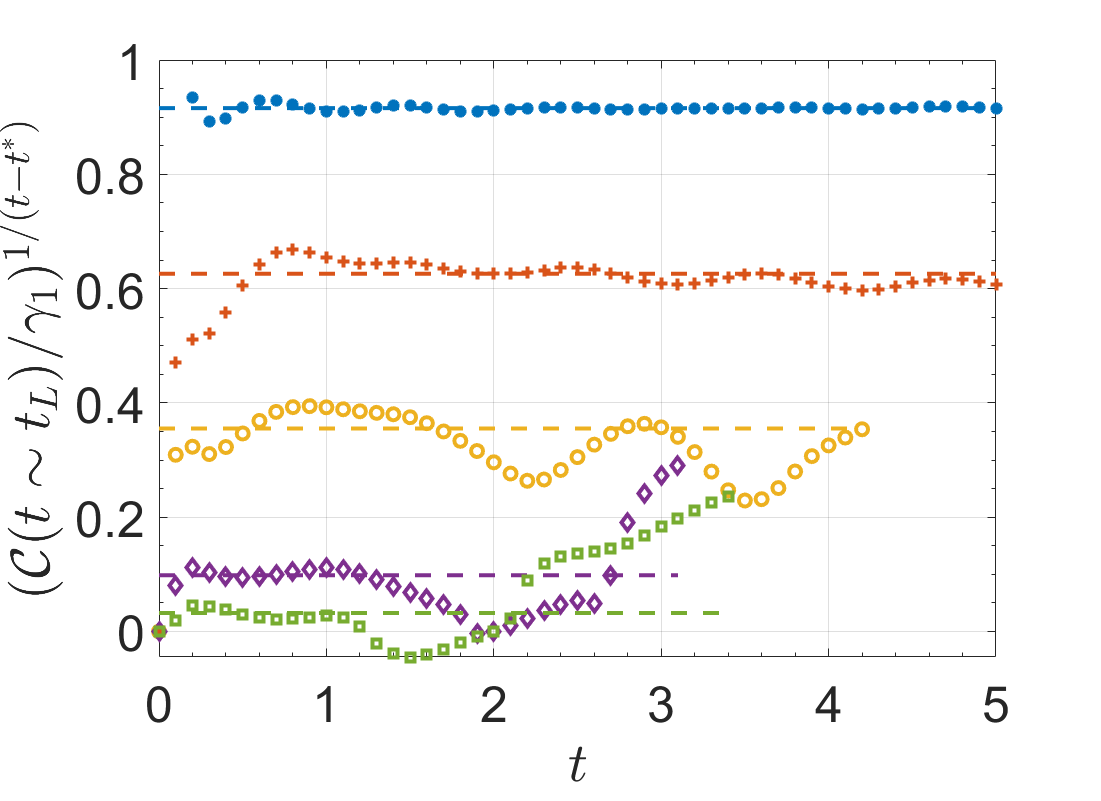}} \hfill
\subfloat[]{\label{fig7}\includegraphics[width=0.33\textwidth]{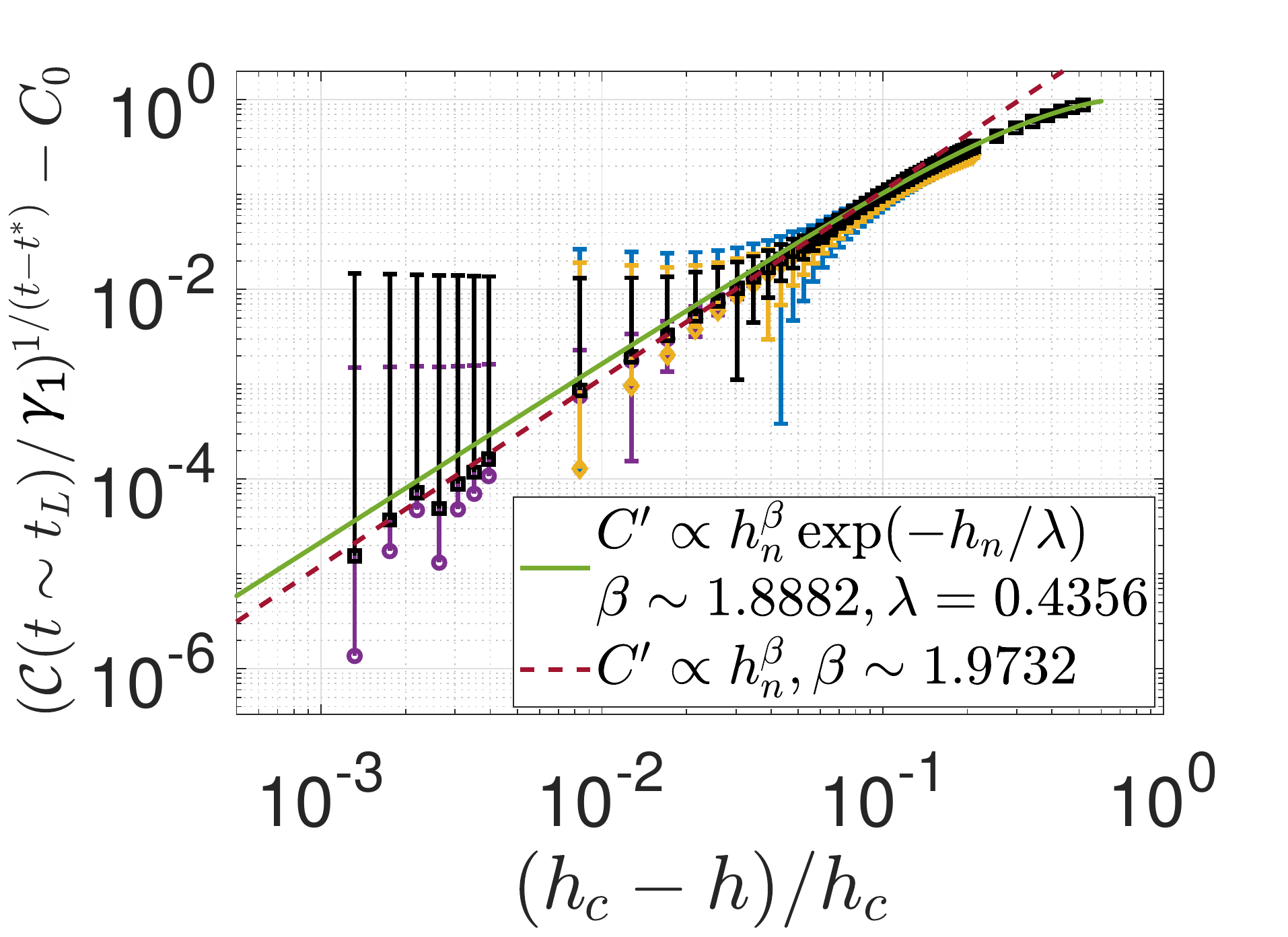}}
\caption{(a) Decay rate of the first term in the fit function Eq.~\eqref{minimal}, $f_{\Phi,1}$ shows a minimum at $h_c=2.278\pm0.001$ signaling a boundary between the ordered regime (yellow-circles) that can be modeled by logarithmic function (black-solid) and crossover region (blue-diamonds). (b) Decay rate functions $\exp(f_{\Phi,1})$ shown with solid flat lines and rescaled observable data according to the method (i) (see text) around the flat lines for $h/J=1.1,1.5,1.8,2.1,2.28$ with blue-dots, red-pluses, yellow-circles, purple-diamonds and green-squares, respectively. Data accumulates around the flat lines. (c) Power-law dynamical scaling in the vicinity of the boundary between regions (1) and (2) with an exponent of $\beta\sim 2$ with blue and yellow data at very early times $t=0.3,0.5$ with the rescaling method (i) and purple data at the nodes of the oscillations motivated by the method (ii) (see text). The black-squares are the decay rate function $\exp(f_{\Phi,1})-C_0$.}
\label{Fig10}
\end{figure*}

The fit function for the nonintegrable TFIM could be tested further with larger system size data and hence, in longer times in the future studies. Additionally, testing the fit function against nonintegrability strength $\Delta/J$ is an interesting direction for future studies. In particular, it would be interesting to study how the regions (1) through (3) change in a near-integrability model. Finally let us note that although there might be other equally accurate models to represent the dynamics of nonintegrable TFIM, the current model has the least amount of free parameters and is physically intuitive.

\subsection{Dynamical crossover region and an OP-like quantity}

We focus on the decay rate of the first term, $f_{\Phi_1}$ since this is the term that governs the exponential decay of the dynamical response, whereas $f_{\Phi_2}$ controls the exponential decay of the oscillations. In this sense, $f_{\Phi_1}$ is the analogous parameter to the decay parameter in the integrable TFIM. In addition to the observation that the system thermalizes the fastest at the crossover in the integrable TFIM, we notice that the minimum of $f_{\Phi_1}$ roughly coincides with the boundary between the dynamically-ordered (1) and the crossover (2) regions. Given that in the integrable TFIM, the cusplike feature emerges in short time dynamics when the nonequilibrium response changes nature, it seems that the minimum of $f_{\Phi_1}$ implies a possible boundary between the regions (1) and (2). In this regard, region (1) is where the nonequilibrium response can be approximated well enough with an exponential decay only; and region (2) is where one cannot ignore the oscillatory term anymore.

Fig.~\ref{fig5} demonstrates this minimum for $f_{\Phi_1}$. We determine the location of the minimum as $h_c=2.278\pm0.001$ which sets the boundary from dynamically-ordered (1) to the crossover (2) regions. The decay rate in the region (1) follows previously introduced logarithmic scaling in $h_n$ (Fig.~\ref{fig5}) giving rise Eq.~\ref{scaling} to hold for the nonintegrable model, as long as the oscillations are taken care of. This could be performed in a couple of different ways, e.g. averaging over a period $\mathcal{T}=2\pi/\omega$, working only at the nodes of the oscillations $(\pi+2\pi n)/2\omega$ where $n\in \mathbb{Z}$ or simply rescaling the observable by substracting the oscillatory term from the observable data. Let us briefly discuss these options.

(i) The first method employed here in the main text is simple rescaling by subtracting the oscillatory term from $C(t) \rightarrow \mathcal{C}(t)=C(t)-\gamma_2 \exp(f_{\Phi,2}t) \cos (\omega t)$. Hence the rescaling of the observable $\mathcal{C}(t)$ follows similarly to Eq.~\eqref{rescaling}. In such an expression, $\gamma_1$, $\gamma_2$, $f_{\Phi,2}$ and $\omega$ are free parameters.  Fig.~\ref{fig6} demonstrates how well the rescaled data can be explained by an exponential decay when the observable is rescaled according to method (i). In the vicinity of the boundary and in early times, data coincides well with the flat lines which are $\exp(f_{\Phi,1})$. Overall, the exponential decay describes the general trend of the data in region (1). 

(ii) The second method is to choose a temporal cutoff at the nodes of the oscillations. This introduces a condition on the temporal cutoff time $t_L$ as
\begin{eqnarray}
t_L=\frac{\pi+2\pi n}{2\omega}, \hspace{5mm} n\in \mathcal{Z}. \notag
\end{eqnarray}
For sufficiently long dynamical response, this condition is not restrictive. When the condition is satisfied, the rescaled observable reduces to Eq.~\eqref{rescaling} with only one free parameter $\gamma_1$. 

(iii) Finally one can think of averaging the observable data over a period of $\mathcal{T}$. Let us first discuss this case for the integrable TFIM. By a time-averaging integral over a period of $\mathcal{T}$ around the temporal cutoff $t_L$, the result reads
\begin{eqnarray}
\frac{1}{\mathcal{T}} &\int_{t_L-\mathcal{T}/2}^{t_L+\mathcal{T}/2}& dt \hspace{1mm} C(t^*) \exp[f_{\Phi}(t-t^*)] \notag \\
&=& C(t^*) \exp[f_{\Phi}(t-t^*)] \frac{\sinh(f_{\Phi}\mathcal{T}/2)}{f_{\Phi}\mathcal{T}/2}.\notag
\end{eqnarray}
In the limit of $\mathcal{T}\rightarrow 0$, we recover the result with no averaging. We note that in case of averaging, one needs to rescale the observable correctly with the averaging interval $\mathcal{T}$ as well in order to construct a dynamical OP-like quantity. Although a similar procedure can be applied for the nonintegrable model, this method requires fine-tuning of $t_L$ and the averaging interval $\mathcal{T}$ based on the free parameters $\omega$ and $f_{\Phi,2}$ to get rid of the oscillatory term in the fit function. Even though there happens to be infinite number of possible pairs of temporal cutoff and averaging interval $(t_L,T)$ in total, there are conditions for viable sets $(t_L,\mathcal{T})$ which introduces fine-tuning. Since such a method is likely to be inconvenient both for computation and experiment, we do not discuss it further.

We plot the rescaled observable with temporal cutoff applied at $t=0.3$ (blue) and $t=0.5$ (yellow) in Fig.~\ref{fig7} in addition to data at a node of the oscillation with angular frequency $\omega$ (purple). The black-squares are the decay rate function $\exp(f_{\Phi,1})-C_0$ where $C_0=\exp(f_{\Phi,1})|_{h_c}$ at the boundary between crossover and dynamically-ordered regions. All data collapses reasonably well and can be described by the power-law scaling of $\beta\sim 2$ in the vicinity $h_n \ll \Lambda = 0.44$, which corresponds to $h \sim 2.23$. The scaling exponent is consistent with the smooth and continuous crossover. The error bars mainly stem from the fitting parameters when we model the dynamical response via the fit function Eq.~\eqref{minimal} applied on our data limited to short times.

Since a single fit function can describe the data in the nonintegrable TFIM, here the observed physics clearly point out to different manifestations of the same quantum phase connected by a smooth crossover. Nevertheless finite-size scaling analysis could be applied to learn about the late-time behavior, since studying larger system sizes would provide larger time intervals remaining in the lightcone. This in turn does not only test the fit function for late times, but could also lead to more precise and accurate predictions on these emerging regions of different nonequilibrium responses as a function of transverse field.

\section{Conclusions}

We studied the decay rates of single-site one-point observables, that is magnetization per site for (non-)integrable TFIM as a function of transverse field. The integrable TFIM exhibited cusplike feature in the decay rates at the dynamical crossover $h_c=1$ between dynamically-ordered and -disordered regimes in early times. In the dynamically-ordered regime, the observable exponentially decays to zero, whereas the nonequilibrium response is an exponential decay superposed with oscillations in the dynamically-disordered regime. By studying the scaling of the decay rates in the vicinity of the crossover, we found a rescaling for the observable and the rescaled observable exhibited a linear dynamical scaling law with $\beta=1$ in the ordered vicinity of the crossover in early times in contrast to $\beta_{\infty}=1/2$ predicted by late time analytical expression. In the dynamically-disordered regime, we showed that both exponents $\beta=1$ and $\delta=0.533$ take up correction factors in early times and differ from the predictions of analytical expression $\beta_{\infty}=0$ and $\delta_{\infty}=1/2$.

Next we wrote down a fit function for the nonequilibrium behavior of the nonintegrable TFIM. Three regions appeared from the model where in (1) $h\lesssim 2.3$ the response is dominated by a smooth exponential decay and hence acting like a dynamically-ordered regime; (3) $h\gtrsim 2.6$ the response is dominated by an oscillatory exponential decay and hence acting like a dynamically-disordered regime; and (2) the intermediate crossover region where none of the terms can be ignored. Hence, we observe that the point-like crossover boundary in the integrable TFIM turns into a region in the nonintegrable model. It is an interesting direction to test this model, its parameters and the region boundaries against different $\Delta/J$. Later we focused on the decay rate of the non-oscillatory term which showed a minimum at the boundary between dynamically-ordered and crossover regions $h_c=2.278\pm0.001$; and found a dynamical OP-like quantity based on temporal cutoffs in the transient regime that can probe this feature of the model after rescaling the observable. The rescaled observable exhibited a dynamical scaling law exponent $\beta\sim 2$.

Our work opens new avenues to explore non-equilibrium order, in particular with local observables, with no need for reaching the saturation regime which might be challenging for experiments \cite{PhysRevA.100.013622}. There are interesting directions for future, such as (i) whether a similar dynamical OP-like quantity could be constructed for other short-range Hamiltonians with exponential decay, e.g. the XXZ model; and (ii) whether long-range interacting TFIM \cite{PhysRevLett.114.157201} could exhibit similar behavior in its transient temporal regimes.

\section{Acknowledgments} 

C.B.D. thanks Jad C. Halimeh, Philipp J. Uhrich and L.-M. Duan for helpful suggestions and discussions. This work was supported by National Science Foundation under Grant EFRI-1741618.

\bibliographystyle{apsrev4-1}

\begin{thebibliography}{48}%
\makeatletter
\providecommand \@ifxundefined [1]{%
 \@ifx{#1\undefined}
}%
\providecommand \@ifnum [1]{%
 \ifnum #1\expandafter \@firstoftwo
 \else \expandafter \@secondoftwo
 \fi
}%
\providecommand \@ifx [1]{%
 \ifx #1\expandafter \@firstoftwo
 \else \expandafter \@secondoftwo
 \fi
}%
\providecommand \natexlab [1]{#1}%
\providecommand \enquote  [1]{``#1''}%
\providecommand \bibnamefont  [1]{#1}%
\providecommand \bibfnamefont [1]{#1}%
\providecommand \citenamefont [1]{#1}%
\providecommand \href@noop [0]{\@secondoftwo}%
\providecommand \href [0]{\begingroup \@sanitize@url \@href}%
\providecommand \@href[1]{\@@startlink{#1}\@@href}%
\providecommand \@@href[1]{\endgroup#1\@@endlink}%
\providecommand \@sanitize@url [0]{\catcode `\\12\catcode `\$12\catcode
  `\&12\catcode `\#12\catcode `\^12\catcode `\_12\catcode `\%12\relax}%
\providecommand \@@startlink[1]{}%
\providecommand \@@endlink[0]{}%
\providecommand \url  [0]{\begingroup\@sanitize@url \@url }%
\providecommand \@url [1]{\endgroup\@href {#1}{\urlprefix }}%
\providecommand \urlprefix  [0]{URL }%
\providecommand \Eprint [0]{\href }%
\providecommand \doibase [0]{http://dx.doi.org/}%
\providecommand \selectlanguage [0]{\@gobble}%
\providecommand \bibinfo  [0]{\@secondoftwo}%
\providecommand \bibfield  [0]{\@secondoftwo}%
\providecommand \translation [1]{[#1]}%
\providecommand \BibitemOpen [0]{}%
\providecommand \bibitemStop [0]{}%
\providecommand \bibitemNoStop [0]{.\EOS\space}%
\providecommand \EOS [0]{\spacefactor3000\relax}%
\providecommand \BibitemShut  [1]{\csname bibitem#1\endcsname}%
\let\auto@bib@innerbib\@empty
\bibitem [{\citenamefont {Landau}(1937)}]{landau1937theory}%
  \BibitemOpen
  \bibfield  {author} {\bibinfo {author} {\bibfnamefont {L.~D.}\ \bibnamefont
  {Landau}},\ }\href@noop {} {\bibfield  {journal} {\bibinfo  {journal} {Zh.
  Eksp. Teor. Fiz.}\ }\textbf {\bibinfo {volume} {11}},\ \bibinfo {pages} {19}
  (\bibinfo {year} {1937})}\BibitemShut {NoStop}%
\bibitem [{\citenamefont {Zinn-Justin}(2002)}]{Zinn-Justin:572813}%
  \BibitemOpen
  \bibfield  {author} {\bibinfo {author} {\bibfnamefont {J.}~\bibnamefont
  {Zinn-Justin}},\ }\href {\doibase 10.1093/acprof:oso/9780198509233.001.0001}
  {\emph {\bibinfo {title} {{Quantum Field Theory and Critical Phenomena; 4th
  ed.}}}},\ Internat. Ser. Mono. Phys.\ (\bibinfo  {publisher} {Clarendon
  Press},\ \bibinfo {address} {Oxford},\ \bibinfo {year} {2002})\BibitemShut
  {NoStop}%
\bibitem [{\citenamefont {Sachdev}(2001)}]{sachdev2001quantum}%
  \BibitemOpen
  \bibfield  {author} {\bibinfo {author} {\bibfnamefont {S.}~\bibnamefont
  {Sachdev}},\ }\href {https://books.google.com/books?id=Ih\_E05N5TZQC} {\emph
  {\bibinfo {title} {Quantum Phase Transitions}}}\ (\bibinfo  {publisher}
  {Cambridge University Press},\ \bibinfo {year} {2001})\BibitemShut {NoStop}%
\bibitem [{\citenamefont {Zurek}\ \emph {et~al.}(2005)\citenamefont {Zurek},
  \citenamefont {Dorner},\ and\ \citenamefont
  {Zoller}}]{PhysRevLett.95.105701}%
  \BibitemOpen
  \bibfield  {author} {\bibinfo {author} {\bibfnamefont {W.~H.}\ \bibnamefont
  {Zurek}}, \bibinfo {author} {\bibfnamefont {U.}~\bibnamefont {Dorner}}, \
  and\ \bibinfo {author} {\bibfnamefont {P.}~\bibnamefont {Zoller}},\ }\href
  {\doibase 10.1103/PhysRevLett.95.105701} {\bibfield  {journal} {\bibinfo
  {journal} {Phys. Rev. Lett.}\ }\textbf {\bibinfo {volume} {95}},\ \bibinfo
  {pages} {105701} (\bibinfo {year} {2005})}\BibitemShut {NoStop}%
\bibitem [{\citenamefont {Damski}\ and\ \citenamefont
  {Zurek}(2007)}]{PhysRevLett.99.130402}%
  \BibitemOpen
  \bibfield  {author} {\bibinfo {author} {\bibfnamefont {B.}~\bibnamefont
  {Damski}}\ and\ \bibinfo {author} {\bibfnamefont {W.~H.}\ \bibnamefont
  {Zurek}},\ }\href {\doibase 10.1103/PhysRevLett.99.130402} {\bibfield
  {journal} {\bibinfo  {journal} {Phys. Rev. Lett.}\ }\textbf {\bibinfo
  {volume} {99}},\ \bibinfo {pages} {130402} (\bibinfo {year}
  {2007})}\BibitemShut {NoStop}%
\bibitem [{\citenamefont {Dziarmaga}(2010)}]{doi:10.1080/00018732.2010.514702}%
  \BibitemOpen
  \bibfield  {author} {\bibinfo {author} {\bibfnamefont {J.}~\bibnamefont
  {Dziarmaga}},\ }\href {\doibase 10.1080/00018732.2010.514702} {\bibfield
  {journal} {\bibinfo  {journal} {Advances in Physics}\ }\textbf {\bibinfo
  {volume} {59}},\ \bibinfo {pages} {1063} (\bibinfo {year}
  {2010})}\BibitemShut {NoStop}%
\bibitem [{\citenamefont {Polkovnikov}\ \emph {et~al.}(2011)\citenamefont
  {Polkovnikov}, \citenamefont {Sengupta}, \citenamefont {Silva},\ and\
  \citenamefont {Vengalattore}}]{RevModPhys.83.863}%
  \BibitemOpen
  \bibfield  {author} {\bibinfo {author} {\bibfnamefont {A.}~\bibnamefont
  {Polkovnikov}}, \bibinfo {author} {\bibfnamefont {K.}~\bibnamefont
  {Sengupta}}, \bibinfo {author} {\bibfnamefont {A.}~\bibnamefont {Silva}}, \
  and\ \bibinfo {author} {\bibfnamefont {M.}~\bibnamefont {Vengalattore}},\
  }\href {\doibase 10.1103/RevModPhys.83.863} {\bibfield  {journal} {\bibinfo
  {journal} {Rev. Mod. Phys.}\ }\textbf {\bibinfo {volume} {83}},\ \bibinfo
  {pages} {863} (\bibinfo {year} {2011})}\BibitemShut {NoStop}%
\bibitem [{\citenamefont {Nicklas}\ \emph {et~al.}(2015)\citenamefont
  {Nicklas}, \citenamefont {Karl}, \citenamefont {H\"ofer}, \citenamefont
  {Johnson}, \citenamefont {Muessel}, \citenamefont {Strobel}, \citenamefont
  {Tomkovi\ifmmode~\check{c}\else \v{c}\fi{}}, \citenamefont {Gasenzer},\ and\
  \citenamefont {Oberthaler}}]{PhysRevLett.115.245301}%
  \BibitemOpen
  \bibfield  {author} {\bibinfo {author} {\bibfnamefont {E.}~\bibnamefont
  {Nicklas}}, \bibinfo {author} {\bibfnamefont {M.}~\bibnamefont {Karl}},
  \bibinfo {author} {\bibfnamefont {M.}~\bibnamefont {H\"ofer}}, \bibinfo
  {author} {\bibfnamefont {A.}~\bibnamefont {Johnson}}, \bibinfo {author}
  {\bibfnamefont {W.}~\bibnamefont {Muessel}}, \bibinfo {author} {\bibfnamefont
  {H.}~\bibnamefont {Strobel}}, \bibinfo {author} {\bibfnamefont
  {J.}~\bibnamefont {Tomkovi\ifmmode~\check{c}\else \v{c}\fi{}}}, \bibinfo
  {author} {\bibfnamefont {T.}~\bibnamefont {Gasenzer}}, \ and\ \bibinfo
  {author} {\bibfnamefont {M.~K.}\ \bibnamefont {Oberthaler}},\ }\href
  {\doibase 10.1103/PhysRevLett.115.245301} {\bibfield  {journal} {\bibinfo
  {journal} {Phys. Rev. Lett.}\ }\textbf {\bibinfo {volume} {115}},\ \bibinfo
  {pages} {245301} (\bibinfo {year} {2015})}\BibitemShut {NoStop}%
\bibitem [{\citenamefont {Heyl}\ \emph {et~al.}(2013)\citenamefont {Heyl},
  \citenamefont {Polkovnikov},\ and\ \citenamefont
  {Kehrein}}]{PhysRevLett.110.135704}%
  \BibitemOpen
  \bibfield  {author} {\bibinfo {author} {\bibfnamefont {M.}~\bibnamefont
  {Heyl}}, \bibinfo {author} {\bibfnamefont {A.}~\bibnamefont {Polkovnikov}}, \
  and\ \bibinfo {author} {\bibfnamefont {S.}~\bibnamefont {Kehrein}},\ }\href
  {\doibase 10.1103/PhysRevLett.110.135704} {\bibfield  {journal} {\bibinfo
  {journal} {Phys. Rev. Lett.}\ }\textbf {\bibinfo {volume} {110}},\ \bibinfo
  {pages} {135704} (\bibinfo {year} {2013})}\BibitemShut {NoStop}%
\bibitem [{\citenamefont {Heyl}(2018)}]{Heyl_2018}%
  \BibitemOpen
  \bibfield  {author} {\bibinfo {author} {\bibfnamefont {M.}~\bibnamefont
  {Heyl}},\ }\href {\doibase 10.1088/1361-6633/aaaf9a} {\bibfield  {journal}
  {\bibinfo  {journal} {Reports on Progress in Physics}\ }\textbf {\bibinfo
  {volume} {81}},\ \bibinfo {pages} {054001} (\bibinfo {year}
  {2018})}\BibitemShut {NoStop}%
\bibitem [{\citenamefont {Mori}\ \emph {et~al.}(2018)\citenamefont {Mori},
  \citenamefont {Ikeda}, \citenamefont {Kaminishi},\ and\ \citenamefont
  {Ueda}}]{Mori_2018}%
  \BibitemOpen
  \bibfield  {author} {\bibinfo {author} {\bibfnamefont {T.}~\bibnamefont
  {Mori}}, \bibinfo {author} {\bibfnamefont {T.~N.}\ \bibnamefont {Ikeda}},
  \bibinfo {author} {\bibfnamefont {E.}~\bibnamefont {Kaminishi}}, \ and\
  \bibinfo {author} {\bibfnamefont {M.}~\bibnamefont {Ueda}},\ }\href {\doibase
  10.1088/1361-6455/aabcdf} {\bibfield  {journal} {\bibinfo  {journal} {Journal
  of Physics B: Atomic, Molecular and Optical Physics}\ }\textbf {\bibinfo
  {volume} {51}},\ \bibinfo {pages} {112001} (\bibinfo {year}
  {2018})}\BibitemShut {NoStop}%
\bibitem [{\citenamefont {\ifmmode \check{Z}\else
  \v{Z}\fi{}unkovi\ifmmode~\check{c}\else \v{c}\fi{}}\ \emph
  {et~al.}(2018)\citenamefont {\ifmmode \check{Z}\else
  \v{Z}\fi{}unkovi\ifmmode~\check{c}\else \v{c}\fi{}}, \citenamefont {Heyl},
  \citenamefont {Knap},\ and\ \citenamefont {Silva}}]{PhysRevLett.120.130601}%
  \BibitemOpen
  \bibfield  {author} {\bibinfo {author} {\bibfnamefont {B.}~\bibnamefont
  {\ifmmode \check{Z}\else \v{Z}\fi{}unkovi\ifmmode~\check{c}\else
  \v{c}\fi{}}}, \bibinfo {author} {\bibfnamefont {M.}~\bibnamefont {Heyl}},
  \bibinfo {author} {\bibfnamefont {M.}~\bibnamefont {Knap}}, \ and\ \bibinfo
  {author} {\bibfnamefont {A.}~\bibnamefont {Silva}},\ }\href {\doibase
  10.1103/PhysRevLett.120.130601} {\bibfield  {journal} {\bibinfo  {journal}
  {Phys. Rev. Lett.}\ }\textbf {\bibinfo {volume} {120}},\ \bibinfo {pages}
  {130601} (\bibinfo {year} {2018})}\BibitemShut {NoStop}%
\bibitem [{\citenamefont {{Zhang}}\ \emph {et~al.}(2017)\citenamefont
  {{Zhang}}, \citenamefont {{Pagano}}, \citenamefont {{Hess}}, \citenamefont
  {{Kyprianidis}}, \citenamefont {{Becker}}, \citenamefont {{Kaplan}},
  \citenamefont {{Gorshkov}}, \citenamefont {{Gong}},\ and\ \citenamefont
  {{Monroe}}}]{2017Natur.551..601Z}%
  \BibitemOpen
  \bibfield  {author} {\bibinfo {author} {\bibfnamefont {J.}~\bibnamefont
  {{Zhang}}}, \bibinfo {author} {\bibfnamefont {G.}~\bibnamefont {{Pagano}}},
  \bibinfo {author} {\bibfnamefont {P.~W.}\ \bibnamefont {{Hess}}}, \bibinfo
  {author} {\bibfnamefont {A.}~\bibnamefont {{Kyprianidis}}}, \bibinfo {author}
  {\bibfnamefont {P.}~\bibnamefont {{Becker}}}, \bibinfo {author}
  {\bibfnamefont {H.}~\bibnamefont {{Kaplan}}}, \bibinfo {author}
  {\bibfnamefont {A.~V.}\ \bibnamefont {{Gorshkov}}}, \bibinfo {author}
  {\bibfnamefont {Z.~X.}\ \bibnamefont {{Gong}}}, \ and\ \bibinfo {author}
  {\bibfnamefont {C.}~\bibnamefont {{Monroe}}},\ }\href {\doibase
  10.1038/nature24654} {\bibfield  {journal} {\bibinfo  {journal} {\nat}\
  }\textbf {\bibinfo {volume} {551}},\ \bibinfo {pages} {601} (\bibinfo {year}
  {2017})},\ \Eprint {http://arxiv.org/abs/1708.01044} {arXiv:1708.01044
  [quant-ph]} \BibitemShut {NoStop}%
\bibitem [{\citenamefont {Heyl}\ \emph {et~al.}(2018)\citenamefont {Heyl},
  \citenamefont {Pollmann},\ and\ \citenamefont
  {D\'ora}}]{PhysRevLett.121.016801}%
  \BibitemOpen
  \bibfield  {author} {\bibinfo {author} {\bibfnamefont {M.}~\bibnamefont
  {Heyl}}, \bibinfo {author} {\bibfnamefont {F.}~\bibnamefont {Pollmann}}, \
  and\ \bibinfo {author} {\bibfnamefont {B.}~\bibnamefont {D\'ora}},\ }\href
  {\doibase 10.1103/PhysRevLett.121.016801} {\bibfield  {journal} {\bibinfo
  {journal} {Phys. Rev. Lett.}\ }\textbf {\bibinfo {volume} {121}},\ \bibinfo
  {pages} {016801} (\bibinfo {year} {2018})}\BibitemShut {NoStop}%
\bibitem [{\citenamefont {Da\ifmmode~\breve{g}\else \u{g}\fi{}}\ \emph
  {et~al.}(2019)\citenamefont {Da\ifmmode~\breve{g}\else \u{g}\fi{}},
  \citenamefont {Sun},\ and\ \citenamefont {Duan}}]{PhysRevLett.123.140602}%
  \BibitemOpen
  \bibfield  {author} {\bibinfo {author} {\bibfnamefont {C.~B.}\ \bibnamefont
  {Da\ifmmode~\breve{g}\else \u{g}\fi{}}}, \bibinfo {author} {\bibfnamefont
  {K.}~\bibnamefont {Sun}}, \ and\ \bibinfo {author} {\bibfnamefont {L.-M.}\
  \bibnamefont {Duan}},\ }\href {\doibase 10.1103/PhysRevLett.123.140602}
  {\bibfield  {journal} {\bibinfo  {journal} {Phys. Rev. Lett.}\ }\textbf
  {\bibinfo {volume} {123}},\ \bibinfo {pages} {140602} (\bibinfo {year}
  {2019})}\BibitemShut {NoStop}%
\bibitem [{\citenamefont {Titum}\ \emph {et~al.}(2019)\citenamefont {Titum},
  \citenamefont {Iosue}, \citenamefont {Garrison}, \citenamefont {Gorshkov},\
  and\ \citenamefont {Gong}}]{PhysRevLett.123.115701}%
  \BibitemOpen
  \bibfield  {author} {\bibinfo {author} {\bibfnamefont {P.}~\bibnamefont
  {Titum}}, \bibinfo {author} {\bibfnamefont {J.~T.}\ \bibnamefont {Iosue}},
  \bibinfo {author} {\bibfnamefont {J.~R.}\ \bibnamefont {Garrison}}, \bibinfo
  {author} {\bibfnamefont {A.~V.}\ \bibnamefont {Gorshkov}}, \ and\ \bibinfo
  {author} {\bibfnamefont {Z.-X.}\ \bibnamefont {Gong}},\ }\href {\doibase
  10.1103/PhysRevLett.123.115701} {\bibfield  {journal} {\bibinfo  {journal}
  {Phys. Rev. Lett.}\ }\textbf {\bibinfo {volume} {123}},\ \bibinfo {pages}
  {115701} (\bibinfo {year} {2019})}\BibitemShut {NoStop}%
\bibitem [{\citenamefont {Wei}\ \emph {et~al.}(2019)\citenamefont {Wei},
  \citenamefont {Sun},\ and\ \citenamefont {Hwang}}]{PhysRevB.100.195107}%
  \BibitemOpen
  \bibfield  {author} {\bibinfo {author} {\bibfnamefont {B.-B.}\ \bibnamefont
  {Wei}}, \bibinfo {author} {\bibfnamefont {G.}~\bibnamefont {Sun}}, \ and\
  \bibinfo {author} {\bibfnamefont {M.-J.}\ \bibnamefont {Hwang}},\ }\href
  {\doibase 10.1103/PhysRevB.100.195107} {\bibfield  {journal} {\bibinfo
  {journal} {Phys. Rev. B}\ }\textbf {\bibinfo {volume} {100}},\ \bibinfo
  {pages} {195107} (\bibinfo {year} {2019})}\BibitemShut {NoStop}%
\bibitem [{\citenamefont {Uhrich}\ \emph {et~al.}(2020)\citenamefont {Uhrich},
  \citenamefont {Defenu}, \citenamefont {Jafari},\ and\ \citenamefont
  {Halimeh}}]{PhysRevB.101.245148}%
  \BibitemOpen
  \bibfield  {author} {\bibinfo {author} {\bibfnamefont {P.}~\bibnamefont
  {Uhrich}}, \bibinfo {author} {\bibfnamefont {N.}~\bibnamefont {Defenu}},
  \bibinfo {author} {\bibfnamefont {R.}~\bibnamefont {Jafari}}, \ and\ \bibinfo
  {author} {\bibfnamefont {J.~C.}\ \bibnamefont {Halimeh}},\ }\href {\doibase
  10.1103/PhysRevB.101.245148} {\bibfield  {journal} {\bibinfo  {journal}
  {Phys. Rev. B}\ }\textbf {\bibinfo {volume} {101}},\ \bibinfo {pages}
  {245148} (\bibinfo {year} {2020})}\BibitemShut {NoStop}%
\bibitem [{\citenamefont {Eckstein}\ \emph {et~al.}(2009)\citenamefont
  {Eckstein}, \citenamefont {Kollar},\ and\ \citenamefont
  {Werner}}]{PhysRevLett.103.056403}%
  \BibitemOpen
  \bibfield  {author} {\bibinfo {author} {\bibfnamefont {M.}~\bibnamefont
  {Eckstein}}, \bibinfo {author} {\bibfnamefont {M.}~\bibnamefont {Kollar}}, \
  and\ \bibinfo {author} {\bibfnamefont {P.}~\bibnamefont {Werner}},\ }\href
  {\doibase 10.1103/PhysRevLett.103.056403} {\bibfield  {journal} {\bibinfo
  {journal} {Phys. Rev. Lett.}\ }\textbf {\bibinfo {volume} {103}},\ \bibinfo
  {pages} {056403} (\bibinfo {year} {2009})}\BibitemShut {NoStop}%
\bibitem [{\citenamefont {Sciolla}\ and\ \citenamefont
  {Biroli}(2013)}]{PhysRevB.88.201110}%
  \BibitemOpen
  \bibfield  {author} {\bibinfo {author} {\bibfnamefont {B.}~\bibnamefont
  {Sciolla}}\ and\ \bibinfo {author} {\bibfnamefont {G.}~\bibnamefont
  {Biroli}},\ }\href {\doibase 10.1103/PhysRevB.88.201110} {\bibfield
  {journal} {\bibinfo  {journal} {Phys. Rev. B}\ }\textbf {\bibinfo {volume}
  {88}},\ \bibinfo {pages} {201110} (\bibinfo {year} {2013})}\BibitemShut
  {NoStop}%
\bibitem [{\citenamefont {Tsuji}\ \emph {et~al.}(2013)\citenamefont {Tsuji},
  \citenamefont {Eckstein},\ and\ \citenamefont
  {Werner}}]{PhysRevLett.110.136404}%
  \BibitemOpen
  \bibfield  {author} {\bibinfo {author} {\bibfnamefont {N.}~\bibnamefont
  {Tsuji}}, \bibinfo {author} {\bibfnamefont {M.}~\bibnamefont {Eckstein}}, \
  and\ \bibinfo {author} {\bibfnamefont {P.}~\bibnamefont {Werner}},\ }\href
  {\doibase 10.1103/PhysRevLett.110.136404} {\bibfield  {journal} {\bibinfo
  {journal} {Phys. Rev. Lett.}\ }\textbf {\bibinfo {volume} {110}},\ \bibinfo
  {pages} {136404} (\bibinfo {year} {2013})}\BibitemShut {NoStop}%
\bibitem [{\citenamefont {Halimeh}\ \emph {et~al.}(2017)\citenamefont
  {Halimeh}, \citenamefont {Zauner-Stauber}, \citenamefont {McCulloch},
  \citenamefont {de~Vega}, \citenamefont {Schollw\"ock},\ and\ \citenamefont
  {Kastner}}]{PhysRevB.95.024302}%
  \BibitemOpen
  \bibfield  {author} {\bibinfo {author} {\bibfnamefont {J.~C.}\ \bibnamefont
  {Halimeh}}, \bibinfo {author} {\bibfnamefont {V.}~\bibnamefont
  {Zauner-Stauber}}, \bibinfo {author} {\bibfnamefont {I.~P.}\ \bibnamefont
  {McCulloch}}, \bibinfo {author} {\bibfnamefont {I.}~\bibnamefont {de~Vega}},
  \bibinfo {author} {\bibfnamefont {U.}~\bibnamefont {Schollw\"ock}}, \ and\
  \bibinfo {author} {\bibfnamefont {M.}~\bibnamefont {Kastner}},\ }\href
  {\doibase 10.1103/PhysRevB.95.024302} {\bibfield  {journal} {\bibinfo
  {journal} {Phys. Rev. B}\ }\textbf {\bibinfo {volume} {95}},\ \bibinfo
  {pages} {024302} (\bibinfo {year} {2017})}\BibitemShut {NoStop}%
\bibitem [{\citenamefont {Halimeh}\ and\ \citenamefont
  {Zauner-Stauber}(2017)}]{PhysRevB.96.134427}%
  \BibitemOpen
  \bibfield  {author} {\bibinfo {author} {\bibfnamefont {J.~C.}\ \bibnamefont
  {Halimeh}}\ and\ \bibinfo {author} {\bibfnamefont {V.}~\bibnamefont
  {Zauner-Stauber}},\ }\href {\doibase 10.1103/PhysRevB.96.134427} {\bibfield
  {journal} {\bibinfo  {journal} {Phys. Rev. B}\ }\textbf {\bibinfo {volume}
  {96}},\ \bibinfo {pages} {134427} (\bibinfo {year} {2017})}\BibitemShut
  {NoStop}%
\bibitem [{\citenamefont {Essler}\ \emph {et~al.}(2012)\citenamefont {Essler},
  \citenamefont {Evangelisti},\ and\ \citenamefont
  {Fagotti}}]{PhysRevLett.109.247206}%
  \BibitemOpen
  \bibfield  {author} {\bibinfo {author} {\bibfnamefont {F.~H.~L.}\
  \bibnamefont {Essler}}, \bibinfo {author} {\bibfnamefont {S.}~\bibnamefont
  {Evangelisti}}, \ and\ \bibinfo {author} {\bibfnamefont {M.}~\bibnamefont
  {Fagotti}},\ }\href {\doibase 10.1103/PhysRevLett.109.247206} {\bibfield
  {journal} {\bibinfo  {journal} {Phys. Rev. Lett.}\ }\textbf {\bibinfo
  {volume} {109}},\ \bibinfo {pages} {247206} (\bibinfo {year}
  {2012})}\BibitemShut {NoStop}%
\bibitem [{\citenamefont {Calabrese}\ \emph {et~al.}(2012)\citenamefont
  {Calabrese}, \citenamefont {Essler},\ and\ \citenamefont
  {Fagotti}}]{Calabrese_2012}%
  \BibitemOpen
  \bibfield  {author} {\bibinfo {author} {\bibfnamefont {P.}~\bibnamefont
  {Calabrese}}, \bibinfo {author} {\bibfnamefont {F.~H.~L.}\ \bibnamefont
  {Essler}}, \ and\ \bibinfo {author} {\bibfnamefont {M.}~\bibnamefont
  {Fagotti}},\ }\href {\doibase 10.1088/1742-5468/2012/07/p07016} {\bibfield
  {journal} {\bibinfo  {journal} {Journal of Statistical Mechanics: Theory and
  Experiment}\ }\textbf {\bibinfo {volume} {2012}},\ \bibinfo {pages} {P07016}
  (\bibinfo {year} {2012})}\BibitemShut {NoStop}%
\bibitem [{\citenamefont {Calabrese}\ \emph {et~al.}(2011)\citenamefont
  {Calabrese}, \citenamefont {Essler},\ and\ \citenamefont
  {Fagotti}}]{PhysRevLett.106.227203}%
  \BibitemOpen
  \bibfield  {author} {\bibinfo {author} {\bibfnamefont {P.}~\bibnamefont
  {Calabrese}}, \bibinfo {author} {\bibfnamefont {F.~H.~L.}\ \bibnamefont
  {Essler}}, \ and\ \bibinfo {author} {\bibfnamefont {M.}~\bibnamefont
  {Fagotti}},\ }\href {\doibase 10.1103/PhysRevLett.106.227203} {\bibfield
  {journal} {\bibinfo  {journal} {Phys. Rev. Lett.}\ }\textbf {\bibinfo
  {volume} {106}},\ \bibinfo {pages} {227203} (\bibinfo {year}
  {2011})}\BibitemShut {NoStop}%
\bibitem [{\citenamefont {Sachdev}\ and\ \citenamefont
  {Young}(1997)}]{PhysRevLett.78.2220}%
  \BibitemOpen
  \bibfield  {author} {\bibinfo {author} {\bibfnamefont {S.}~\bibnamefont
  {Sachdev}}\ and\ \bibinfo {author} {\bibfnamefont {A.~P.}\ \bibnamefont
  {Young}},\ }\href {\doibase 10.1103/PhysRevLett.78.2220} {\bibfield
  {journal} {\bibinfo  {journal} {Phys. Rev. Lett.}\ }\textbf {\bibinfo
  {volume} {78}},\ \bibinfo {pages} {2220} (\bibinfo {year}
  {1997})}\BibitemShut {NoStop}%
\bibitem [{\citenamefont {Calabrese}\ and\ \citenamefont
  {Cardy}(2006)}]{PhysRevLett.96.136801}%
  \BibitemOpen
  \bibfield  {author} {\bibinfo {author} {\bibfnamefont {P.}~\bibnamefont
  {Calabrese}}\ and\ \bibinfo {author} {\bibfnamefont {J.}~\bibnamefont
  {Cardy}},\ }\href {\doibase 10.1103/PhysRevLett.96.136801} {\bibfield
  {journal} {\bibinfo  {journal} {Phys. Rev. Lett.}\ }\textbf {\bibinfo
  {volume} {96}},\ \bibinfo {pages} {136801} (\bibinfo {year}
  {2006})}\BibitemShut {NoStop}%
\bibitem [{\citenamefont {Rossini}\ \emph {et~al.}(2009)\citenamefont
  {Rossini}, \citenamefont {Silva}, \citenamefont {Mussardo},\ and\
  \citenamefont {Santoro}}]{PhysRevLett.102.127204}%
  \BibitemOpen
  \bibfield  {author} {\bibinfo {author} {\bibfnamefont {D.}~\bibnamefont
  {Rossini}}, \bibinfo {author} {\bibfnamefont {A.}~\bibnamefont {Silva}},
  \bibinfo {author} {\bibfnamefont {G.}~\bibnamefont {Mussardo}}, \ and\
  \bibinfo {author} {\bibfnamefont {G.~E.}\ \bibnamefont {Santoro}},\ }\href
  {\doibase 10.1103/PhysRevLett.102.127204} {\bibfield  {journal} {\bibinfo
  {journal} {Phys. Rev. Lett.}\ }\textbf {\bibinfo {volume} {102}},\ \bibinfo
  {pages} {127204} (\bibinfo {year} {2009})}\BibitemShut {NoStop}%
\bibitem [{\citenamefont {Barmettler}\ \emph {et~al.}(2009)\citenamefont
  {Barmettler}, \citenamefont {Punk}, \citenamefont {Gritsev}, \citenamefont
  {Demler},\ and\ \citenamefont {Altman}}]{PhysRevLett.102.130603}%
  \BibitemOpen
  \bibfield  {author} {\bibinfo {author} {\bibfnamefont {P.}~\bibnamefont
  {Barmettler}}, \bibinfo {author} {\bibfnamefont {M.}~\bibnamefont {Punk}},
  \bibinfo {author} {\bibfnamefont {V.}~\bibnamefont {Gritsev}}, \bibinfo
  {author} {\bibfnamefont {E.}~\bibnamefont {Demler}}, \ and\ \bibinfo {author}
  {\bibfnamefont {E.}~\bibnamefont {Altman}},\ }\href {\doibase
  10.1103/PhysRevLett.102.130603} {\bibfield  {journal} {\bibinfo  {journal}
  {Phys. Rev. Lett.}\ }\textbf {\bibinfo {volume} {102}},\ \bibinfo {pages}
  {130603} (\bibinfo {year} {2009})}\BibitemShut {NoStop}%
\bibitem [{\citenamefont {Karl}\ \emph {et~al.}(2017)\citenamefont {Karl},
  \citenamefont {Cakir}, \citenamefont {Halimeh}, \citenamefont {Oberthaler},
  \citenamefont {Kastner},\ and\ \citenamefont
  {Gasenzer}}]{PhysRevE.96.022110}%
  \BibitemOpen
  \bibfield  {author} {\bibinfo {author} {\bibfnamefont {M.}~\bibnamefont
  {Karl}}, \bibinfo {author} {\bibfnamefont {H.}~\bibnamefont {Cakir}},
  \bibinfo {author} {\bibfnamefont {J.~C.}\ \bibnamefont {Halimeh}}, \bibinfo
  {author} {\bibfnamefont {M.~K.}\ \bibnamefont {Oberthaler}}, \bibinfo
  {author} {\bibfnamefont {M.}~\bibnamefont {Kastner}}, \ and\ \bibinfo
  {author} {\bibfnamefont {T.}~\bibnamefont {Gasenzer}},\ }\href {\doibase
  10.1103/PhysRevE.96.022110} {\bibfield  {journal} {\bibinfo  {journal} {Phys.
  Rev. E}\ }\textbf {\bibinfo {volume} {96}},\ \bibinfo {pages} {022110}
  (\bibinfo {year} {2017})}\BibitemShut {NoStop}%
\bibitem [{\citenamefont {Yang}\ \emph {et~al.}(2019)\citenamefont {Yang},
  \citenamefont {Tian}, \citenamefont {Yang}, \citenamefont {Qiu},
  \citenamefont {Liang}, \citenamefont {Chu}, \citenamefont
  {Da\ifmmode~\breve{g}\else \u{g}\fi{}}, \citenamefont {Xu}, \citenamefont
  {Liu},\ and\ \citenamefont {Duan}}]{PhysRevA.100.013622}%
  \BibitemOpen
  \bibfield  {author} {\bibinfo {author} {\bibfnamefont {H.-X.}\ \bibnamefont
  {Yang}}, \bibinfo {author} {\bibfnamefont {T.}~\bibnamefont {Tian}}, \bibinfo
  {author} {\bibfnamefont {Y.-B.}\ \bibnamefont {Yang}}, \bibinfo {author}
  {\bibfnamefont {L.-Y.}\ \bibnamefont {Qiu}}, \bibinfo {author} {\bibfnamefont
  {H.-Y.}\ \bibnamefont {Liang}}, \bibinfo {author} {\bibfnamefont {A.-J.}\
  \bibnamefont {Chu}}, \bibinfo {author} {\bibfnamefont {C.~B.}\ \bibnamefont
  {Da\ifmmode~\breve{g}\else \u{g}\fi{}}}, \bibinfo {author} {\bibfnamefont
  {Y.}~\bibnamefont {Xu}}, \bibinfo {author} {\bibfnamefont {Y.}~\bibnamefont
  {Liu}}, \ and\ \bibinfo {author} {\bibfnamefont {L.-M.}\ \bibnamefont
  {Duan}},\ }\href {\doibase 10.1103/PhysRevA.100.013622} {\bibfield  {journal}
  {\bibinfo  {journal} {Phys. Rev. A}\ }\textbf {\bibinfo {volume} {100}},\
  \bibinfo {pages} {013622} (\bibinfo {year} {2019})}\BibitemShut {NoStop}%
\bibitem [{\citenamefont {Tian}\ \emph {et~al.}(2020)\citenamefont {Tian},
  \citenamefont {Yang}, \citenamefont {Qiu}, \citenamefont {Liang},
  \citenamefont {Yang}, \citenamefont {Xu},\ and\ \citenamefont
  {Duan}}]{PhysRevLett.124.043001}%
  \BibitemOpen
  \bibfield  {author} {\bibinfo {author} {\bibfnamefont {T.}~\bibnamefont
  {Tian}}, \bibinfo {author} {\bibfnamefont {H.-X.}\ \bibnamefont {Yang}},
  \bibinfo {author} {\bibfnamefont {L.-Y.}\ \bibnamefont {Qiu}}, \bibinfo
  {author} {\bibfnamefont {H.-Y.}\ \bibnamefont {Liang}}, \bibinfo {author}
  {\bibfnamefont {Y.-B.}\ \bibnamefont {Yang}}, \bibinfo {author}
  {\bibfnamefont {Y.}~\bibnamefont {Xu}}, \ and\ \bibinfo {author}
  {\bibfnamefont {L.-M.}\ \bibnamefont {Duan}},\ }\href {\doibase
  10.1103/PhysRevLett.124.043001} {\bibfield  {journal} {\bibinfo  {journal}
  {Phys. Rev. Lett.}\ }\textbf {\bibinfo {volume} {124}},\ \bibinfo {pages}
  {043001} (\bibinfo {year} {2020})}\BibitemShut {NoStop}%
\bibitem [{\citenamefont {{Bakr}}\ \emph {et~al.}(2009)\citenamefont {{Bakr}},
  \citenamefont {{Gillen}}, \citenamefont {{Peng}}, \citenamefont
  {{F{\"o}lling}},\ and\ \citenamefont {{Greiner}}}]{2009Natur.462...74B}%
  \BibitemOpen
  \bibfield  {author} {\bibinfo {author} {\bibfnamefont {W.~S.}\ \bibnamefont
  {{Bakr}}}, \bibinfo {author} {\bibfnamefont {J.~I.}\ \bibnamefont
  {{Gillen}}}, \bibinfo {author} {\bibfnamefont {A.}~\bibnamefont {{Peng}}},
  \bibinfo {author} {\bibfnamefont {S.}~\bibnamefont {{F{\"o}lling}}}, \ and\
  \bibinfo {author} {\bibfnamefont {M.}~\bibnamefont {{Greiner}}},\ }\href
  {\doibase 10.1038/nature08482} {\bibfield  {journal} {\bibinfo  {journal}
  {\nat}\ }\textbf {\bibinfo {volume} {462}},\ \bibinfo {pages} {74} (\bibinfo
  {year} {2009})},\ \Eprint {http://arxiv.org/abs/0908.0174} {arXiv:0908.0174
  [cond-mat.quant-gas]} \BibitemShut {NoStop}%
\bibitem [{\citenamefont {Swingle}\ \emph {et~al.}(2016)\citenamefont
  {Swingle}, \citenamefont {Bentsen}, \citenamefont {Schleier-Smith},\ and\
  \citenamefont {Hayden}}]{PhysRevA.94.040302}%
  \BibitemOpen
  \bibfield  {author} {\bibinfo {author} {\bibfnamefont {B.}~\bibnamefont
  {Swingle}}, \bibinfo {author} {\bibfnamefont {G.}~\bibnamefont {Bentsen}},
  \bibinfo {author} {\bibfnamefont {M.}~\bibnamefont {Schleier-Smith}}, \ and\
  \bibinfo {author} {\bibfnamefont {P.}~\bibnamefont {Hayden}},\ }\href
  {\doibase 10.1103/PhysRevA.94.040302} {\bibfield  {journal} {\bibinfo
  {journal} {Phys. Rev. A}\ }\textbf {\bibinfo {volume} {94}},\ \bibinfo
  {pages} {040302} (\bibinfo {year} {2016})}\BibitemShut {NoStop}%
\bibitem [{\citenamefont {Da\ifmmode~\breve{g}\else \u{g}\fi{}}\ and\
  \citenamefont {Duan}(2019)}]{PhysRevA.99.052322}%
  \BibitemOpen
  \bibfield  {author} {\bibinfo {author} {\bibfnamefont {C.~B.}\ \bibnamefont
  {Da\ifmmode~\breve{g}\else \u{g}\fi{}}}\ and\ \bibinfo {author}
  {\bibfnamefont {L.-M.}\ \bibnamefont {Duan}},\ }\href {\doibase
  10.1103/PhysRevA.99.052322} {\bibfield  {journal} {\bibinfo  {journal} {Phys.
  Rev. A}\ }\textbf {\bibinfo {volume} {99}},\ \bibinfo {pages} {052322}
  (\bibinfo {year} {2019})}\BibitemShut {NoStop}%
\bibitem [{\citenamefont {Zhu}\ \emph {et~al.}(2016)\citenamefont {Zhu},
  \citenamefont {Hafezi},\ and\ \citenamefont {Grover}}]{PhysRevA.94.062329}%
  \BibitemOpen
  \bibfield  {author} {\bibinfo {author} {\bibfnamefont {G.}~\bibnamefont
  {Zhu}}, \bibinfo {author} {\bibfnamefont {M.}~\bibnamefont {Hafezi}}, \ and\
  \bibinfo {author} {\bibfnamefont {T.}~\bibnamefont {Grover}},\ }\href
  {\doibase 10.1103/PhysRevA.94.062329} {\bibfield  {journal} {\bibinfo
  {journal} {Phys. Rev. A}\ }\textbf {\bibinfo {volume} {94}},\ \bibinfo
  {pages} {062329} (\bibinfo {year} {2016})}\BibitemShut {NoStop}%
\bibitem [{\citenamefont {{Yao}}\ \emph {et~al.}(2016)\citenamefont {{Yao}},
  \citenamefont {{Grusdt}}, \citenamefont {{Swingle}}, \citenamefont {{Lukin}},
  \citenamefont {{Stamper-Kurn}}, \citenamefont {{Moore}},\ and\ \citenamefont
  {{Demler}}}]{2016arXiv160701801Y}%
  \BibitemOpen
  \bibfield  {author} {\bibinfo {author} {\bibfnamefont {N.~Y.}\ \bibnamefont
  {{Yao}}}, \bibinfo {author} {\bibfnamefont {F.}~\bibnamefont {{Grusdt}}},
  \bibinfo {author} {\bibfnamefont {B.}~\bibnamefont {{Swingle}}}, \bibinfo
  {author} {\bibfnamefont {M.~D.}\ \bibnamefont {{Lukin}}}, \bibinfo {author}
  {\bibfnamefont {D.~M.}\ \bibnamefont {{Stamper-Kurn}}}, \bibinfo {author}
  {\bibfnamefont {J.~E.}\ \bibnamefont {{Moore}}}, \ and\ \bibinfo {author}
  {\bibfnamefont {E.~A.}\ \bibnamefont {{Demler}}},\ }\href@noop {} {\bibfield
  {journal} {\bibinfo  {journal} {arXiv e-prints}\ ,\ \bibinfo {eid}
  {arXiv:1607.01801}} (\bibinfo {year} {2016})},\ \Eprint
  {http://arxiv.org/abs/1607.01801} {arXiv:1607.01801 [quant-ph]} \BibitemShut
  {NoStop}%
\bibitem [{\citenamefont {Vermersch}\ \emph {et~al.}(2019)\citenamefont
  {Vermersch}, \citenamefont {Elben}, \citenamefont {Sieberer}, \citenamefont
  {Yao},\ and\ \citenamefont {Zoller}}]{PhysRevX.9.021061}%
  \BibitemOpen
  \bibfield  {author} {\bibinfo {author} {\bibfnamefont {B.}~\bibnamefont
  {Vermersch}}, \bibinfo {author} {\bibfnamefont {A.}~\bibnamefont {Elben}},
  \bibinfo {author} {\bibfnamefont {L.~M.}\ \bibnamefont {Sieberer}}, \bibinfo
  {author} {\bibfnamefont {N.~Y.}\ \bibnamefont {Yao}}, \ and\ \bibinfo
  {author} {\bibfnamefont {P.}~\bibnamefont {Zoller}},\ }\href {\doibase
  10.1103/PhysRevX.9.021061} {\bibfield  {journal} {\bibinfo  {journal} {Phys.
  Rev. X}\ }\textbf {\bibinfo {volume} {9}},\ \bibinfo {pages} {021061}
  (\bibinfo {year} {2019})}\BibitemShut {NoStop}%
\bibitem [{ITe()}]{ITensor}%
  \BibitemOpen
  \href@noop {} {\bibinfo  {journal} {\mbox{ITensor Library} (version 2.0.11)
  http://itensor.org}\ }\BibitemShut {NoStop}%
\bibitem [{\citenamefont {Lieb}\ and\ \citenamefont
  {Robinson}(1972)}]{lieb1972}%
  \BibitemOpen
\bibfield  {journal} {  }\bibfield  {author} {\bibinfo {author} {\bibfnamefont
  {E.~H.}\ \bibnamefont {Lieb}}\ and\ \bibinfo {author} {\bibfnamefont {D.~W.}\
  \bibnamefont {Robinson}},\ }\href@noop {} {\bibfield  {journal} {\bibinfo
  {journal} {Communications in Mathematical Physics}\ }\textbf {\bibinfo
  {volume} {28}},\ \bibinfo {pages} {251} (\bibinfo {year} {1972})}\BibitemShut
  {NoStop}%
\bibitem [{\citenamefont {Calabrese}\ and\ \citenamefont
  {Cardy}(2005)}]{Calabrese_2005}%
  \BibitemOpen
  \bibfield  {author} {\bibinfo {author} {\bibfnamefont {P.}~\bibnamefont
  {Calabrese}}\ and\ \bibinfo {author} {\bibfnamefont {J.}~\bibnamefont
  {Cardy}},\ }\href {\doibase 10.1088/1742-5468/2005/04/p04010} {\bibfield
  {journal} {\bibinfo  {journal} {Journal of Statistical Mechanics: Theory and
  Experiment}\ }\textbf {\bibinfo {volume} {2005}},\ \bibinfo {pages} {P04010}
  (\bibinfo {year} {2005})}\BibitemShut {NoStop}%
\bibitem [{\citenamefont {{Wang}}\ \emph {et~al.}(2020)\citenamefont {{Wang}},
  \citenamefont {{Foss-Feig}},\ and\ \citenamefont
  {{Hazzard}}}]{2020arXiv200912032W}%
  \BibitemOpen
  \bibfield  {author} {\bibinfo {author} {\bibfnamefont {Z.}~\bibnamefont
  {{Wang}}}, \bibinfo {author} {\bibfnamefont {M.}~\bibnamefont {{Foss-Feig}}},
  \ and\ \bibinfo {author} {\bibfnamefont {K.~R.~A.}\ \bibnamefont
  {{Hazzard}}},\ }\href@noop {} {\bibfield  {journal} {\bibinfo  {journal}
  {arXiv e-prints}\ ,\ \bibinfo {eid} {arXiv:2009.12032}} (\bibinfo {year}
  {2020})},\ \Eprint {http://arxiv.org/abs/2009.12032} {arXiv:2009.12032
  [quant-ph]} \BibitemShut {NoStop}%
\bibitem [{\citenamefont {Clauset}\ \emph {et~al.}(2009)\citenamefont
  {Clauset}, \citenamefont {Shalizi},\ and\ \citenamefont
  {Newman}}]{doi:10.1137/070710111}%
  \BibitemOpen
  \bibfield  {author} {\bibinfo {author} {\bibfnamefont {A.}~\bibnamefont
  {Clauset}}, \bibinfo {author} {\bibfnamefont {C.~R.}\ \bibnamefont
  {Shalizi}}, \ and\ \bibinfo {author} {\bibfnamefont {M.~E.~J.}\ \bibnamefont
  {Newman}},\ }\href {\doibase 10.1137/070710111} {\bibfield  {journal}
  {\bibinfo  {journal} {SIAM Review}\ }\textbf {\bibinfo {volume} {51}},\
  \bibinfo {pages} {661} (\bibinfo {year} {2009})},\ \Eprint
  {http://arxiv.org/abs/https://doi.org/10.1137/070710111}
  {https://doi.org/10.1137/070710111} \BibitemShut {NoStop}%
\bibitem [{\citenamefont {Foss-Feig}\ \emph {et~al.}(2015)\citenamefont
  {Foss-Feig}, \citenamefont {Gong}, \citenamefont {Clark},\ and\ \citenamefont
  {Gorshkov}}]{PhysRevLett.114.157201}%
  \BibitemOpen
  \bibfield  {author} {\bibinfo {author} {\bibfnamefont {M.}~\bibnamefont
  {Foss-Feig}}, \bibinfo {author} {\bibfnamefont {Z.-X.}\ \bibnamefont {Gong}},
  \bibinfo {author} {\bibfnamefont {C.~W.}\ \bibnamefont {Clark}}, \ and\
  \bibinfo {author} {\bibfnamefont {A.~V.}\ \bibnamefont {Gorshkov}},\ }\href
  {\doibase 10.1103/PhysRevLett.114.157201} {\bibfield  {journal} {\bibinfo
  {journal} {Phys. Rev. Lett.}\ }\textbf {\bibinfo {volume} {114}},\ \bibinfo
  {pages} {157201} (\bibinfo {year} {2015})}\BibitemShut {NoStop}%
\bibitem [{\citenamefont {Budich}\ and\ \citenamefont
  {Heyl}(2016)}]{PhysRevB.93.085416}%
  \BibitemOpen
  \bibfield  {author} {\bibinfo {author} {\bibfnamefont {J.~C.}\ \bibnamefont
  {Budich}}\ and\ \bibinfo {author} {\bibfnamefont {M.}~\bibnamefont {Heyl}},\
  }\href {\doibase 10.1103/PhysRevB.93.085416} {\bibfield  {journal} {\bibinfo
  {journal} {Phys. Rev. B}\ }\textbf {\bibinfo {volume} {93}},\ \bibinfo
  {pages} {085416} (\bibinfo {year} {2016})}\BibitemShut {NoStop}%
\bibitem [{\citenamefont {Da\ifmmode~\breve{g}\else \u{g}\fi{}}\ \emph
  {et~al.}(2020)\citenamefont {Da\ifmmode~\breve{g}\else \u{g}\fi{}},
  \citenamefont {Duan},\ and\ \citenamefont {Sun}}]{PhysRevB.101.104415}%
  \BibitemOpen
  \bibfield  {author} {\bibinfo {author} {\bibfnamefont {C.~B.}\ \bibnamefont
  {Da\ifmmode~\breve{g}\else \u{g}\fi{}}}, \bibinfo {author} {\bibfnamefont
  {L.-M.}\ \bibnamefont {Duan}}, \ and\ \bibinfo {author} {\bibfnamefont
  {K.}~\bibnamefont {Sun}},\ }\href {\doibase 10.1103/PhysRevB.101.104415}
  {\bibfield  {journal} {\bibinfo  {journal} {Phys. Rev. B}\ }\textbf {\bibinfo
  {volume} {101}},\ \bibinfo {pages} {104415} (\bibinfo {year}
  {2020})}\BibitemShut {NoStop}%
\bibitem [{\citenamefont {{Lieb}}\ \emph {et~al.}(1961)\citenamefont {{Lieb}},
  \citenamefont {{Schultz}},\ and\ \citenamefont
  {{Mattis}}}]{1961AnPhy..16..407L}%
  \BibitemOpen
  \bibfield  {author} {\bibinfo {author} {\bibfnamefont {E.}~\bibnamefont
  {{Lieb}}}, \bibinfo {author} {\bibfnamefont {T.}~\bibnamefont {{Schultz}}}, \
  and\ \bibinfo {author} {\bibfnamefont {D.}~\bibnamefont {{Mattis}}},\ }\href
  {\doibase 10.1016/0003-4916(61)90115-4} {\bibfield  {journal} {\bibinfo
  {journal} {Annals of Physics}\ }\textbf {\bibinfo {volume} {16}},\ \bibinfo
  {pages} {407} (\bibinfo {year} {1961})}\BibitemShut {NoStop}%
\end{thebibliography}

%


\appendix

\section{Periodic vs. open boundaries}

\begin{figure}[H]
\centering \includegraphics[width=0.4\textwidth]{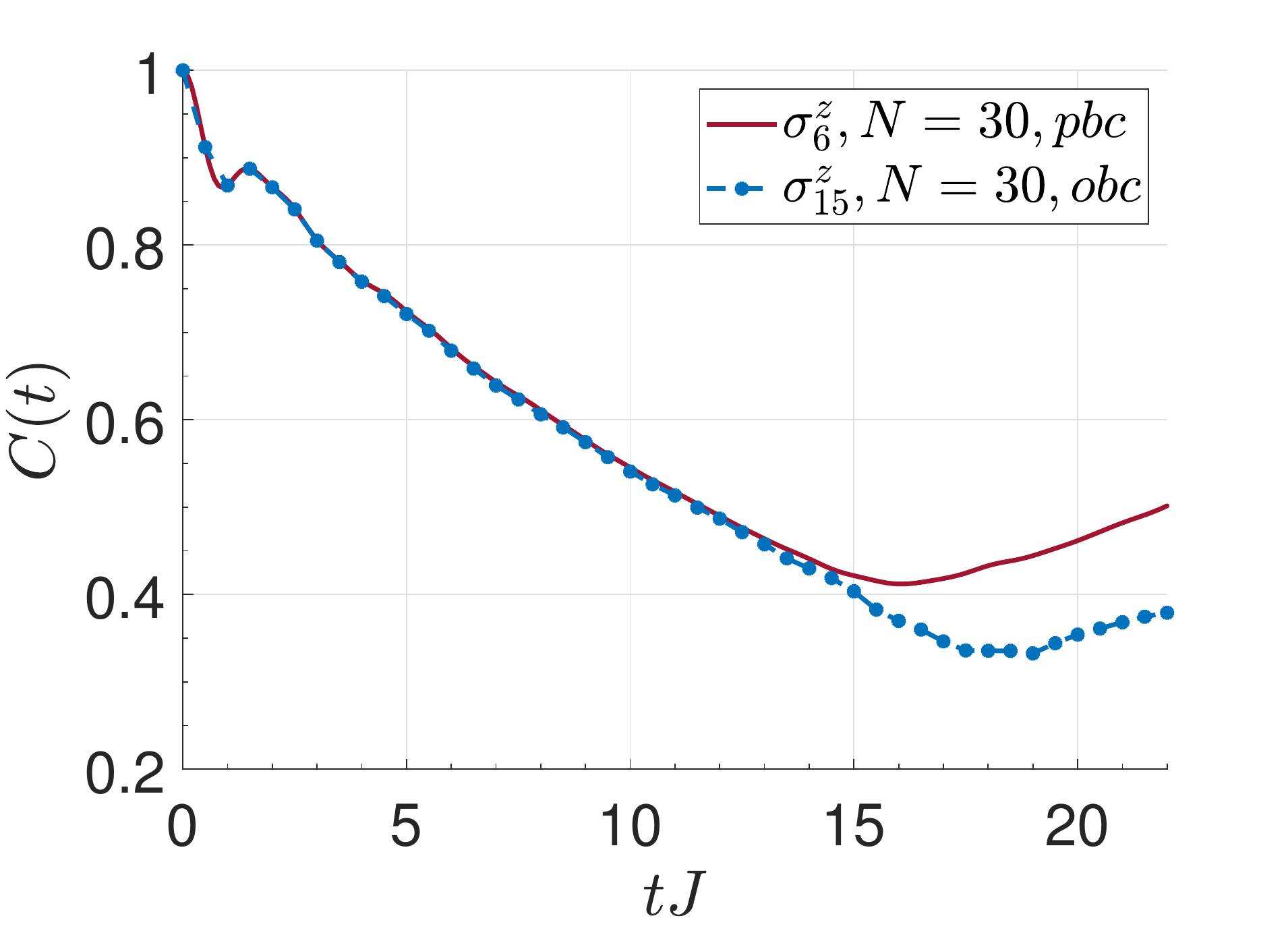}\hfill 
\caption{$C(t)$ nonequilibrium response of the middle spin of an open-boundary $N=30$, $\sigma^z_{15}$ (blue-dotted) and a spin at an arbitrary location $\sigma^z_{6}$ in a periodic chain $N=30$ (red-solid).}
\label{FigS1}
\end{figure}

In this section, we demonstrate how the spin operator (longitudinal magnetization per site) in the middle of an open-boundary chain exhibits exponential decay comparable with a spin operator at an arbitrary site in a periodic chain. Fig.~\ref{FigS1} compares the nonequilibrium responses of these two spins and as observed, the responses match with each other until the finite-size effects appear. This is reasonable, because a spin in the middle of the chain is equally distant to both edges, and hence it should exhibit behavior closest to a spin in a periodic chain. Therefore, based on this equivalence we can argue that the middle spin of an open-boundary chain behaves similar to total magnetization in exhibiting an exponential decay. This is simply because the total magnetization is an average over all spin operators $\sigma^z_i$.

\section{Mapping to noninteracting fermions in quench dynamics}

We map the integrable TFIM to noninteracting fermionic model in 1D via the transformation \cite{sachdev2001quantum},
\begin{eqnarray}
\sigma^z_i &=& - \prod_{j<i} \left(1-2c_j^{\dagger}c_j \right)\left(c_i+c_i^{\dagger}\right),\label{mapping} \\
\sigma^x_i &=& 1-2c_i^{\dagger}c_i,\notag \\
\sigma^y_i &=& -i \prod_{j<i} \left(1-2c_j^{\dagger}c_j \right)\left(c_i-c_i^{\dagger}\right),\notag
\end{eqnarray}
to obtain the noninteracting Hamiltonian
\begin{eqnarray}
H &=& -J \sum_i \bigg ( c^{\dagger}_i c_{i+1} + c^{\dagger}_{i+1} c_i + c_i^{\dagger} c_{i+1}^{\dagger} \notag \\
&+& c_{i+1}c_i - 2 h c_i^{\dagger} c_i \bigg ). \label{nonInteractingH}
\end{eqnarray}
One can immediately see that calculating the dynamical evolution of a bulk spin $\Braket{\sigma_i^z(t)}$ in the noninteracting picture brings a string of operators and is not really tractable. Hence we instead calculate equal-time two-point correlators and invoke the cluster theorem \cite{Calabrese_2012},
\begin{eqnarray}
\Braket{\sigma_{i}^z(t)\sigma_{i+N/2}^z(t)} \sim \Braket{\sigma_{i}^z(t)} \Braket{\sigma_{i+N/2}^z(t)}. \label{clusterTheorem}
\end{eqnarray}
Cluster theorem holds in the lightcone, meaning for a time interval up until two sites $i$ and $i+N/2$ start getting correlated with each other due to operator spreading. The time when the theorem breaks down can be estimated based on the maximum quasiparticle velocities $v_q$, $t < \Delta x/(2v_q)$ where $\Delta x = N/2$ is the distance between two spins that are selected symmetrically around the symmetry center of a periodic chain in Eq.~\eqref{clusterTheorem}. Since each site in a periodic chain experiences identical dynamical response due to translational symmetry, one can write
\begin{eqnarray}
\Braket{\sigma_{i}^z(t)} = \sqrt{\Braket{\sigma_{i}^z(t)\sigma_{i+N/2}^z(t)}}.
\end{eqnarray}
Therefore, we need to calculate equal-time two-point correlators. Via introducing auxiliary operators, $\phi^{\pm}_i=c_i^{\dagger}\pm c_i$, one can write
\begin{eqnarray}
\Braket{\sigma_{i}^z(t)\sigma_{i+N/2}^z(t)} = \Braket{C_2(t)} &=& \label{SOP} \\
\phi_i^-(t) \bigg( \prod_{j=i+1}^{i+N/2-1}  \phi_j^+(t) \phi_j^-(t) \bigg) & \phi_{i+N/2}^+(t) &. \notag
\end{eqnarray}
This is called string order parameter (SOP) \cite{PhysRevB.93.085416}, and to calculate it one needs to invoke Wick's theorem, write Eq.~\eqref{SOP} with two-point contractions and construct a Pfaffian matrix $T(t)$ at time $t$ to calculate $\Braket{C_2(t)}= |\text{Pf}(T(t))|$ \cite{sachdev2001quantum,Calabrese_2012}. Now we only need to calculate all possible elementary two-point contractions $\Braket{\phi_a^p(t) \phi_b^q(t)}$ where $p,q=\pm$ and $i\leq a \leq b \leq i+N/2$. Additionally we need to incorporate the mechanism of sudden quench in this picture. For this, we mainly follow the procedure outlined in Ref.~\cite{PhysRevB.101.245148}. Let us briefly review this procedure here.

Since we would like to quench from a polarized state, this quench point corresponds to $h_i=0$ where $h_i$ is the transverse field of the initial Hamiltonian in quench procedure. Therefore, we first solve the initial Hamiltonian $H_i$,
\begin{eqnarray}
H_i = \sum_k E_k^i \alpha_k^{\dagger}\alpha_k, 
\end{eqnarray}
where $E_k$ and $\alpha_k$ are the single particle eigenenergies and eigenstates, respectively. The solution reads in general terms,
\begin{eqnarray}
\left(\begin{array}{c}
\alpha \\
\alpha^{\dagger}
\end{array} \right) = \left( \begin{array}{c c}
G_i & F_i \\
F_i & G_i
\end{array} \right) \left(\begin{array}{c}
c_i \\
c_i^{\dagger}
\end{array} \right),
\end{eqnarray}
where $c_i=(c_1, c_2, \cdots, c_N)^{T}$ and similarly for the creation operator $c_i^{\dagger}$. Note that one can work in this Bogoliubov-de Gennes (BdG) basis with the size doubled \cite{PhysRevB.101.104415}, however here we work with the block matrices $G$ and $F$ \cite{1961AnPhy..16..407L} which is computationally more efficient. By solving the eigensystem of
\begin{eqnarray}
\left[(A_i - B_i)(A_i + B_i) \right] \Ket{\Phi_k^i} = (E_k^i)^2 \Ket{\Phi_k^i},
\end{eqnarray}
we obtain the eigenenergies $E_k^i$ and eigenvectors $\Ket{\Phi_k^i}$. Here $A_i$ and $B_i$ are the nearest neighbor hopping and the pairing terms in the Hamiltonian, respectively, so that the Hamiltonian could be written as,
\begin{eqnarray}
H_i = \left( \begin{array}{c c}
A_i & B_i \\
B_i^{\dagger} & -A_i
\end{array} \right),
\end{eqnarray}
in $(c \hspace{1mm} c^{\dagger})^T$ basis. Then we use the eigensystem $(E_k^i,\Ket{\Phi_k^i})$ to find
\begin{eqnarray}
\Ket{\Psi_k^i} = \frac{1}{E_k^i}\left[\Bra{\Phi_k^i} (A_i - B_i) \right]^T.
\end{eqnarray}
Now we can calculate the $G_i$ and $F_i$ in terms of $\Ket{\Phi_k^i}$ and $\Ket{\Psi_k^i}$. Noting that
\begin{eqnarray}
\Phi_i &=& \left[\Ket{\Phi_1^i} \hspace{1mm} \Ket{\Phi_2^i} \cdots \Ket{\Phi_N^i} \right], \notag \\
\Psi_i &=& \left[\Ket{\Psi_1^i} \hspace{1mm} \Ket{\Psi_2^i} \cdots \Ket{\Psi_N^i} \right], \notag
\end{eqnarray}
The block matrices follow
\begin{eqnarray}
G_i = \frac{1}{2}\left(\Phi_i^T + \Psi_i^T\right), \hspace{1mm} F_i = \frac{1}{2}\left(\Phi_i^T - \Psi_i^T\right).
\end{eqnarray}
A similar procedure follows for the final Hamiltonian $H_f$ with
\begin{eqnarray}
\left(\begin{array}{c}
\beta \\
\beta^{\dagger}
\end{array} \right) = \left( \begin{array}{c c}
G_f & F_f \\
F_f & G_f
\end{array} \right) \left(\begin{array}{c}
c_f \\
c_f^{\dagger}
\end{array} \right),
\end{eqnarray}
and corresponding $\Phi_f$ and $\Psi_f$. Based on the pairs of block matrices, we calculate the transfer matrices,
\begin{eqnarray}
T_1=G_f G_i^T + F_f F_i^T, \notag \\
T_2=G_f F_i^T + F_f G_i^T. \notag
\end{eqnarray}
Now we want to calculate the Pfaffian matrix elements, ${}_{\alpha}\Bra{\psi_0} \left[\phi^p_a \phi^q_b\right]_{\beta}\Ket{\psi_0}_{\alpha}$ where subscripts imply in which basis we have the states and the operators. Since we would like to make use of $\alpha \Ket{\psi_0}_{\alpha}=0$, we write $\left[\phi^p_a \phi^q_b\right]_{\beta}$ in the $\alpha$ basis.
\begin{eqnarray}
\left[\phi^{\pm}_b\right]_{\beta} \Ket{\psi_0}_{\alpha}  &=& \left[c_b^{\dagger}(t)\pm c_b(t)  \right]_{\beta}\Ket{\psi_0}_{\alpha}, \notag \\
&=& \left[(G_f^T \pm F_f^T)\left( e^{i \mathcal{E} t} T_1 \pm e^{-i \mathcal{E} t} T_2 \right) \alpha^{\dagger}\right]_b\Ket{\psi_0}_{\alpha}, \notag
\end{eqnarray}
where $\mathcal{E}$ is a diagonal matrix with eigenenergies of the final Hamiltonian as the entries, $\mathcal{E}=\text{diag}[E_1^f \hspace{1mm} E_2^f \cdots E_N^f]$. Based on this formulation, we construct matrices $\mathcal{M}_{q}(t)$ in an explicit form,
\begin{eqnarray}
\mathcal{M}_+(t) &=& \Phi_f \left( e^{-i \mathcal{E} t} T_1 + e^{i \mathcal{E} t} T_2 \right),\notag \\
\mathcal{M}_-(t) &=& \left(T_1^T e^{i \mathcal{E} t} - T_2^T e^{-i \mathcal{E} t} \right)\Psi_f^T,
\end{eqnarray}
to utilize in the following contractions,
\begin{eqnarray}
\Braket{\phi^+_a(t) \phi^+_b (t)} &=& [\mathcal{M}_+(t)\mathcal{M}_+(t)^{\dagger}]_{ab}, \notag\\
\Braket{\phi^-_a(t) \phi^-_b(t)} &=& -[\mathcal{M}_-^{\dagger}(t)\mathcal{M}_-(t)]_{ab}, \notag\\
\Braket{\phi^+_a(t) \phi^-_b(t)} &=& [\mathcal{M}_+(t)\mathcal{M}_-(t)]_{ab}, \notag \\
\Braket{\phi^-_a(t)\phi^+_b(t)} &=& -[\mathcal{M}_-^{\dagger}(t)\mathcal{M}_+^{\dagger}(t)]_{ab}. \label{contractions}
\end{eqnarray}
With these contractions, one can now construct the Pfaffian matrix $T(t)$ \cite{PhysRevB.101.245148}.

\section{Vanishing dynamical order for one-point observables}

Here we compare the nonequilibrium responses of a one-point observable and an OTOC, defined at the same site, middle of an open-boundary and quenched from a polarized state $\Ket{\psi_0}=\Ket{\uparrow \uparrow ... \uparrow}$. OTOC is defined as, $F(t)=\Bra{\psi_0}\sigma^z_{r}(t)\sigma^z_{r} \sigma^z_{r}(t) \sigma^z_{r}\Ket{\psi_0}$. 

\begin{figure}
\centering
\subfloat[]{\label{figS1a}\includegraphics[width=0.24\textwidth]{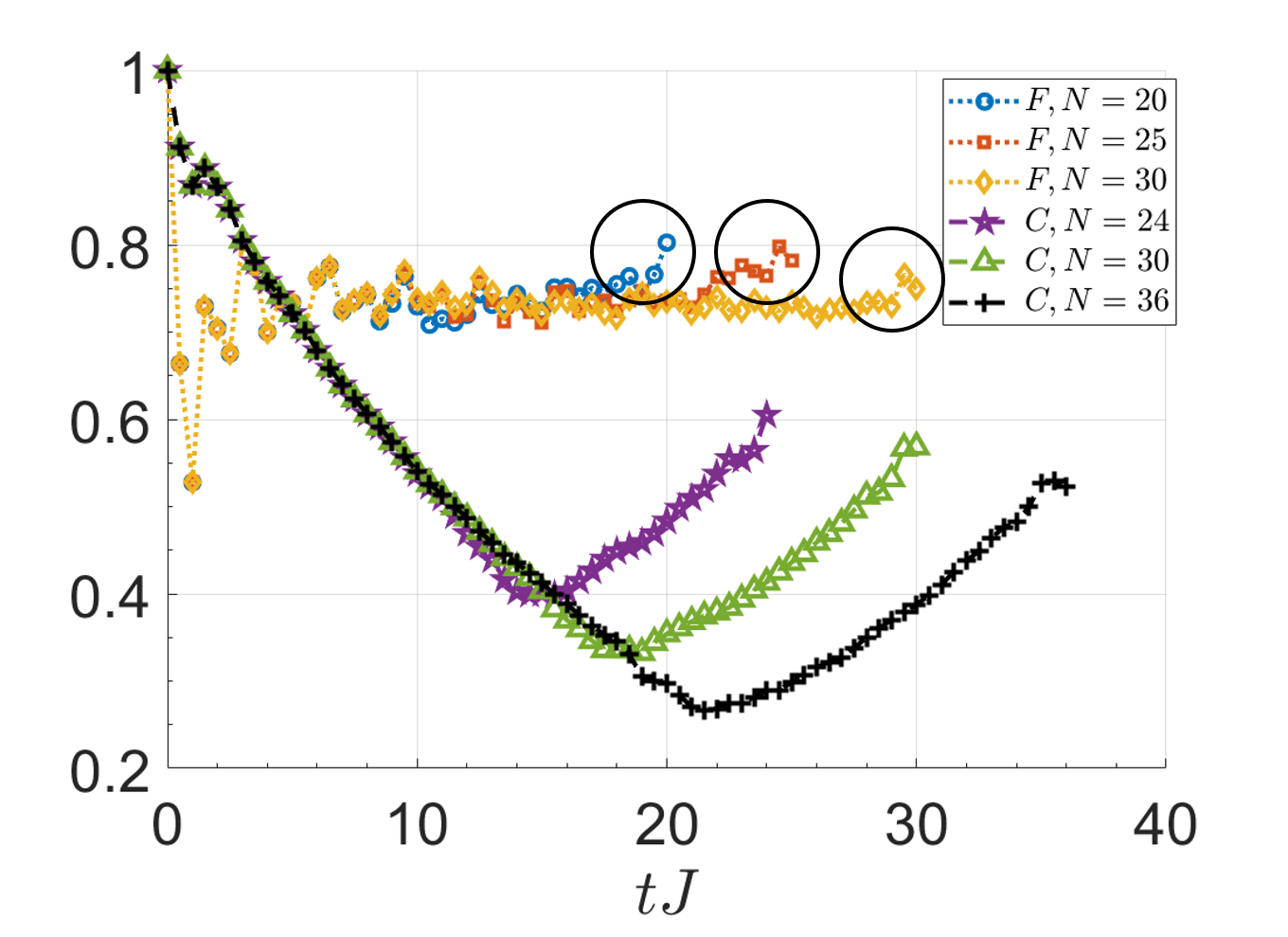}} \hfill
\subfloat[]{\label{figS1b}%
\includegraphics[width=0.24\textwidth]{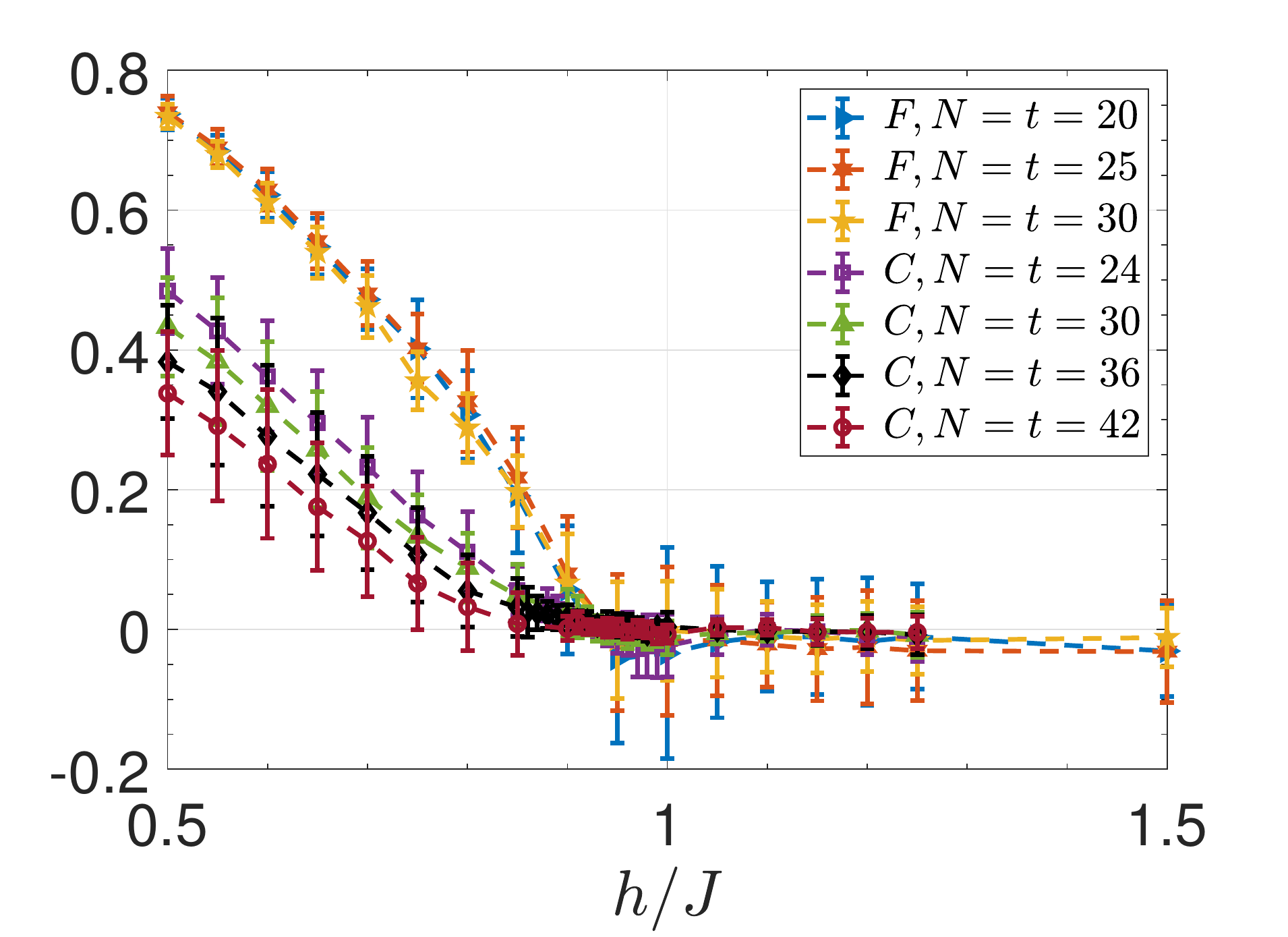}}
\hfill 
\caption{(a) One-point observable $C(t)$ and OTOC $F(t)$, both defined at a single site, for different system sizes $N$ for integrable TFIM at $h/J=0.5$; (b) A system-size dependent temporal cutoff is applied to $C(t)$ and $F(t)$ for a time interval of $t=N$ resulting in $\bar{C}$ and $\bar{F}$ with respect to control parameter $h/J$.}
\end{figure}

Fig.~\ref{figS1a} compares $F(t)$ and $C(t)$ for different system sizes $N$ computed via time-dependent density matrix renormalization group (t-DMRG) at transverse field $h/J=0.5$ in a time interval of $t=N$. When we apply a temporal cutoff of $t=N$, for OTOC, we observe that the dynamical order persists indefinitely resulting in a well-defined dynamical phase boundary for the time-average or long-time saturation value $\bar{F}$ in Fig.~\ref{figS1b}. Note that at $t \sim N$, $F(t)$ in Fig.~\ref{figS1a} starts to demonstrate finite-size effects, illustrated with black circles, which justifies the argument that $t \sim N$ is a sufficiently long-time limit $t\rightarrow \infty$ for chosen system sizes. With the same reasoning, one can plot $C(t)$ in a time interval of $t=N$ in Fig.~\ref{figS1a} and observe the decay of initial magnetization which dramatically becomes more pronounced as the system size increases, resulting in featureless long-time dynamics as well as a vanishing DPT-I boundary for $\bar{C}$ as seen in Fig.~\ref{figS1b}. 
The error bars in Fig.~\ref{figS1b} are $1\sigma$ standard deviation of the nonequilibrium response in time (due to oscillations) around the average of the response.

\section{Comparison between fixed and parametric temporal cutoffs in the open-boundary}

\begin{figure}
\centering
\subfloat[]{\label{figS9a}\includegraphics[width=0.24\textwidth]{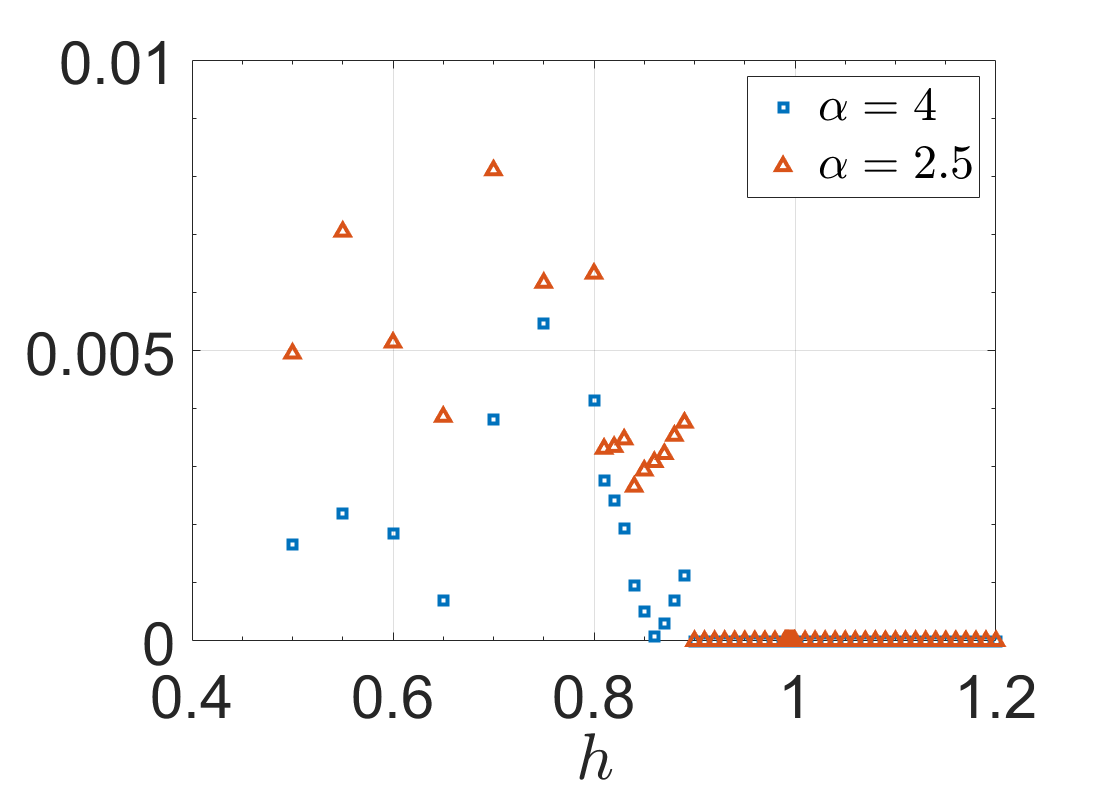}} \hfill
 \subfloat[]{\label{figS9b}%
\includegraphics[width=0.24\textwidth]{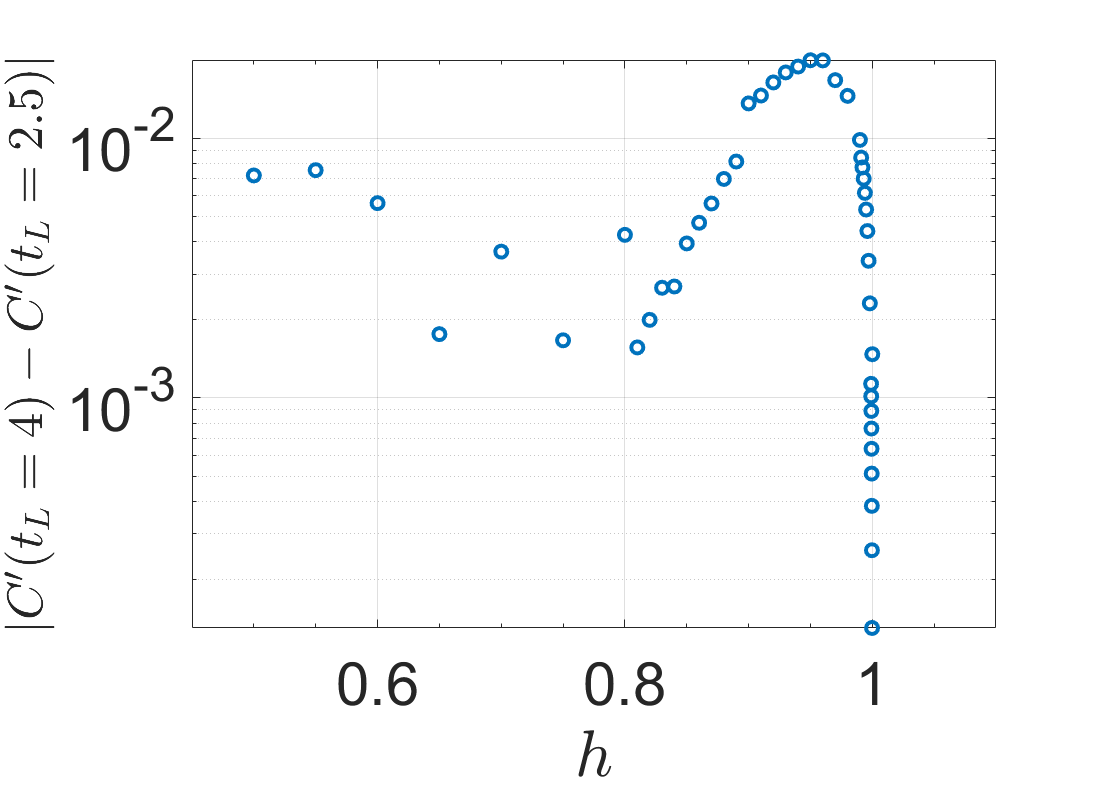}} \hfill
\caption{(a) The differences between rescaled observables with two different temporal cutoffs, parametric $2\alpha/v_q$ and fixed $\alpha$ for different $\alpha$ values (see legend). (b) The difference between the rescaled observables with two different fixed temporal cutoffs.}
\label{FigS9}
\end{figure}
\begin{figure*}
\centering
\subfloat[]{\label{figS17a}\includegraphics[width=0.3\textwidth]{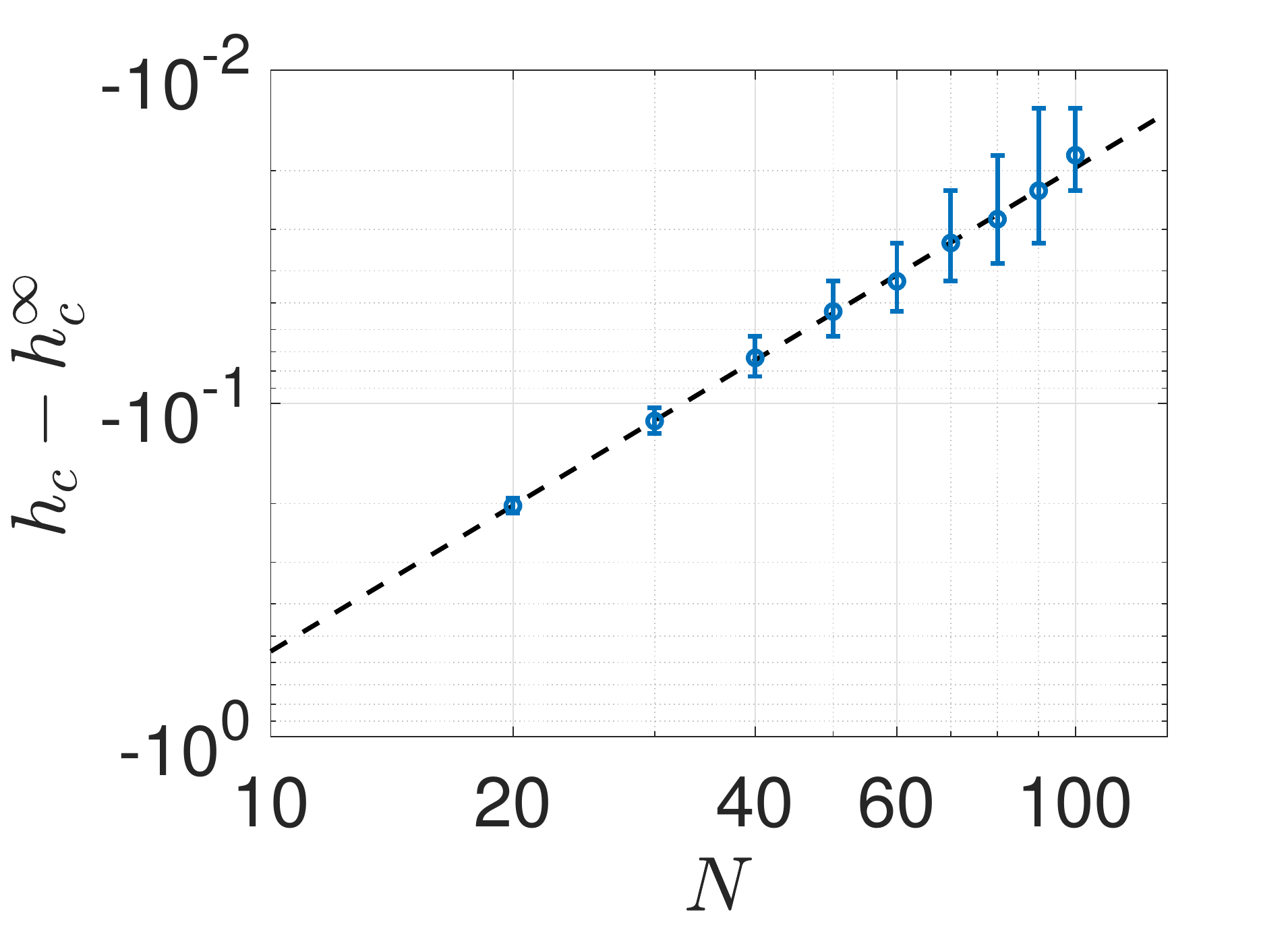}} \hfill
\subfloat[]{\label{figS17b}%
\includegraphics[width=0.3\textwidth]{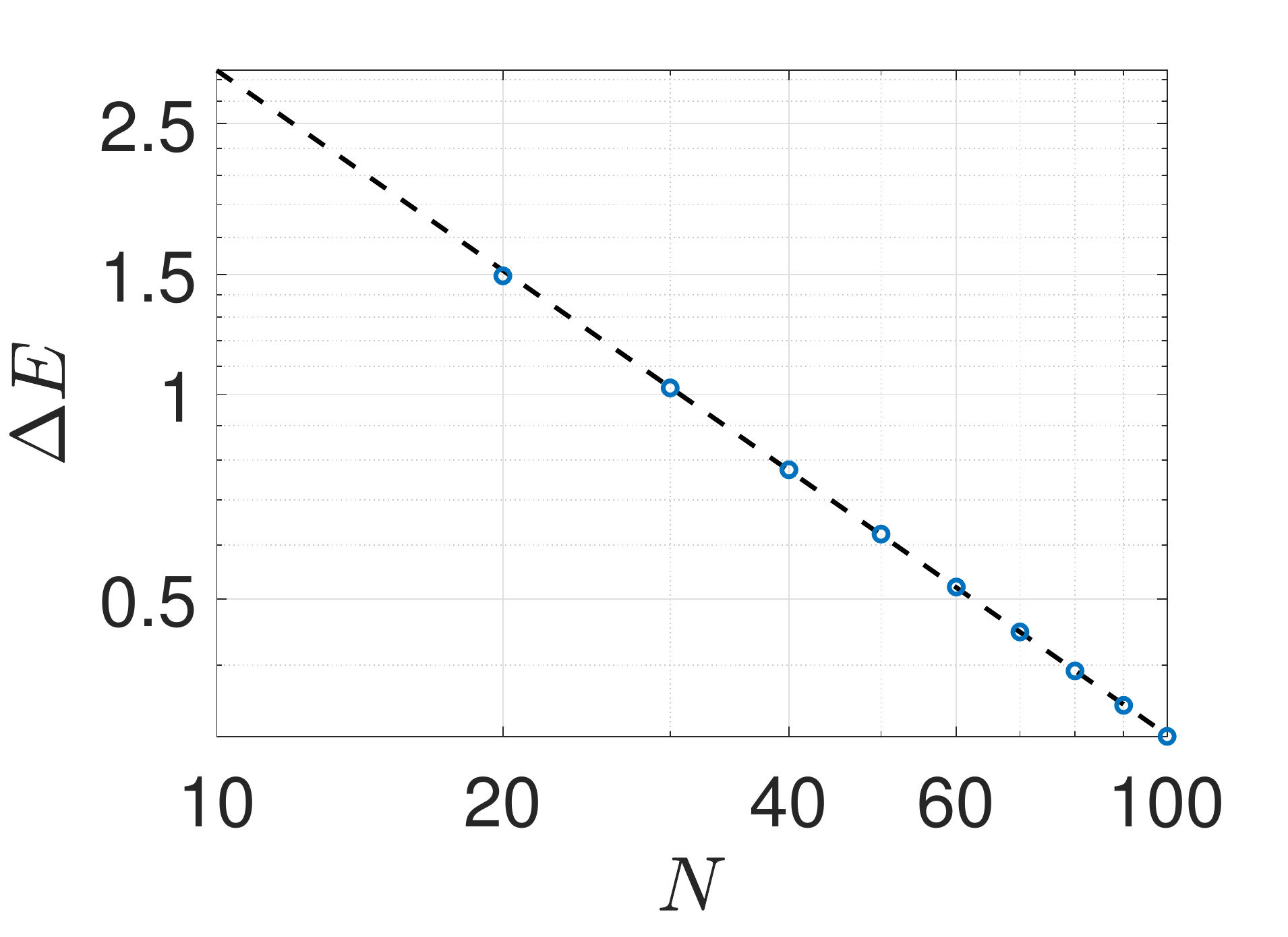}} \hfill
\subfloat[]{\label{figS17c}%
\includegraphics[width=0.3\textwidth]{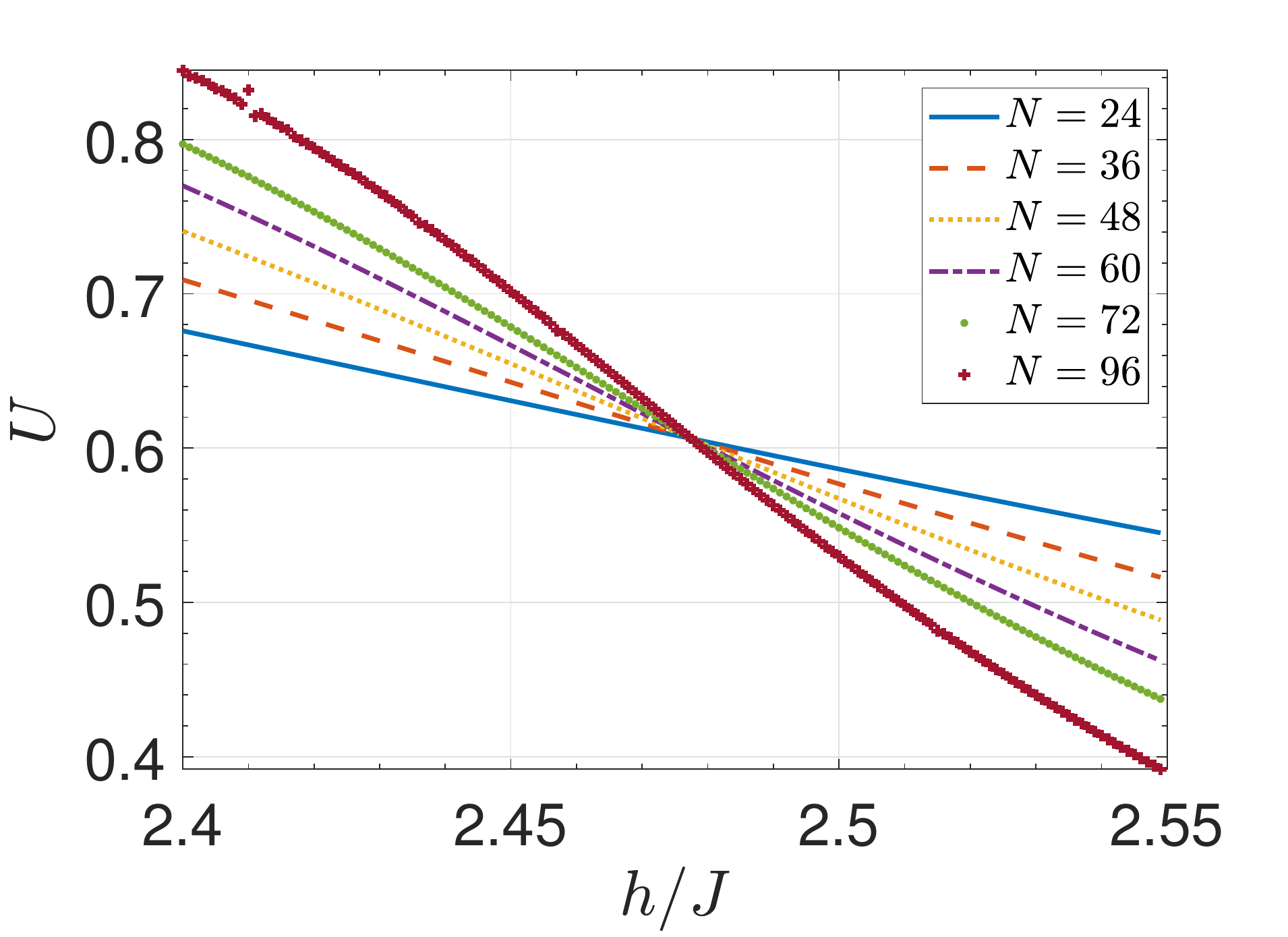}}
\hfill 
\caption{(a-b) Ground state energy gap analysis with respect to system size $N$ to determine the equilibrium QPT. (a) The critical point is marked as $h_c^{\infty}=2.463$ in thermodynamic limit via scaling analysis. (b) Energy gap $\Delta E$ closes as we approach the QPT boundary. The scaling exponent is $\Delta E = N^{-1}$. (c) Binder cumulant $U$ for different system sizes ranging between $N=24-96$, all crossing at $h_c=2.477\pm 0.001$.}
\label{FigS17}
\end{figure*}

In this section, we plot the difference between rescaled observable values with different choices of temporal cutoffs: (i) fixed $\alpha$ and parametric $2\alpha/v_q$ (ii) two fixed cutoffs in integrable TFIM. Even though these are clearly distinct temporal cutoffs, the differences are bounded for all $h/J$ values in the dynamically-ordered regime and more importantly the differences steadily decrease as we approach the crossover. Fig.~\ref{figS9a} demonstrates the differences between rescaled observable values generated with two types of temporal cutoffs for different $\alpha$ values. They are exactly zero in the vicinity of the crossover. This is likely because two types of temporal cutoffs converge to each other as we approach the crossover boundary. Fig.~\ref{figS9b} shows  the difference between rescaled observable values for two fixed temporal cutoffs. In Fig.~\ref{Fig2} in the main text, these differences seem to be the largest. Here we explicitly plot the difference and show that it steadily decreases as we approach the crossover boundary.

\section{Equilibrium QPT boundary for the nonintegrable TFIM}

In this section, we present the equilibrium phase transition boundary via both an analysis of ground state energy gap and Binder ratio for the nonintegrable TFIM with $\Delta/J=-1$. Figs.~\ref{figS17a}-\ref{figS17b} shows the determination of the QPT via energy gap analysis. We find that the equilibrium transition happens at $h_c\sim 2.463$ and the scaling exponent of the energy gap closing is $\delta \sim -1$. Further, we compute the Binder cumulant in Fig.~\ref{figS17c}, 
\begin{eqnarray}
U=\frac{3}{2}\left(1-\frac{1}{3}\frac{\Braket{S_z^4}}{\Braket{S_z^2}^2}  \right),
\end{eqnarray}
where $S_z=\sum_i^N \sigma^z_i$, the total magnetization operator. This method marks the QPT as $h_c^{\infty}=2.477\pm 0.001$. The equilibrium transition boundaries determined by these two different methods are very close.

\section{Error bar calculations}

\begin{figure}
\centering
 \subfloat[]{\label{figS10b}
\includegraphics[width=0.24\textwidth]{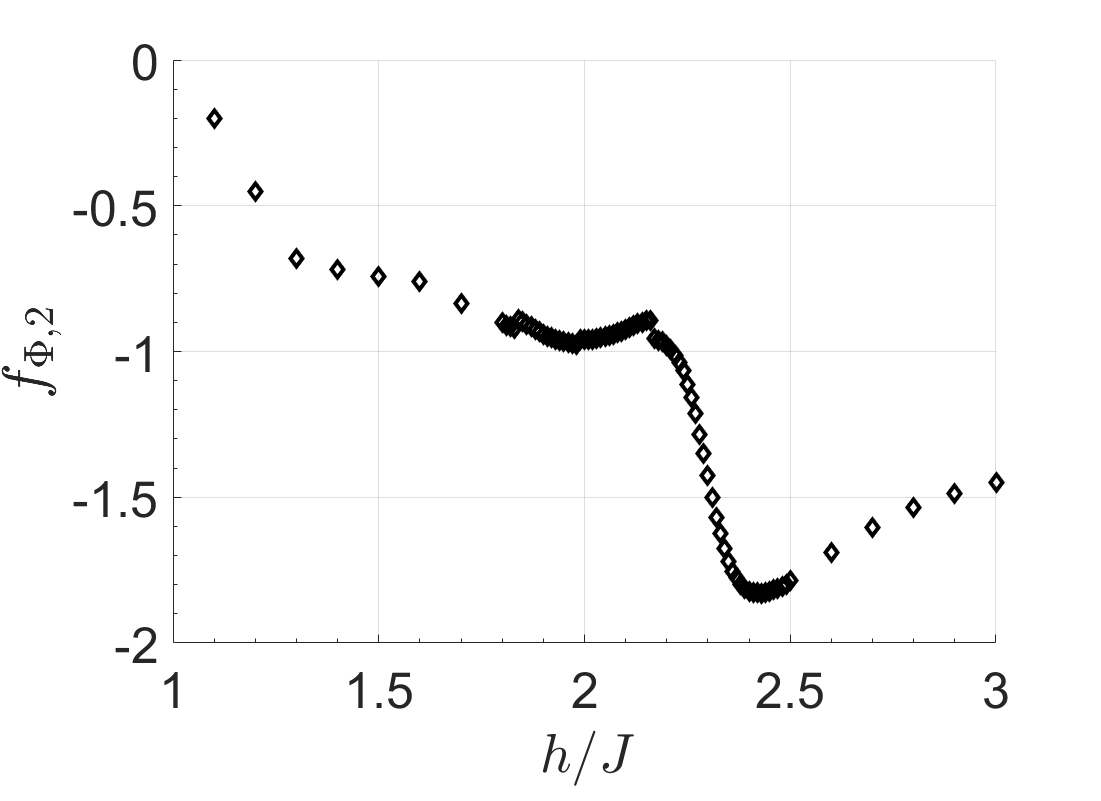}}
 \subfloat[]{\label{figS10c}
\includegraphics[width=0.24\textwidth]{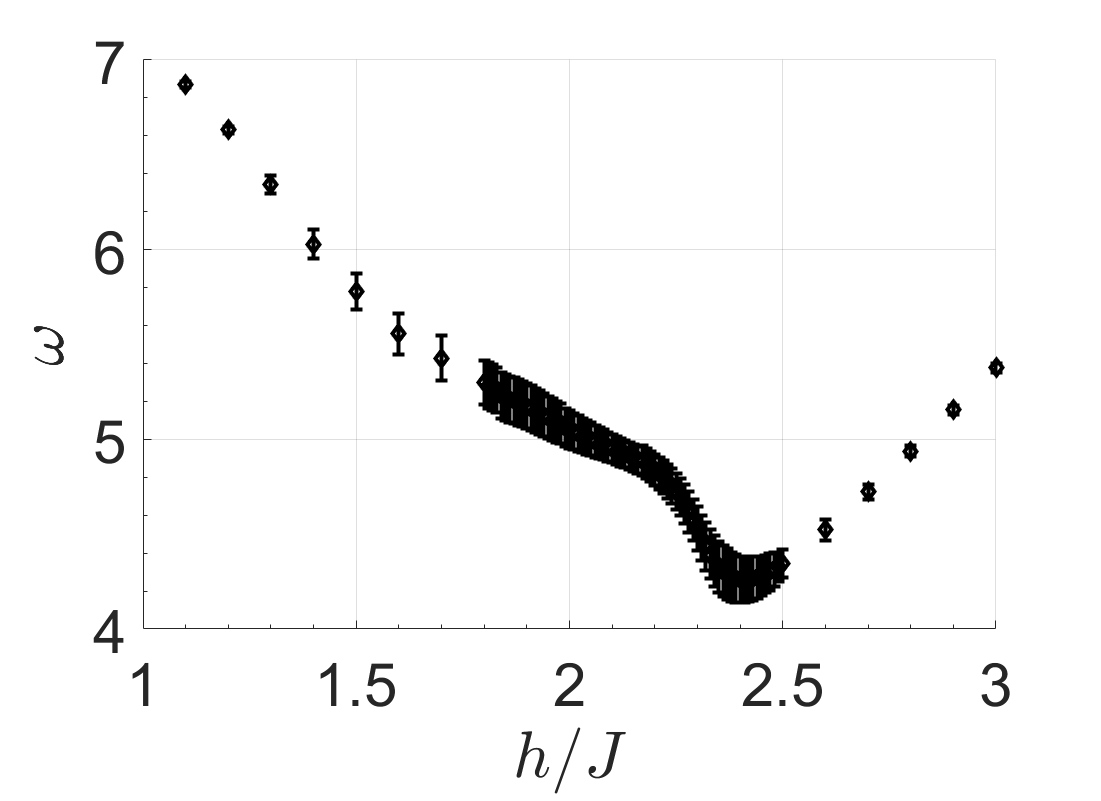}}
\caption{(a) The decay rate $f_{\Phi,2}$ and (b) the angular frequency $\omega$ of the fit function for the dynamics of nonintegrable TFIM at $\Delta/J=-1$.}
\label{FigS10}
\end{figure}

The error bars in Figs.~\ref{fig6B}, \ref{fig6C} and \ref{fig7} are calculated via error propagation and in Figs.~\ref{fig3a}, \ref{fig8a} and \ref{fig5}, they are 1$\sigma$ error bars computed via the confidence intervals of the fittings. $C_0$ is fixed parameter in Eq.~\eqref{rescaling}. In the case where one uses $\gamma_1$ parameter in the rescaling expression instead of $C(t^*)$ data, the free parameter $\gamma_1$ brings an uncertainty of $\Delta \gamma_1$ that can be computed via the confidence intervals of the fitting. Based on the data points, one can have an uncertainty from $t_L$ too: $\Delta t$ denotes this uncertainty which is calculated as the difference between $t_L$ and the available data point. Hence, we can calculate the propagation of error as,
\begin{eqnarray}
E^2=\left(\frac{\partial \text{OP}}{\partial t} \right)^2 (\Delta t)^2 + \left(\frac{\partial \text{OP}}{\partial \gamma_1} \right)^2 (\Delta \gamma_1)^2,
\end{eqnarray}
where $\text{OP}$ stands for rescaled observable, or in other words the dynamical OP-like quantity. Note that if one uses the rescaling method (i) for nonintegrable TFIM, additional terms should be added to the expression. The terms in the expression above reads
\begin{eqnarray}
\frac{\partial \text{OP}}{\partial \gamma_1} &=& -\frac{C(t)^{1/t}}{t }\gamma_1^{-1/t-1}. \notag \\
 \frac{\partial \text{OP}}{\partial t} &=& -t^{-2} \left(\frac{C(t)}{\gamma_1}\right)^{1/t} \log\left(\frac{C(t)}{\gamma_1}\right). \notag
\end{eqnarray}

\section{The rest of the fit parameters of the nonintegrable TFIM}

In this appendix section, we plot the decay rate $f_{\Phi,2}$ and angular frequency $\omega$ with respect to $h/J$ based on the fit function utilized for the nonintegrable TFIM. Fig.~\ref{FigS10} shows these fit parameters. Interestingly, both plots dip around $h/J\sim 2.41$. Whether these parameters could signal crossover physics is a question for future research. 

\end{document}